\documentclass[%
reprint,
superscriptaddress,
preprint numbers,
nofootinbib,
amsmath,amssymb,
aps,
prd
]{revtex4-2}

\usepackage{graphicx}
\usepackage{dcolumn}
\usepackage{bm}
\usepackage{mathtools}
\usepackage{hyperref}
\usepackage{tensor}
\usepackage{comment}
\usepackage{xcolor}
\hypersetup{colorlinks=true, citecolor=blue, urlcolor=blue, linkcolor=blue}
\begin{document}

\newcommand{\jvr}[1]{\textcolor{orange}{[{\bf JVR}: #1]}}
\newcommand{\gui}[1]{\textcolor{teal}{[{\bf Gui}: #1]}}

\newcommand{\aap}{Astronomy \& Astrophysics}
\newcommand{\jcap}{Journal of Cosmology and Astroparticle Physics}
\newcommand{\physrep}{Physics Reports}
\newcommand{\mnras}{Monthly Notices of the Royal Astronomical Society}

\preprint{APS/123-QED}

\title{The sound of dynamical dark energy and modified gravity}

\author{João Rebouças}
\affiliation{Department of Astronomy/Steward Observatory, University of Arizona, 933 North Cherry Avenue, Tucson, AZ 85721, USA}
\affiliation{CBPF - Brazilian Center for Research in Physics, Xavier Sigaud st. 150, zip 22290-180, Rio de Janeiro, RJ, Brazil}

\author{Guilherme Brando}
\affiliation{CBPF - Brazilian Center for Research in Physics, Xavier Sigaud st. 150, zip 22290-180, Rio de Janeiro, RJ, Brazil}

\author{Felipe T. Falciano}
\affiliation{CBPF - Brazilian Center for Research in Physics, Xavier Sigaud st. 150, zip 22290-180, Rio de Janeiro, RJ, Brazil}

\author{Vivian Miranda}
\affiliation{C. N. Yang Institute for Theoretical Physics, Stony Brook University, Stony Brook, NY, 11794, USA}

\date{\today}% It is always \today, today,

\begin{abstract}
Different candidate models are able to reproduce the dynamical dark energy signal preferred by combinations of recent distance measurements. These models may be distinguished by the behavior of their perturbations, which are controlled by the effective sound speed $c_s^2(k,a)$. To explore correlations between the dark energy sound speed and perturbative behavior, we test modified gravity (MG) scenarios in which the dark energy equation of state and sound speed determine modifications to the clustering of matter. We investigate the impact of varying the dark energy sound speed on several cosmological quantities in both General Relativity (GR) and MG. We constrain the dark energy and modified gravity parameters using measurements of the Cosmic Microwave Background (CMB) from Planck PR4, type Ia supernova luminosity distances (SN) from Pantheon+, Baryon Acoustic Oscillations (BAO) from DESI DR2, and cosmic shear from DES-Y3. Using the combination of CMB+BAO+SN, we find that, in the MG scenarios, the preference for dynamical dark energy is correlated with deviations from GR over redshifts $z < 2$ at over 95\% confidence level. The significance of these deviations is not degraded when considering a dynamical or superluminal sound speed, but vanishes if we assume a cosmological constant. The inclusion of cosmic shear and CMB lensing data significantly shifts the constraints towards GR. Our framework enables the exploration of modified gravity models using the dark energy sound speed as a physically meaningful free parameter.
\end{abstract}

\maketitle

%\tableofcontents

\section{Introduction}
\label{sec:introduction}

One of the main scientific goals of modern cosmology is to uncover the nature of dark energy. In the $\Lambda$CDM model, the current paradigm, dark energy is assumed to be a cosmological constant driving the present accelerated expansion of the Universe. Despite its simplicity, $\Lambda$CDM provides a remarkably good fit to a wide range of independent measurements, including Cosmic Microwave Background (CMB) anisotropies \cite{planck_2018_cmb, planck_pr4_results, act_cmb, act_dr6_results, spt_cmb, spt_3g_2025}, correlation functions of galaxy positions and shapes \cite{kids_legacy, desi_y1_bao, desi_dr2_bao, desi_full_shape, des_y1_3x2, des_y3_3x2, des_y6_3x2, kids-1000-shear, hsc-shear}, and the luminosity distance-redshift relation of type Ia supernovae \cite{pantheon, pantheonplus, desy5, desy5_cosmo, des_dovekie, union3, union3-1}. In recent years, with the rise of precision cosmology, $\Lambda$CDM has faced several challenges, most notably the Hubble tension between CMB and cosmic distance ladder measurements of the $H_0$ parameter \cite{sh0es_hubble, local_dist_network}. A milder discrepancy has also been reported by galaxy and CMB surveys involving the $S_8$ parameter, which describes the variance of matter fluctuations, although its significance depends on the specific datasets and systematic uncertainty modelling \cite{kids_legacy, cosmology_intertwined, s8_tension_nonlinear_solution, flamingo_s8, growth_modification_s8}.

In recent years, new distance measurements from type Ia supernovae~\cite{pantheonplus, union3, desy5_cosmo, des_dovekie} and galaxy surveys using the Baryon Acoustic Oscillation (BAO) feature~\cite{desi_y1_bao, desi_dr2_bao, desi_full_shape} have reported another inconsistency in the $\Lambda$CDM model, this time favoring a dynamical dark energy fluid over a cosmological constant. These results are based on the $w_0w_a$ model~\cite{linder_w0wa, chevallier_polarski}, in which the dark energy equation of state is a linear function of the scale factor, $w(a) = w_0 + w_a(1-a)$. This scenario has been widely used as a null test of $\Lambda$CDM, since the model reduces to a cosmological constant when $w_0 = -1$ and $w_a = 0$. While past analyses were able to recover the cosmological constant within the error bars~\cite{pantheon, planck_2018_cmb}, combinations of state-of-the-art datasets from the DESI, Union3, DES, Pantheon+, and Planck collaborations exclude the cosmological constant with statistical significance ranging from $2.8\sigma$ to $4.2\sigma$~\cite{desi_dr2_bao, des_dovekie}. The preferred dark energy behavior has a phantom ($w_{\rm DE} < -1$) phase in the past, smoothly transitioning to a non-phantom ($w_{\rm DE} > -1$) phase at redshift $z \approx 0.5$~\cite{desi_crossing, desi_joao, desi_dr2_extensions}. Whether these results represent the first hint of dynamical dark energy remains under debate \cite{liddle_desi, vivian_mirage}.

Connecting the phenomenological $w_0w_a$ model to a fundamental theory is challenging, and multiple candidates have been proposed to explain the phantom crossing behavior~\cite{desi_constraints_ide, desi_coupled_de_dm, desi_general_param, desi_interacting_de, desi_mg, desi_monodromic_de, desi_non_minimal_grav, desi_quintessential, desi_sign_switching}. The simplest fundamental models assume dark energy is an additional, minimally coupled scalar field, as in quintessence~\cite{Ratra1988, Caldwell1998} and k-essence~\cite{kessence} theories. The former cannot have a phantom equation of state, while the latter is plagued by instabilities in their perturbations when the equation of state is phantom~\cite{quintom, de_phantom_instability, de_phantom_instability_2}. More general scalar-tensor theories, such as the Effective Field Theory of Dark Energy (EFTofDE)~\cite{EFTofDE_1, EFTofDE_2, EFTofDE_review} and Horndeski theory~\cite{horndeski}, naturally introduce new couplings between the metric tensor and the scalar field in the action, and the phenomenology associated with these extra couplings can be characterized by free functions of time~\cite{bellini_sawicky}. At the same time, this extra freedom significantly enlarges the theory space, making it difficult to explore viable dark energy candidates.

While investigating modified gravity theories as possible candidates for the current acceleration of our Universe is a more first-principles-oriented approach, model-agnostic investigations remain highly valuable, especially in the era of precision cosmology, since they allow us to perform more targeted stress tests of the concordance model and to search for possible cracks within it. Within this approach, it is useful to adopt a fluid description of dark energy while going beyond the smooth dynamical dark energy paradigm, in which the dark energy fluid is described only by its equation of state. Instead, one may also allow the sound speed of the fluid to vary. It is well known that, at the perturbative level, a general perfect fluid can be characterized by its sound speed, $c_s^2$, which may depend on both time and scale, as well as by its viscosity parameter, $c_v^2(k,a)$~\cite{gdm_hu,Kodama:1984ziu}. Quintessence models are characterized by $c_s^2 = 1$, which suppresses perturbations within the cosmological horizon, and $c_v^2 = 0$, implying dark energy has no anisotropic stress. These are standard assumptions adopted in most analyses of fluid dark energy models.

In k-essence and EFTofDE scenarios, however, the sound speed may deviate significantly from $c_s^2 = 1$, allowing dark energy perturbations to contribute nontrivially to the gravitational potential. Consequently, the impact of dark energy perturbations on large-scale structure (LSS) can be used to distinguish between candidate models that otherwise predict the same background expansion history of the Universe~\cite{kevolution, nefertiti, de_perturbations_simulations, growth_mg,Ballesteros:2010ks}. The difficulty in constraining the dark energy sound speed stems from the fact that its effects become relevant only on very large scales compared to models with $c_s^2 = 1$. Previous studies have shown that the impact of the late-time dark energy sound speed on cosmological observables is generally small~\cite{cs2_wmap_bean, cs2_cmb_forecasts, cs2_xia_2007, cs2_hannestad, cs2_huterer_linder}. This conclusion, however, may not hold in modified gravity scenarios, where the effective dark energy sound speed can be sourced by additional terms in the gravitational action that modify the clustering properties of matter~\cite{bellini_sawicky, mg_sims, kazuya_theoretical_priors_mg}.

In this work, we propose a new parametrized approach to testing the dark energy sound speed by investigating its impact on the growth of cosmic structure. As mentioned above, simple minimally coupled scalar-field models such as quintessence and k-essence are generically prone to instabilities when crossing the phantom divide, $w_{\rm DE}=-1$. In Ref.~\cite{Deffayet:2010qz}, it was shown that introducing self-interaction terms in the scalar-field action, commonly referred to as kinetic gravity braiding (KGB), causes the stress-energy tensor to deviate from the perfect-fluid form. This feature allows the scalar field to cross the phantom divide without developing instabilities. 

In this class of models, deviations from General Relativity are characterized by two time-dependent functions, $\alpha_{\rm K}$ and $\alpha_{\rm B}$, which are directly related to the scalar-field sound speed. Our goal is therefore to move away from the standard covariant approach to modified gravity, in which one specifies functional forms for $\alpha_{\rm K}$ and $\alpha_{\rm B}$, and instead adopt a phenomenological perspective in which the dark energy sound speed is treated as the fundamental varying quantity. In practice, we still parameterize the kineticity function, $\alpha_{\rm K}$, while deriving $\alpha_{\rm B}$ from the imposed sound-speed evolution. A similar approach was recently employed in~\cite{kazuya_mg_phantom}, where a general time evolution for $c_s^2$ was assumed. This approach ensures that dark energy perturbations remain stable even with a phantom equation of state. In this modified gravity scenario, we use data from CMB, BAO, supernovae, and cosmic shear, to obtain novel constraints on the dark energy equation of state, sound speed, and deviations from General Relativity (GR).

This paper is organized as follows: Section~\ref{sec:models} discusses the theoretical background of the sound speed in scalar-field theories and presents our phenomenological relation; Section~\ref{sec:analysis} describes the datasets used in our work and the methodology used to extract cosmological parameter constraints from data; Section~\ref{sec:results} presents and discusses the results of the data analysis; and Section~\ref{sec:conclusions} summarizes the discussion and our findings.

\section{Theory}
\label{sec:models}

\subsection{Fluid Dark Energy}

We assume a perturbed FRW Universe in the scalar Newtonian gauge, neglecting vector and tensor perturbations,
\begin{equation}
    ds^2 = a^2(\tau)\left[ -(1 + 2\Psi)d\tau^2 + (1 - 2\Phi)\delta_{ij}dx^idx^j \right].
\end{equation}
We parameterize the dark energy energy-momentum tensor as
\begin{subequations}
    \begin{align}
        T\indices{^0_0} &= -\rho - \delta\rho, \\
        T\indices{^i_0} &= -(\rho + P)v^i, \\
        T\indices{^i_j} &= \delta\indices{^i_j}(P + \delta P) + P\Pi\indices{^i_j},
    \end{align}
\end{subequations}
where $\rho$ and $P = w_\mathrm{DE}\rho$ are the background density and pressure, $\delta\rho$ is the density perturbation, $\delta P$ is the pressure perturbation, $v^i$ is the bulk velocity, and $\Pi^{i}_j$ is the anisotropic stress. The energy-momentum conservation equations do not determine the evolution of $\delta P$, so an additional parameterization is needed to fully specify the microphysics of the dark energy fluid. At the linear perturbation level, one such parameterization is defined as follows. The pressure perturbation can be decomposed into adiabatic and non-adiabatic (or entropic) parts,
\begin{equation}
    \delta P = \frac{P'}{\rho'}\delta\rho + \delta P_s,
\end{equation}
where the first term on the right-hand side is the adiabatic pressure perturbation, and the second is the entropic pressure perturbation, which is gauge-independent. Therefore, $\delta P_s$ must be parameterized in terms of gauge-independent perturbations.
One simple but widely adopted model is~\cite{gdm_hu}
\begin{equation}\label{eq:effective_cs}
    \delta P_s = c_s^2\left[ \delta\rho + 3\mathcal{H}(\rho + P)\frac{v}{k} \right],
\end{equation}
where $c_s^2$ is a free function of the scale factor $a$ and Fourier wavenumber $k$, and the term inside square brackets is the comoving density perturbation, which is gauge-independent. This effective sound speed parameterization is recovered in scalar field theories.

%\jvr{TODO: trocar essa discussão para PPF.} With this definition, the energy-momentum conservation equations read~\cite{gdm_hu}
%\begin{align}
%    \delta' = &-(1 + w)\left( \theta - 3\Phi'\right) - 3\mathcal{H}(c_a^2 + c_s^2 - w) \nonumber\\ &- 9\mathcal{H}^2(1 + w)c_s^2\frac{v}{k}, \label{eq:continuity} \\
%    \theta' = &-(1 - 3c_s^2)\mathcal{H}\theta + \frac{c_s^2k^2}{1 + w}\delta + k^2\Psi \label{eq:euler}
%\end{align}
%where $\theta = kv$ and $c_a^2 = P'/\rho'$. This effective sound speed parameterization, given in Equation~\ref{eq:effective_cs}, is recovered in scalar field theories.
%Non-adiabatic perturbations modify the initial conditions of the dark energy perturbations~\cite{cs2_initial_conditions}. In terms of the initial condition of the synchronous gauge density perturbations, $\delta_c = C(k\tau)^2$, the initial conditions for dark energy perturbations are
%\begin{subequations}
%    \begin{align}
%        \delta &= C(1 + w)\frac{4 - 3c_s^2}{4 - 6w + 3c_s^2}(k\tau)^2, \\
%        \theta &= C\frac{c_s^2}{4 - 6w + 3c_s^2}(k\tau)^3k.
%    \end{align}
%\end{subequations}
%Reference~\cite{cs2_initial_conditions} claims that the initial conditions may be set to zero without loss of accuracy, since the perturbations quickly reach an attractor. We verified that this is indeed the case, but we follow the recommendation of including this initial condition.

The standard energy-momentum conservation equations are unstable if the equation of state is dynamical and happens to cross $w = -1$ at some redshift. An alternative for the fluid perturbation equations is the Parameterized Post-Friedmann (PPF) approach~\cite{ppf_phantom_crossing}, a reparameterization of the perturbation variables that has been shown to agree with the fluid equations with great precision and can be applied to phantom crossing fluids. The new variable $\Gamma$ is a proxy to the dark energy density perturbations in the matter rest frame, and its evolution equation is given by
\begin{equation}
    (1 + c_\Gamma^2 k_H^2)[\Gamma' + (1 + c_\Gamma^2 k_H^2)\Gamma] = \frac{4\pi G(1 + w_\mathrm{DE})\rho_\mathrm{DE}v_T}{k_HH^2},
\end{equation}
where $k_H = k/(aH)$, $v_T$ is the total bulk velocity of all species except for dark energy, and $c_\Gamma$ is a parameter controlling the transition scale where dark energy becomes smooth with respect to dark matter. Ref.~\cite{ppf_phantom_crossing} sets $c_\Gamma = 0.4c_s$ to match predictions from phantom crossing scalar field models.

\subsection{Modified Gravity}
\label{sec:cs2mg}

Horndeski theory provides the most general scalar-tensor gravity model with second-order equations of motion~\cite{horndeski}, encompassing many models for dark energy from quintessence~\cite{Frieman1995, Ratra1988, Caldwell1998} and k-essence~\cite{kessence} to more extreme modifications of gravity such as galileons~\cite{galileon, galileon_planck} or $f(R)$~\cite{buchdahl, hu_sawicki_fr}. The theory is summarized by the general lagrangian,

\begin{equation}
    S = \int d^4x\sqrt{-g}\left( \sum_{i=2}^5 \mathcal{L}_i + \mathcal{L}_m \right),
\end{equation}
where $\mathcal{L}_m$ is the matter lagrangian and
\begin{subequations}
    \begin{align}
        \mathcal{L}_2 &= G_2,\\
        \mathcal{L}_3 &= -G_3\Box\phi,\\
        \mathcal{L}_4 &= G_4R + G_{4, X}[(\Box\phi)^2 - \nabla_\mu \nabla_\nu\phi\nabla^\mu \nabla^\nu\phi],\\
        \mathcal{L}_5 &= G_5G_{\mu\nu}\nabla^\mu \nabla^\nu\phi - \frac{1}{6}G_{5,X}[(\Box\phi)^3 \nonumber\\
        &+ 2\nabla^\mu\nabla_\nu\phi \nabla^\nu\nabla_\rho\phi\nabla^\rho\nabla_\mu\phi\nonumber\\
        &- 3\nabla_\mu \nabla_\nu\phi\nabla^\mu \nabla^\nu\phi\Box\phi],
    \end{align}
\end{subequations}
where $G_i(\phi, X)$ are arbitrary functions of the scalar field $\phi$ and its kinetic term $X = -\partial_\mu\phi\partial^\mu\phi/2$, and commas represent partial derivatives. We adopt the parameterization from Ref.~\cite{bellini_sawicky}, where the linear cosmological perturbations can be determined by the background expansion history and four additional functions of time: $\alpha_{\rm K}$, $\alpha_{\rm B}$, $\alpha_{\rm M}$ and $\alpha_{\rm T}$; their specific definitions can be found in Appendix A of~\cite{bellini_sawicky}. In this case, the dark energy sound speed is given by~\cite{bellini_sawicky, kazuya_mg_phantom, kazuya_theoretical_priors_mg, mg_sims, growth_mg}
\begin{align}\label{eq:cs2_mg}
    c_s^2 =  \frac{1}{D_\mathrm{kin}} \biggr[ &(2 - \alpha_{\rm B})\left(-\frac{H'}{aH^2} + \frac{\alpha_{\rm B}}{2}(1 + \alpha_{\rm T}) + \alpha_{\rm M} - \alpha_{\rm T}\right) \nonumber\\ &- \frac{8\pi G(\rho_\mathrm{no \, DE} + P_\mathrm{no \, DE})}{H^2M_*^2} + \frac{\alpha_{\rm B}'}{aH}\biggr],
\end{align}
where $D_\mathrm{kin} = \alpha_{\rm K} + 3\alpha_{\rm B}^2/2$, $\alpha_{\rm M} = d\ln M_*^2/d\ln a$ and the subscript ``$\mathrm{no \, DE}$" refers to the sum of all species excluding dark energy.

The function $\alpha_{\rm T}$ controls the deviation of the sound speed of gravitational waves in these theories, and it has been tightly constrained~\cite{alpha_T_constraints}. Within our work, we will assume $\alpha_{\rm T} = 0$. For simplicity, we further assume $\alpha_{\rm M} = 0$: this assumption is compatible with a minimally coupled scalar field with $G_4 = 1/2$ and $G_5 = 0$. With these assumptions, the slip parameter $\eta = \Phi/\Psi$ is equal to unity, indicating the absence of dark energy anisotropic stress. While anisotropic stress is an important component of the dark energy microphysics, we leave the case of $\alpha_{\rm M} \neq 0$ for future work.

\subsection{Sound Speed and Growth}

While standard analyses of modified gravity parameterize the functions $\alpha_i$~\cite{scalar_tensor_future}, we take a different approach: we invert Equation~\ref{eq:cs2_mg} to obtain a differential equation for the braiding function $\alpha_{\rm B}$ in terms of $\alpha_{\rm K}$ and $c_s^2$. A similar approach was employed in Ref.~\cite{kazuya_mg_phantom}, where both $c_s^2$ and $D_\mathrm{kin}$ were parameterized using a Gaussian Process model on top of power-law parametrizations. After setting $\alpha_{\rm M} = \alpha_{\rm T} = 0$, we obtain
\begin{align}\label{eq:braiding_ode}
    \frac{d\alpha_{\rm B}}{d\ln a} = c_s^2\left(\alpha_{\rm K} + \frac{3}{2}\alpha_{\rm B}^2\right)& + \frac{\alpha_{\rm B}^2}{2} - \alpha_{\rm B}\left(\frac{d\ln H}{d \ln a} + 1\right)\nonumber\\
    &- 3(1+w_\mathrm{DE})\Omega_\mathrm{DE}(a),
\end{align}
where $\Omega_\mathrm{DE}(a) = 8\pi G \rho_\mathrm{DE}(a)/3H^2$. Given an initial condition $\alpha_{{\rm B},i}$ at a high redshift $z_i$ and specific functional forms for $c_s^2$ and $\alpha_{\rm K}$, one can fully determine the function $\alpha_{\rm B}$. Throughout this work, we choose an initial redshift $z_i = 10^5$ with initial condition $\alpha_{\rm B}(z_i) = 0$.

We consider four parametrizations for $\alpha_{\rm K}$:
\begin{itemize}
    \item Proportional to $\alpha_{\rm B}$:
    \begin{align}
        \alpha_{\rm K} = 3\lambda_{\rm MG}\,\alpha_{\rm B};
    \end{align}

    \item Proportional to the dark energy abundance:
    \begin{align}
        \alpha_{\rm K} =
        \lambda_{\rm MG}
        \frac{\Omega_{\rm DE}(a)}{\Omega_{\rm DE,0}};
    \end{align}

    \item K-essence-like:
    \begin{align}
        \alpha_{\rm K} =
        \frac{\Omega_{\rm DE}(a)(1+w_{\rm DE})}{c_s^2};
    \end{align}
    inspired by k-essence models~\cite{bellini_sawicky};

    \item Cubic Galileon-like:
    \begin{align}
        \alpha_{\rm K} =
        6\lambda_{\rm MG}
        \frac{H_0^4\,\Omega_{\rm DE}(a=1)}{H(a)^4},
    \end{align}
    motivated by Cubic Galileon theories~\cite{galileon, galileon_planck, mochi_class}.
\end{itemize}

The first two parametrizations are purely phenomenological, while the latter two are motivated by scalar--tensor theories, since they reproduce the functional form of $\alpha_{\rm K}$ appearing in their covariant formulations. We refer to them as ``k-essence-like'' and ``Cubic Galileon-like'' because only the form of $\alpha_{\rm K}$ is inherited from these models, while the functions $\alpha_{\rm B}$ and $w_{\rm DE}$ differ fundamentally from those of their covariant counterparts. In particular, k-essence models satisfy $\alpha_{\rm B}=0$, while in Cubic Galileon theories one has $\alpha_{\rm B}=\alpha_{\rm K}/3$; the latter motivates us to parameterize the first case as $\alpha_{\rm K} = 3\lambda_{\rm MG}\,\alpha_{\rm B}$. Furthermore, in the Cubic Galileon-like parametrization we also introduce the amplitude parameter $\lambda_{\rm MG}$ in order to test possible deviations from the theoretical prediction, such that $\lambda_{\rm MG}=1$ recovers the standard Cubic Galileon form for $\alpha_{\rm K}$. We do not introduce an additional rescaling parameter in the k-essence-like case, since variations in $c_s^2$ already control the overall amplitude of $\alpha_{\rm K}$.

For the dark energy sound speed, we consider two cases: a constant sound speed and a time-dependent parametrization,
\begin{align}
    c_s^2(a) = c_{s,0}^2 + c_{s,a}^2(1-a),
\end{align}
analogous to the commonly adopted CPL parametrization for the dark energy equation of state,
\begin{align}
    w_\mathrm{DE}(a) = w_0 + w_a(1-a).
\end{align}
Within this modified gravity framework, we may also allow for superluminal dark energy sound speeds, \textit{i.e.} $c_s^2 > 1$. Although in the present framework the dark energy stress-energy tensor takes the form of an imperfect fluid~\cite{Deffayet:2010qz}, we still expect this parametrization to capture the overall time evolution of $c_s^2$ in a sufficiently general manner.

Given our assumptions of $\alpha_{\rm T} = \alpha_{\rm M} = 0$, the braiding and kineticity functions alter the Einstein equations for the metric as~\cite{mg_sims}
\begin{subequations}\label{eq:mg_mu_sigma}
    \begin{align}
        k^2\Psi &= -4\pi G_N \mu(a) a^2\rho\delta, \label{eq:mu}\\
        k^2(\Phi + \Psi) &= -8\pi G_N \Sigma(a)a^2\rho\delta\label{eq:sigma},
    \end{align}
\end{subequations}
where $G_N$ is the usual Newton's constant. While the general form of $\mu$ and $\Sigma$ is scale-dependent, we assume the quasi-static approximation~\cite{mg_sims} holds in the scales of interest. In this approximation, if $\alpha_\mathrm{M} = \alpha_\mathrm{T} = 0$, we have
\begin{equation}\label{eq:mg_mu_from_alpha}
    \mu(a) = \Sigma(a) = 1 + \frac{\alpha_{\rm B}^2}{2c^2_{s, \mathrm N}},
\end{equation}
where $c^2_{s, \mathrm N}$ is the numerator from Equation~\ref{eq:cs2_mg} (\textit{i.e.} the term inside square brackets).

Perturbations are stable if and only if $c_s^2 > 0$ and $D_\mathrm{kin} = \alpha_{\rm K} + 3\alpha_{\rm B}^2/2 > 0$~\cite{bellini_sawicky, mg_sims}. The first condition can be straightforwardly enforced by imposing the prior $c_s^2 > 0$. The second condition is trivially satisfied if $\alpha_{\rm K} > 0$, which holds if we assume $\alpha_{\rm K} \propto \Omega_\mathrm{DE}$ or the Cubic Galileon-like parametrization with $\lambda_\mathrm{MG} > 0$. For the k-essence-like parametrization, perturbations are stable if $w > -1$. Stability can still be achieved in the phantom regime if $3\alpha_{\rm B}^2/2 > |\alpha_{\rm K}|$, but this condition is automatically violated for the initial condition $\alpha_{\rm B} = 0$. Therefore, when analyzing k-essence-like models, we impose $w > -1$. In the case $\alpha_{\rm K} \propto \alpha_{\rm B}$, it is not trivial to ensure stability a priori. Therefore, during the numerical integration of Equation~\ref{eq:braiding_ode}, we check the value of $D_\mathrm{kin}$ and raise an error if it violates the stability condition, effectively excluding unstable models from the analysis.

We implemented Equations~\ref{eq:braiding_ode} and~\ref{eq:mg_mu_sigma} in a modified version of CAMB%
\footnote{\url{https://github.com/joaoreboucas1/CAMB-cs2}}.
After solving Equation~\ref{eq:braiding_ode}, we compute $\mu$ using Equation~\ref{eq:mg_mu_from_alpha}, with the replacement
\[
c^2_{s,\mathrm{N}} = c_s^2\left(\alpha_{\rm K}+\frac{3}{2}\alpha_{\rm B}^2\right).
\]

An important consequence of Equation~\ref{eq:mg_mu_from_alpha} is that the GR value, $\mu(a)=1$, can only be recovered if either $\alpha_{\rm B}=0$, corresponding to a pure k-essence theory, or if $c^2_{s,\mathrm{N}}\to\infty$. In this work, however, we treat $c_s^2$ as a free parameter, while $D_{\rm kin}$ is fixed by the evolution of $\alpha_{\rm K}$ and $\alpha_{\rm B}$. Therefore, in all physical scenarios considered here, the combination $c_s^2D_{\rm kin}$ remains finite. As a result, the recovery of GR-like growth should be understood as the limit in which deviations from $\mu=1$ become small, rather than as an exact return to GR.

\begin{figure*}
    \centering
    \includegraphics[width=\linewidth]{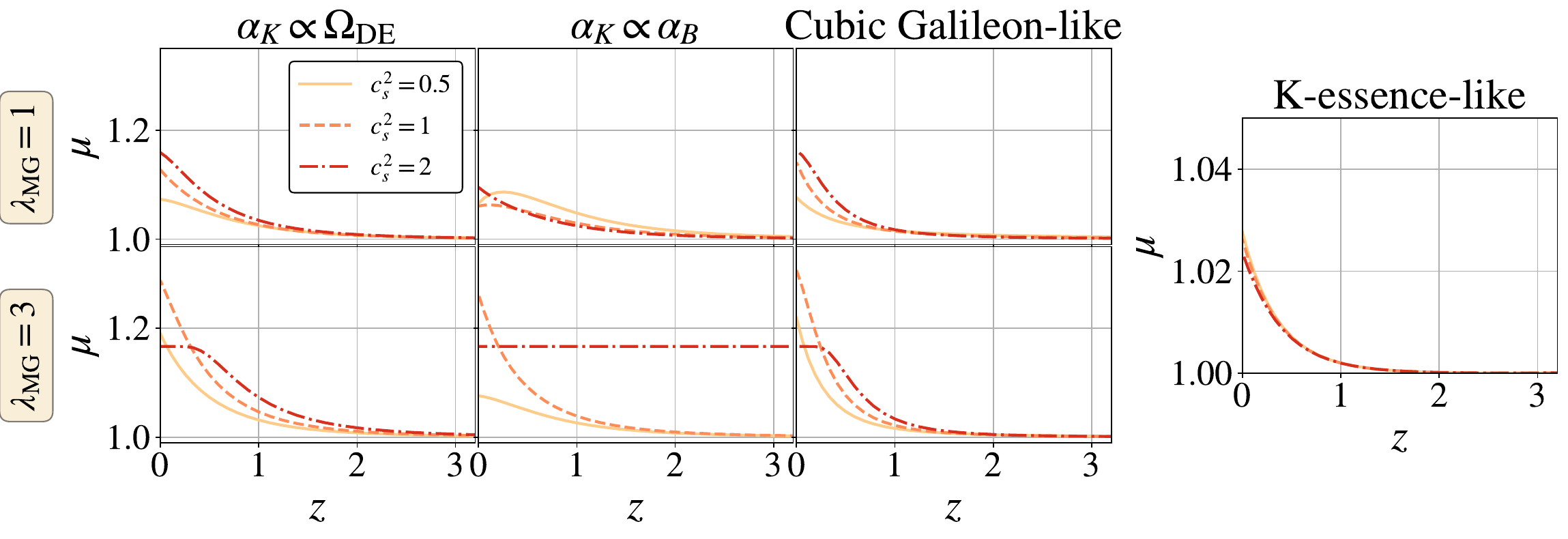}
    \caption{Solutions for $\mu$ from Equation~\ref{eq:braiding_ode} for several values of $c_s^2$ and $\lambda_\mathrm{MG}$, for the four parameterizations of $\alpha_{\rm K}$ considered, assuming $w_0 = -0.838$ and $w_a = -0.62$. The k-essence-like case is not parameterized by $\lambda_\mathrm{MG}$, and we assume $w_a = -0.162$ to stabilize the perturbations. In the case of $\alpha_\mathrm{K}$ with $c_s^2 = 2$ and $\lambda_\mathrm{MG} = 3$, $\mu$ starts deviating from unity around $z = 20$.}
    \label{fig:mu_examples}
\end{figure*}

Figure~\ref{fig:mu_examples} shows solutions for $\mu$ for several values of $\lambda_\mathrm{MG}$ and $c_s^2$, assuming $w_0 = -0.838$ and $w_a = -0.62$. For the k-essence case, we assume $w_a = -0.162$ in order to stabilize the perturbations. In typical cases, $\mu$ is close to unity at redshifts $z > 5$, similar to other phenomenological models for modified gravity in which $\mu \propto \Omega_\mathrm{DE}$~\cite{mu_sigma_amendola, mu_sigma_caldwell, planck_2015_extensions, des_y3_extensions}. Furthermore, since our priors strongly favor $\alpha_{\rm K} > 0$, it follows from Equation~\ref{eq:mg_mu_from_alpha} that $\mu \geq 1$. This qualitative behavior is shared by the four parameterizations of $\alpha_{\rm K}$. The strength of the deviation from $\mu = 1$ and the redshift at which this deviation becomes significant depend on the values of $c_s^2$ and $\lambda_\mathrm{MG}$. In general, higher values of $\lambda_\mathrm{MG}$ induce stronger and earlier deviations from GR. Figure~\ref{fig:mu_examples} also shows that increasing $c_s^2$ leads to earlier deviations from GR. On the other hand, Equation~\ref{eq:mg_mu_from_alpha} implies that $\mu$ is inversely proportional to $c_s^2$.

From Equation~\ref{eq:braiding_ode}, the late-time evolution of $\alpha_{\rm B}$ is tied to the dark energy equation of state. Interestingly, for the $\alpha_{\rm K} \propto \alpha_{\rm B}$ and k-essence-like parameterizations, $\alpha_{\rm B}$ is equal to zero if dark energy is a cosmological constant, leading to $\mu = 1$ and no deviations from GR. For the other parametrizations, the deviations from GR are reduced when we consider a cosmological constant rather than dynamical dark energy. This is why the k-essence-like case in Figure~\ref{fig:mu_examples} shows weaker deviations from GR.

In more extreme cases, such as the one with $\alpha_\mathrm{K} \propto \alpha_\mathrm{B}$, $c_s^2 = 2$ and $\lambda_\mathrm{MG} = 3$ shown in Figure~\ref{fig:mu_examples}, $\mu$ may deviate from unity at higher redshifts. In that specific case, this happens at $z \approx 20$. For higher values of $c_s^2$ and $\lambda_\mathrm{MG}$, the evolution of $\mu$ may happen even earlier, potentially affecting recombination. In the following analysis, we confirmed that the models preferred by data do not alter recombination.

\subsection{Impact on observables}
We assess the impact of the dark energy sound speed in the CMB and matter power spectra, considering both the PPF and modified gravity scenarios, assuming a dynamical dark energy cosmology with $w_0 = -0.838$ and $w_a = -0.62$, representative of the observed signal~\cite{desi_dr2_bao}. For the modified gravity effects, we consider the parameterization $\alpha_{\rm K} \propto \Omega_\mathrm{DE}$ with $\lambda_\mathrm{MG} = 1$ shown in Figure~\ref{fig:mu_examples}.

\begin{figure*}
    \centering
    \includegraphics[width=0.5\linewidth]{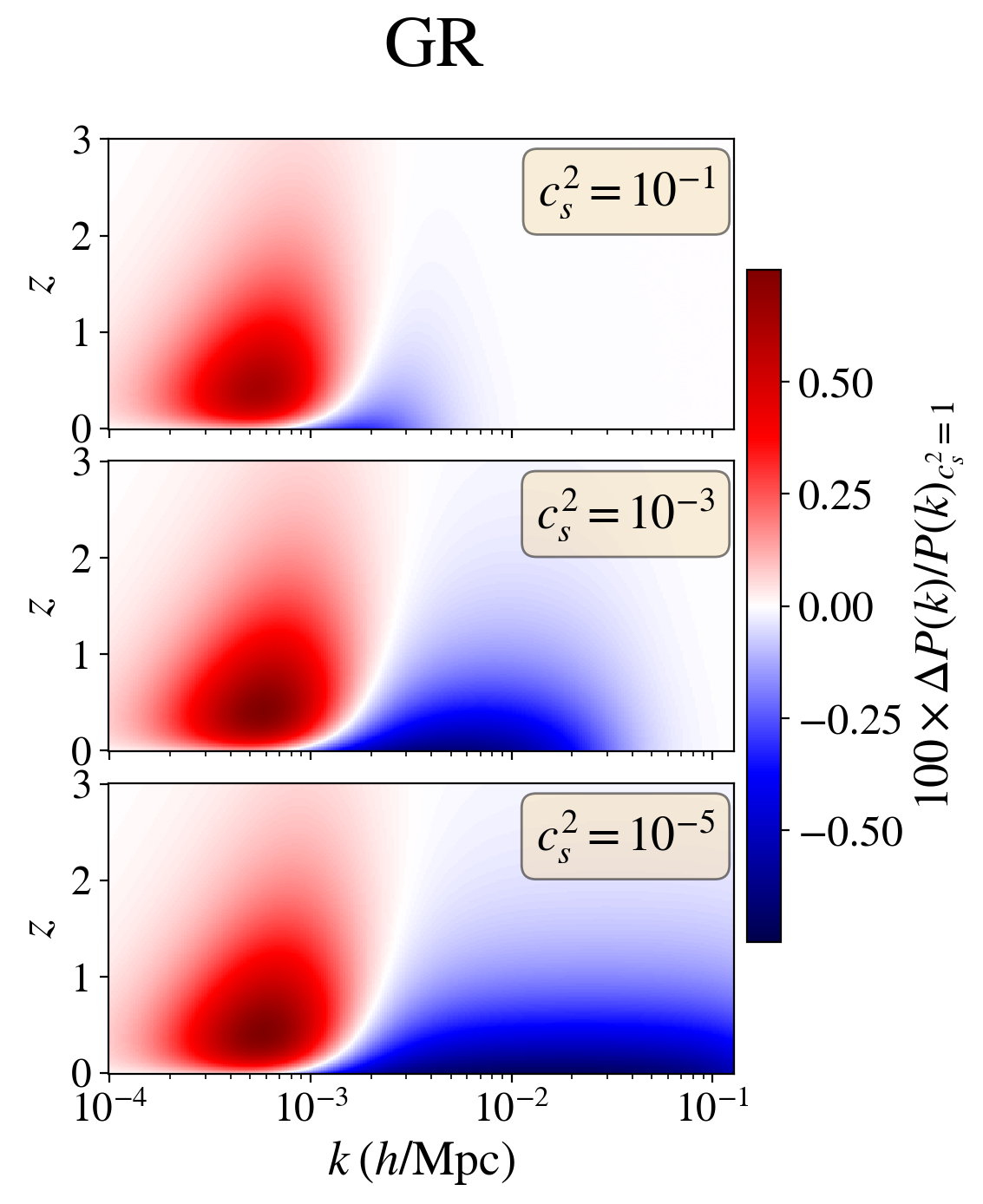}
    \includegraphics[width=0.48\linewidth]{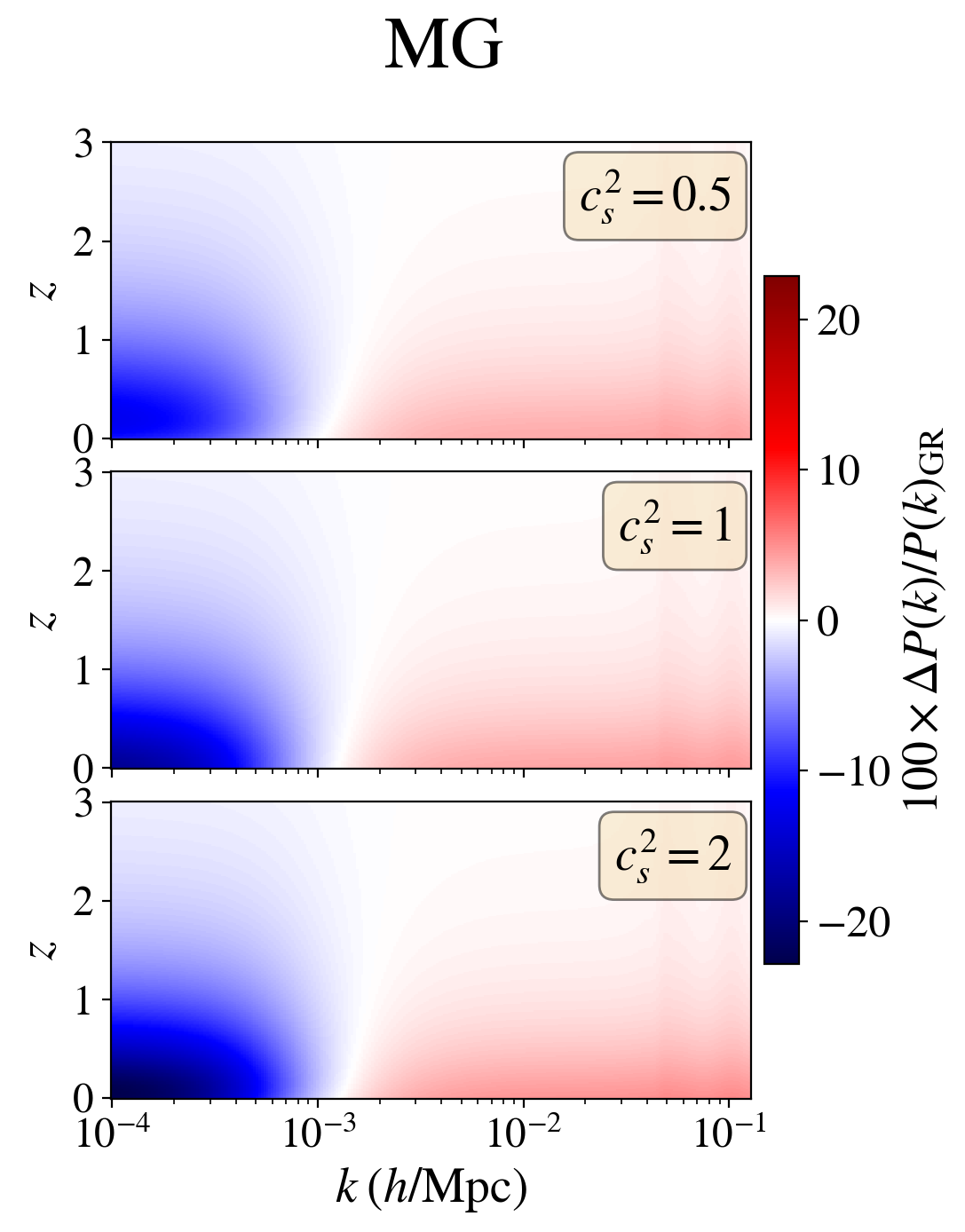}
    \caption{Relative changes in the matter power spectrum for PPF dark energy in GR (left panels) and MG with $\alpha_{\rm K} \propto \Omega_\mathrm{DE}$ (right panels), compared to the GR case where $c_s^2 = 1$. Here, we assume $w_0 = -0.878$ and $w_a = -0.67$.}
    \label{fig:cs2_impact_pk}
\end{figure*}

In Figure~\ref{fig:cs2_impact_pk}, we show the impact of varying the dark energy sound speed on the matter power spectrum $P(k, z)$. For the GR case shown in the left panels, lowering the dark energy sound speed impacts the matter power spectrum by less than 1\% with respect to the case where $c_s^2 = 1$. Matter perturbations at scales $k > 10^{-3} h/\mathrm{Mpc}$ are suppressed, while those at scales $k < 10^{-3} h/\mathrm{Mpc}$ are enhanced. As the the sound speed gets lower, the suppression affects increasingly smaller scales. The MG case shows an opposite effect on the matter power spectrum, enhancing it at small scales, $k > 10^{-3} h/\mathrm{Mpc}$, by $\approx 5\%$, while suppressing it at large scales, $k < 10^{-3} h/\mathrm{Mpc}$, by $10-20\%$ depending on redshift and $c_s^2$.

\begin{figure*}
    \centering
    \includegraphics[width=0.49\linewidth]{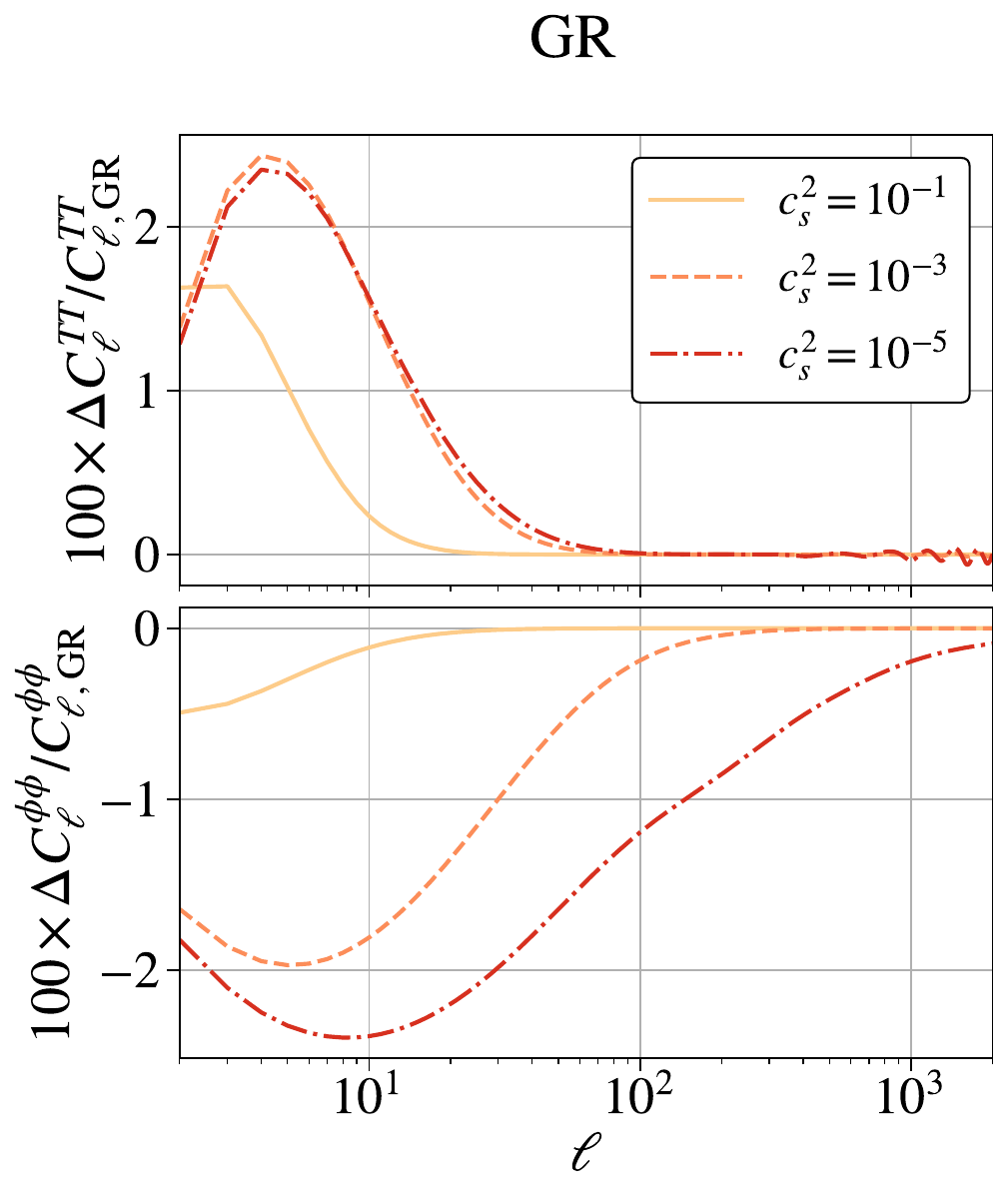}
    \includegraphics[width=0.485\linewidth]{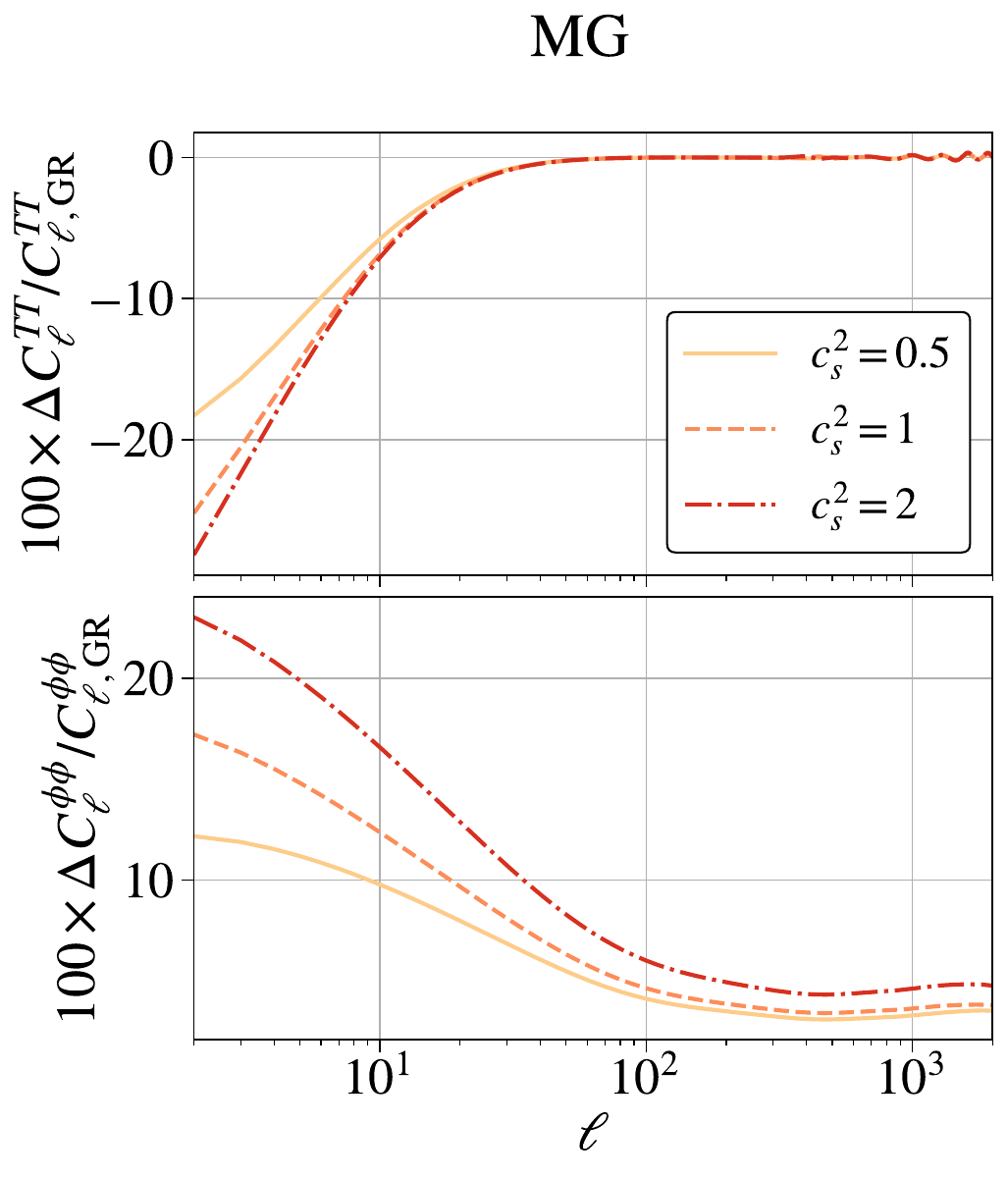}
    \caption{Relative changes on the CMB anisotropy power spectra for PPF dark energy in GR (left panels) and and MG with $\alpha_{\rm K} \propto \Omega_\mathrm{DE}$ (right panels), compared to the GR case where $c_s^2 = 1$. We show the effects on TT (top panels) and $\phi\phi$ (bottom panels) spectra.}
    \label{fig:cs2_impact_cl}
\end{figure*}

The impact of lowering the sound speed on the CMB anisotropies is similar, as shown in Figure~\ref{fig:cs2_impact_cl}. In GR, the temperature power spectrum is enhanced at large scales dominated by cosmic variance, varying by at most $2.5\%$, while the lensing spectrum is suppressed by the same amount. In the MG case, the effect on the temperature power spectrum is larger, suppressing multipoles $\ell < 100$ by around $20\%-30\%$. In the lensing power spectrum, we find an overall $\lesssim 5\%$ increase at multipoles $\ell > 100$, which can become as large as $25\%$ at small multipoles. Therefore, as already assessed by previous works~\cite{cs2_cmb_forecasts, cs2_hannestad, cs2_huterer_linder, cs2_wmap_bean, cs2_xia_2007, cs2_sergijenko}, the small impact of the dark energy sound speed in GR hinders its detectability with current and future observations. However, in MG models, the relation between $w_\mathrm{DE}$, $c_s^2$, and the MG functions $\mu$ and $\Sigma$ has the potential to provide useful constraints, as we investigate in the following.

\section{Data Analysis}
\label{sec:analysis}

\subsection{Datasets}
\label{sec:datasets}

\subsubsection{Planck PR4 CMB}

We use data from Planck PR4, implemented in the \texttt{HiLLiPoP} and \texttt{LoLLiPoP} likelihoods~\cite{planck_pr4_results}. The \texttt{HiLLiPoP} likelihood contains TT, TE and EE angular power spectra from multipoles $\ell = 30$ up to $\ell = 2500$ (for TT) and $\ell = 2000$ (for TE and EE). The \texttt{LoLLiPoP} likelihood contains EE angular power spectra for multipoles ranging from $\ell \in [2, 29]$. For the large-scale TT power spectrum, we use the \texttt{Commander} likelihood. We additionally include CMB lensing from Planck PR4 \texttt{NPIPE} likelihood~\cite{planck_pr4_lensing} as well as the cross-correlation between lensing and the integrated Sachs-Wolfe effect (ISWL, \cite{planck_iswl}).

In the following, the CMB primary spectra dataset will be referred to simply as CMB, and the combination of primary spectra, CMB lensing and ISWL will be denoted as CMBL.

\subsubsection{DESI DR2 BAO}

The DESI collaboration has measured the Baryon Acoustic Oscillation scale relative to the baryon drag epoch for several tracers of the large-scale structure, such as galaxies, quasars and the Lyman-$\alpha$ forest. The effective redshifts of the tracers samples span the redshift range $0.30 < z < 2.33$. In the following, this dataset will be simply referred to as BAO.

\subsubsection{DES-Dovekie}

We use luminosity distances from $\approx 1600$ type Ia supernovae catalogued by the DES collaboration in 5 years of observations. In particular, we use the updated magnitudes published by DES-Dovekie, a recalibration of the photometric redshifts of the DES-Y5 supernovae catalog~\cite{des_dovekie}. While this recalibration has reduced the statistical significance of preference for dynamical dark energy from $4.2\sigma$ to $3.2\sigma$, it also improved the agreement with other supernovae datasets such as Pantheon+ and Union3.1~\cite{pantheonplus, union3-1}. This dataset will be referred to as SN.

\subsubsection{DES-Y3 Cosmic Shear}

We use cosmic shear real-space angular correlation functions $\xi_\pm(\theta)$ calculated from the Dark Energy Survey Year 3 dataset~\cite{des_y3_shear_amon, des_y3_shear_secco}. In the following, this dataset will be referred to as CS. The correlations are measured from the \textsc{metacalibration} shape catalog~\cite{des_y3_metacal}, consisting of approximately 100 million galaxies, with an effective number density of $n_\mathrm{eff} = 5.59 \; \mathrm{arcmin}^{-1}$. The galaxies in this catalog are divided in 4 tomographic redshift bins with number density per redshift $n^i(z)$. The angular correlation functions are measured in 20 logarithmically-spaced angular bins in the range $2.5' < \theta < 250'$.

To make theoretical predictions for $\xi_\pm$, we use the \texttt{CosmoLike} code~\cite{cosmolike}. After calling \texttt{CAMB}, we can obtain the matter power spectrum $P_m(k, z)$, the comoving distances $\chi(z)$, and the MG functions $\mu(z) = \Sigma(z)$. With that information, we can compute the Fourier-space correlations using the Limber approximation,
\begin{equation}\label{eq:cl_kk}
    C_{\kappa\kappa, \ell}^{ij} = \int_0^{\chi_\mathrm{H}} d\chi \frac{q^i_\kappa(\chi)q^j_\kappa(\chi)}{\chi^2}\Sigma^2(\chi)P_m \left( \frac{\ell + 1/2}{\chi}, z(\chi) \right),
\end{equation}
where $\chi_\mathrm{H}$ is the comoving horizon and $q^i_\kappa(\chi)$ is the lensing efficiency for the sources in tomographic bin $i$,
\begin{equation}
    q^i_\kappa(\chi) = \frac{3H_0^2\Omega_m}{2}\frac{\chi}{a(\chi)}\int_\chi^{\chi_\mathrm{H}} d\chi' \frac{\chi' - \chi}{\chi'}n^i(z(\chi'))\frac{dz}{d\chi'}.
\end{equation}
Note that the Weyl potential $\Phi + \Psi$ and matter density are related by Equation~\ref{eq:sigma}, thus warranting an extra factor of $\Sigma^2(\chi)$ in Equation~\ref{eq:cl_kk} compared to the GR case.

With the harmonic space angular power spectrum, we find the real space power spectrum by convolving with Legendre polynomials,
\begin{equation}
    \xi_\pm^{ij}(\theta) = \sum_\ell \frac{2\ell + 1}{4\pi}C_{\kappa\kappa, \ell}^{ij} G_{\ell,2}^{\pm}(\cos\theta),
\end{equation}
where $G^{\pm}_{\ell,2}$ are defined in \textit{e.g.} Ref.~\cite{stebbins}.

The theoretical prediction is modified to account for several systematic effects:
\begin{itemize}
    \item Photometric redshift uncertainties: to account for uncertainties in the photometric redshift calibration, we distort the galaxy number densities with a shift parameter $\Delta z^i$ such that $n^i(z) \rightarrow n^i(z + \Delta z^i)$. The shift parameters, one for each tomographic bin, are sampled in the MCMC.
    \item Multiplicative shear bias: the measurement of the shear from the galaxy catalog is subject to systematic errors. To account for these errors, we scale our shear predictions as $\xi^{ij}_\pm \rightarrow (1 + m^i)(1 + m^j)\xi^{ij}_\pm$ and marginalize over the parameters $m^i$.
    \item Intrinsic alignment (IA): unlensed galaxies' shapes can be correlated due to local density and tidal fields. This accounts for extra contributions in the harmonic space shear power spectrum: $C_{\kappa\kappa, \ell}^{ij} \rightarrow C_{\kappa\kappa, \ell}^{ij} + C_{\kappa I, \ell}^{ij} + C_{I\kappa, \ell}^{ij} + C_{II, \ell}^{ij}$, where $C_{\kappa I, \ell}^{ij}$ are the correlations between cosmic shear at bin $i$ and IA at bin $j$ and $C_{II, \ell}^{ij}$ is the IA autocorrelation. We use the Nonlinear Linear Alignment (NLA) model, where the intrinsic shape field is sourced by the gravitational tidal field. Thus, the extra correlations are given by
    \begin{subequations}
    \begin{align}
    C_{\kappa I,\ell}^{ij}
    &=
    \int_0^{\chi_H} d\chi\,
    \frac{n^i(\chi)q^j(\chi)}{\chi^2}\,
    A_{\rm IA}(\chi)\Sigma(\chi)\mu(\chi)
    \nonumber\\
    &\hspace{2.0cm}\times
    P_m\!\left(
    \frac{\ell+1/2}{\chi},z(\chi)
    \right),
    \\
    C_{II,\ell}^{ij}
    &=
    \int_0^{\chi_H} d\chi\,
    \frac{n^i(\chi)n^j(\chi)}{\chi^2}\,
    A_{\rm IA}^2(\chi)\mu^2(\chi)
    \nonumber\\
    &\hspace{2.0cm}\times
    P_m\!\left(
    \frac{\ell+1/2}{\chi},z(\chi)
    \right).
    \end{align}
    \end{subequations}
    where $A_\mathrm{IA}$ is a redshift-dependent amplitude of the intrinsic alignment effect. Again, note that the gravitational potential $\Psi$ and matter density are related by Equation~\ref{eq:mu}, and thus we include an additional $\mu(\chi)$ factor when relating the tidal field $s_{ij} = (k_ik_j/k^2 - \delta_{ij}/3)\Psi$ to the matter density~\cite{hirata_ia}
    
    We model the time-dependence of the IA amplitude as a power law given by
    \begin{equation}
        A_\mathrm{IA}(z) = A_1C_1\rho_\mathrm{cr}\left(\frac{1 + z}{1 + z_p}\right)^{\eta_1},
    \end{equation}
    where $A_1$ and $\eta_1$ are parameters characterizing the IA redshift evolution, $C_1 = 5 \times 10^{-14} M_\odot h^{-2} \mathrm{Mpc}^2$ is a conventional constant to scale the amplitude based on IA measurements from the SuperCOSMOS survey~\cite{measuring_ia}, $\rho_\mathrm{cr}$ is the critial energy density today, and $z_p$ is an arbitrary pivot redshift.
\end{itemize}

Priors for the nuisance parameters $\Delta z^i$, $m^i$, $A_1$ and $\eta_1$ are shown in Table~\ref{tab:priors}.

\subsection{Modelling}

We assume dark energy is a cosmological fluid whose background dynamics is defined by an equation of state parameterized by
\begin{equation}
    w_\mathrm{DE}(a) = w_0 + w_a(1-a).
\end{equation}

For the perturbative dynamics, we investigate the following scenarios involving the sound speed $c_s^2$:
\begin{itemize}
    \item Fiducial: we use the PPF equations for dark energy perturbations assuming $c_s^2 = 1$.
    \item Modified Gravity: we solve Equation~\ref{eq:braiding_ode}, finding $\alpha_{\rm B}$ and $\mu$, and explicitly changing the metric equations according to Equations~\ref{eq:mg_mu_sigma}. We assume the four parameterizations for $\alpha_{\rm K}$ described in Section~\ref{sec:cs2mg}. Our standard choice is to assume a constant, subluminal sound speed $c_s^2 < 1$. We also test the effect of a superluminal sound speed, where we allow $c_s^2 < 10$, and a dynamical sound speed, where $c_s^2 = c_{s,0}^2 + c_{s,a}^2(1-a)$.
\end{itemize}

We do not use any nonlinear matter power spectrum prescription in our analysis. For DES-Y3 cosmic shear, we make use of linear scale cuts following the methodology of~\cite{planck_2015_extensions, des_y3_extensions}. Furthermore, the CMB primary spectra are mostly insensitive to nonlinear effects. For CMB lensing, while neglecting nonlinear effects can lead to small biases in $\Lambda$CDM parameters, erroneously using the nonlinear corrections from \texttt{HMCode2020}~\cite{hmcode2020} can lead to numerical artifacts in the power spectrum prediction~\cite{cmb_pheno_hard}. We have performed our analysis with both cases and confirmed that the cosmological constraints agree, implying that our results have limited sensitivity to nonlinear effects. However, we have noticed that MCMCs ran using \texttt{HMCode2020} have been plagued with stuck walkers and slow convergence attributed to the erroneous nonlinear prescription.

\subsection{Analysis}

We use \texttt{Cocoa}\footnote{\url{https://github.com/CosmoLike/cocoa}}, the Cobaya-Cosmolike Joint Architecture~\cite{cosmolike}, as our framework for cosmological analysis. For sampling the posterior distributions, we use the Metropolis-Hastings algorithm implemented in \texttt{Cobaya}, using the Gelman-Rubin criterion with $|R-1| < 0.03$ to assess convergence. We sample over the standard cosmological parameters $\{\theta_*, \Omega_b h^2, \Omega_c h^2,\ln(10^{10}A_s), n_s, \tau\}$, as well as the dark energy parameters $\{w_0, w_a, c_s^2 \}$. For the modified gravity scenarios where $\alpha_{\rm K}$ is parameterized, we additionally sample over $\lambda_\mathrm{MG}$. The cosmological parameter priors are described in Table~\ref{tab:priors}. We also sample over the CMB and supernovae nuisance parameters following the collaborations' recommendations. We consider two dataset combinations: CMB+BAO+SN, and CMBL+BAO+SN+CS.

\begin{table}[t]
    \centering
    \begin{tabular}{| c | c || c | c |}
        \hline
        \multicolumn{2}{|c||}{Cosmology} & \multicolumn{2}{c|}{DES-Y3 Nuisance} \\\hline
        \hline
         Parameter  &  Prior & Parameter & Prior \\\hline
        $100\times\theta_*$ & $\mathcal{U}[0.5, 10]$ & $\Delta z^1$ & $\mathcal{N}[0, 0.018]$\\\hline
        $\Omega_bh^2$ & $\mathcal{U}[0.005, 0.1]$ & $\Delta z^2$ & $\mathcal{N}[0, 0.015]$\\\hline
        $\Omega_ch^2$ & $\mathcal{U}[0.001, 0.99]$ & $\Delta z^3$ & $\mathcal{N}[0, 0.011]$\\\hline
        $\ln(10^{10}A_s)$ & $\mathcal{U}[1.61, 3.91]$ & $\Delta z^4$ & $\mathcal{N}[0, 0.017]$\\\hline
        $n_s$ & $\mathcal{U}[0.8, 1.2]$ & $100\times m^1$ & $\mathcal{U}[-0.6, 0.9]$\\\hline
        $\tau$ & $\mathcal{U}[0.01, 0.8]$ & $100\times m^2$ & $\mathcal{U}[-2.0, 0.8]$\\\hline
        $w_0$ & $\mathcal{U}[-3(-1), -1/3]$ & $100\times m^3$ & $\mathcal{U}[-2.4, 0.8]$\\\hline
        $w_0 + w_a$ & $\mathcal{U}[-3(-1), 1]$ & $100\times m^4$ & $\mathcal{U}[-3.7, 0.8]$\\\hline
        $c_s^2$ & $\mathcal{U}[0, 1(10)]$ & $A_1$ & $\mathcal{U}[-5, 5]$\\\hline
        $c_{s,0}^2 + c_{s,a}^2$ & $\mathcal{U}[0, 1(10)]$ & $\eta_1$ & $\mathcal{U}[-5, 5]$\\\hline
        $\lambda_\mathrm{MG}$ & $\mathcal{U}[0, 3]$ & \multicolumn{2}{|c|}{--} \\\hline
    \end{tabular}
    \caption{Priors for cosmological and nuisance parameters sampled in the MCMCs for different modelling choices. Values in parenthesis denote different prior choices tested in this work.}
    \label{tab:priors}
\end{table}

\section{Results and Discussion}
\label{sec:results}

\begin{figure}
    \centering
    \includegraphics[width=\linewidth]{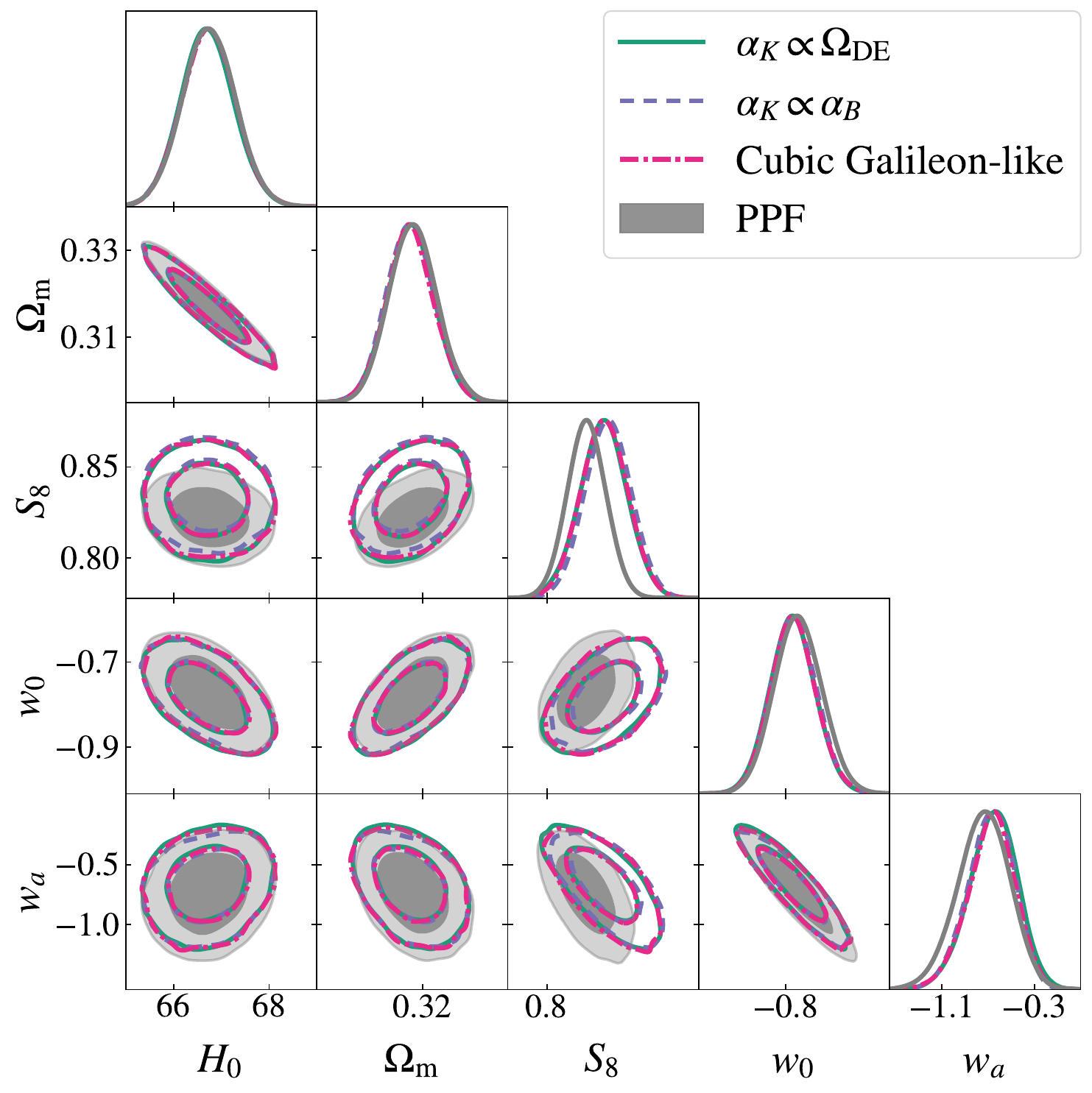}
    \caption{Cosmological parameter posterior contours plots (68\% and 95\% credible regions). The gray contour shows the GR case and the colored contours show the MG cases with three different parameterizations for $\alpha_{\rm K}$: proportional to $\Omega_\mathrm{DE}$ (green, solid), proportional to $\alpha_{\rm B}$ (purple, dashed), and the cubic galileon (pink, dotted-dashed).}
    \label{fig:triangle_mg_subluminal}
\end{figure}

\begin{figure}
    \centering
    \includegraphics[width=\linewidth]{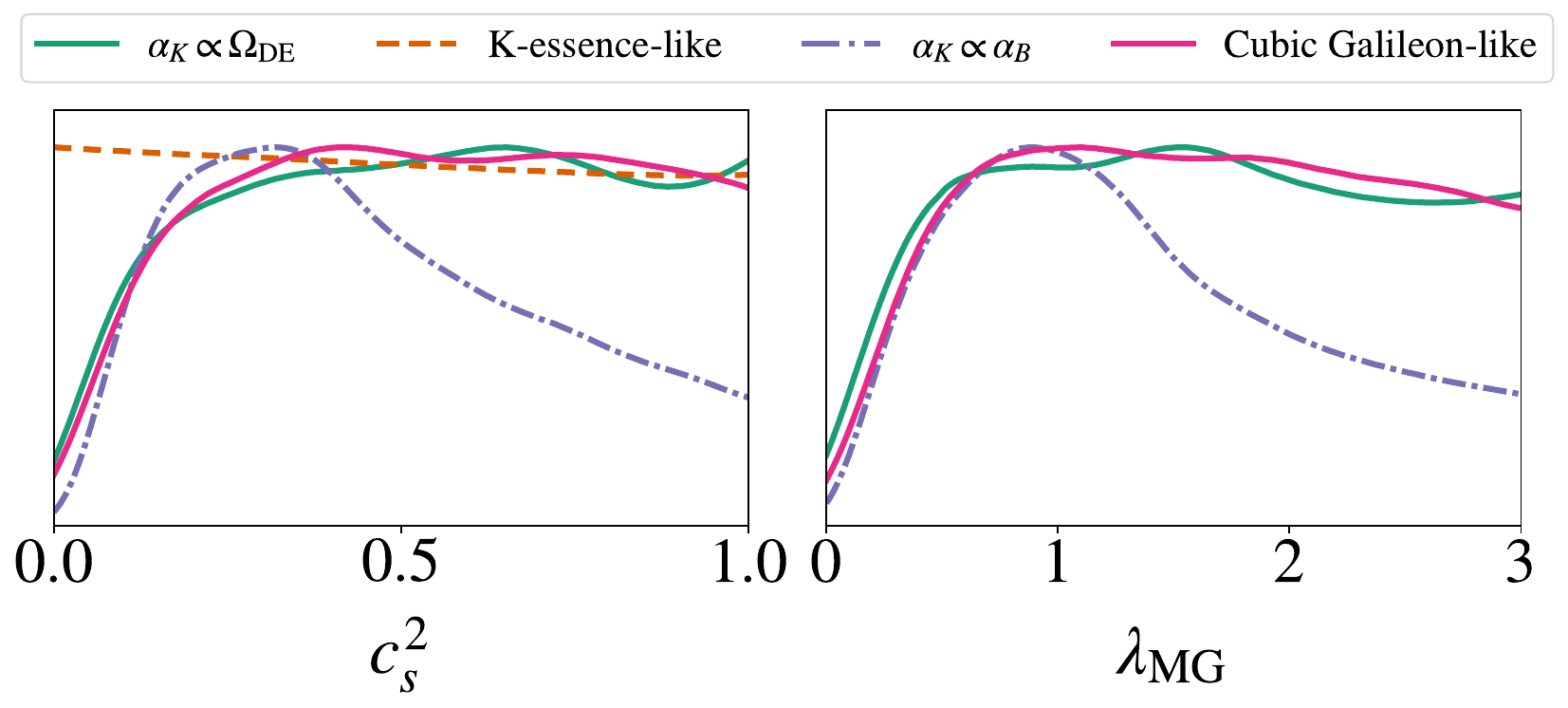}
    \caption{1D marginalized constraints on the perturbative parameters $c_s^2$ and $\lambda_\mathrm{MG}$.}
    \label{fig:cs2_lambda_ds1}
\end{figure}

\begin{figure}
    \centering
    \includegraphics[width=\linewidth]{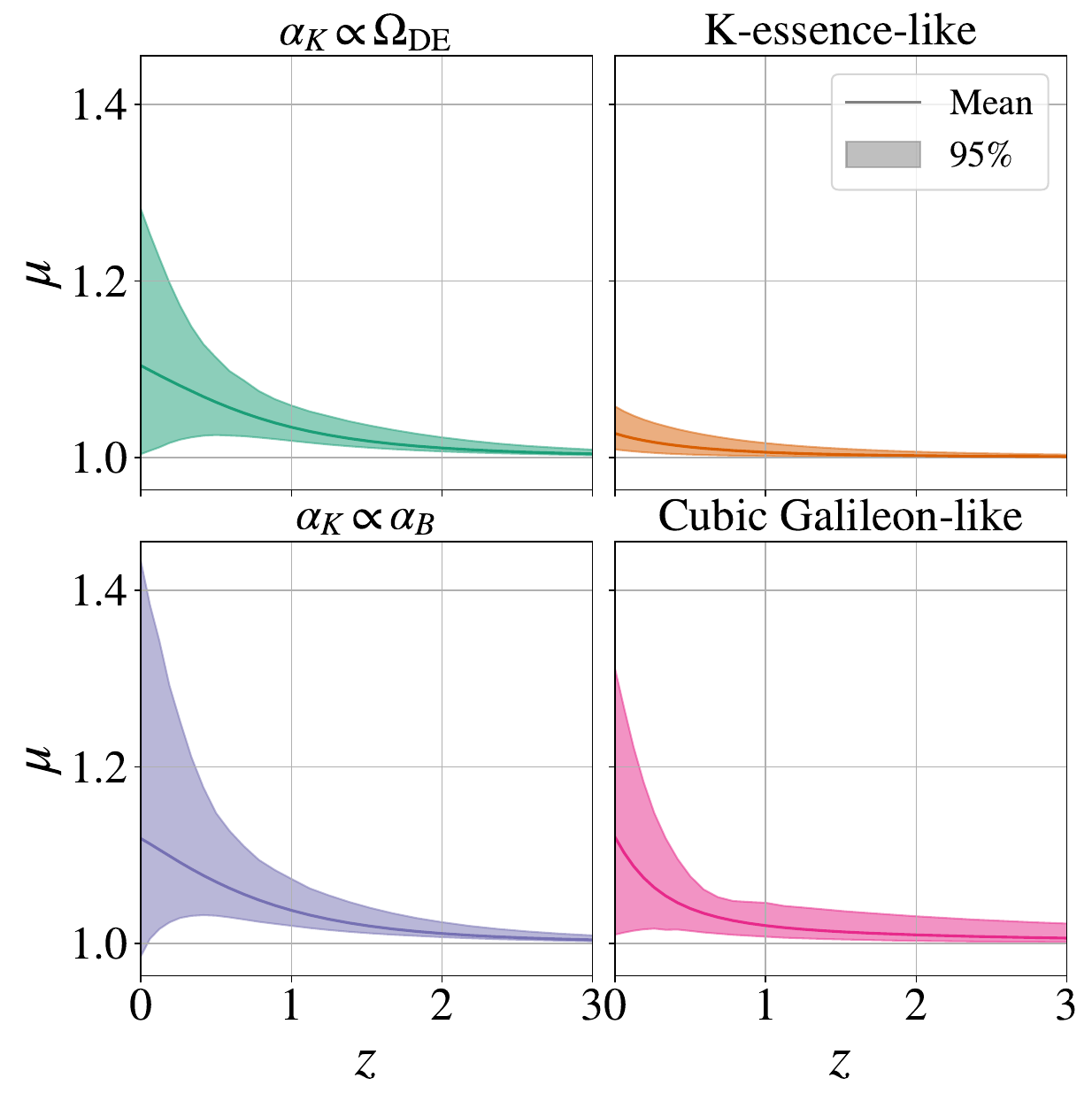}
    \caption{Constraints on $\mu(a)$ for the combination of CMB+BAO+SN, for all parameterizations of $\alpha_{\rm K}$ considered. The thick line shows the average while the bands show 95\% confidence intervals.}
    \label{fig:mu_constraint}
\end{figure}

We begin by discussing cosmological parameter constraints for the CMB+BAO+SN dataset combination. Figure~\ref{fig:triangle_mg_subluminal} shows constraints on the standard cosmological parameters, as well as on $w_0$ and $w_a$. Results for the k-essence-like parameterization are shown in Appendix~\ref{app:kessence_results}, since this case requires the additional priors $w_0 > -1$ and $w_0 + w_a > -1$. The marginalized constraints, reported as means and 68\% confidence intervals, are shown in Table~\ref{tab:all_constraints}.

We first note that the cosmological parameter constraints in the MG cases agree with one another, showing that the three different parametrizations have similar qualitative behavior with respect to the $\mu$ and $\Sigma$ functions. The MG and GR cases yield consistent constraints on $H_0$ and $\Omega_m$. The $S_8$ constraints obtained in MG are slightly shifted towards higher values because $\mu \geq 1$ in our MG models, enhancing the formation of structure. The marginalized constraints, reported as means and 68\% confidence intervals, are
\begin{itemize}
    \item $S_8 = 0.823 \pm 0.011$ (GR),
    \item $S_8 = 0.832 \pm 0.013$ (MG, $\alpha_{\rm K} \propto \Omega_\mathrm{DE}$),
    \item $S_8 = 0.836 \pm 0.013$ (MG, $\alpha_{\rm K} \propto \alpha_{\rm B}$),
    \item $S_8 = 0.832 \pm 0.013$ (MG, Cubic Galileon-like).
\end{itemize}

Interestingly, due to the direct influence of the dark energy equation of state on the growth of structure, the constraints on $w_0$ and $w_a$ are slightly tighter in the MG case than in the GR case. The dark energy figure of merit, defined as the reciprocal of the area of the 95\% confidence region in the $w_0-w_a$ posterior~\cite{detf}, is increased by approximately $20\%$ in the MG models. We find 68\% confidence intervals of
\begin{itemize}
    \item $w_0 = -0.77 \pm 0.06, \; w_a = -0.74 \pm 0.24$ (GR),
    \item $w_0 = -0.78 \pm 0.05, \; w_a = -0.66^{+0.22}_{-0.19}$ (MG, $\alpha_{\rm K} \propto \Omega_\mathrm{DE}$),
    \item $w_0 = -0.78 \pm 0.05, \; w_a = -0.66^{+0.20}_{-0.17}$ (MG, $\alpha_{\rm K} \propto \alpha_{\rm B}$),
    \item $w_0 = -0.79\pm 0.05, \; w_a = -0.65^{+0.22}_{-0.19}$ (MG, Cubic Galileon-like).
\end{itemize}

Figure~\ref{fig:cs2_lambda_ds1} shows marginalized constraints on $c_s^2$ and $\lambda_\mathrm{MG}$. The $c_s^2$ posteriors are mostly flat, indicating weak constraining power. As shown in Figure~\ref{fig:cs2_impact_cl}, the impact of the MG models on the CMB is concentrated at low multipoles dominated by cosmic variance. In all cases except the k-essence-like model, the posterior is mostly flat across the prior, but disfavors low values of $c_s^2$ that tend to increase $\mu$. These cases, however, remain consistent with $c_s^2 = 0$ at 95\% confidence level. Finally, the $c_s^2$ posterior is completely flat for the k-essence-like case. Similar remarks apply to the posterior distribution of the phenomenological $\lambda_\mathrm{MG}$ parameter. Interestingly, for the case where $\alpha_{\rm K} \propto \alpha_{\rm B}$, purple dot-dashed curves, the 1D posterior is consistent with $\lambda_{\rm MG}=1$, which is the limit where this parametrization recovers the relation $\alpha_\mathrm{K} = 3\alpha_\mathrm{B} $ from the Cubic Galileon model~\cite{mochi_class}.

In Figure~\ref{fig:mu_constraint}, we show constraints (mean and 95\% limits) on $\mu(a)$ for the four parameterizations of $\alpha_{\rm K}$. Interestingly, for all models, we find a deviation from $\mu = 1$ at over $2\sigma$ confidence level at redshifts $z < 2$. Except for the k-essence-like model, which has non-phantom priors on $w_0$ and $w_a$, all phenomenological parameterizations agree very well with one another, with present-day mean values $\mu_0 \coloneq \mu(z = 0) \approx 1.1$; the k-essence-like parameterization instead gives $\mu_0 \approx 1.04$.

\begin{figure}
    \centering
    \includegraphics[width=\linewidth]{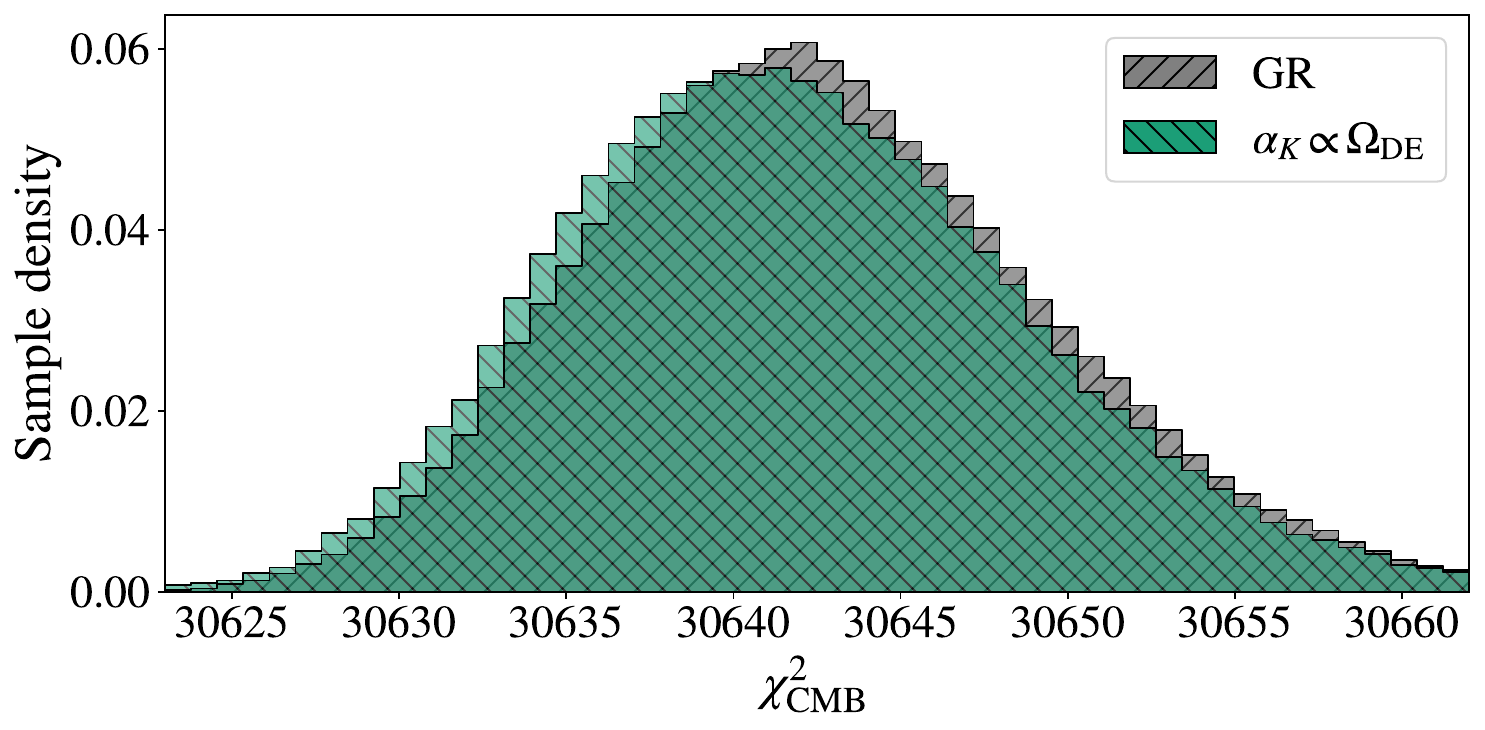}
    \caption{Distribution of $\chi^2_\mathrm{CMB}$ in the MCMCs. The blue histogram shows the MG case, and the orange histogram shows the GR case.}
    \label{fig:chi2_cmb}
\end{figure}

We interpret the evidence for $\mu > 1$ as being driven by the preference for a phantom dynamical dark energy rather than by an improvement in the CMB fit. First, we remark that our theory space is mostly restricted to $\mu \geq 1$. The growth of $\alpha_{\rm B}$ is driven by dark energy at low redshifts when $\Omega_\mathrm{DE}(a)$ is non-negligible and $w_\mathrm{DE} \neq -1$. Since the CMB+BAO+SN data combination has a significant preference for dynamical dark energy, this drives the late-time evolution of $\alpha_{\rm B}$, leading to deviations from GR. This also explains why the k-essence-like parameterization shows a weaker deviation from $\mu = 1$: because of its restriction to $w_{\rm DE} \geq -1$, its constraints are compatible with $\Lambda$CDM within the $2\sigma$ confidence level, and dark energy remains constant until thawing at $z \approx 0.5$. To test this hypothesis, we first compare the distribution of $\chi^2_\mathrm{CMB}$ between the GR and MG cases, using $\alpha_{\rm K} \propto \Omega_\mathrm{DE}$ as a representative model. Figure~\ref{fig:chi2_cmb} shows the distribution of $\chi^2_\mathrm{CMB}$ in both cases. While the MG $\chi^2_\mathrm{CMB}$ is shifted towards lower $\chi^2_\mathrm{CMB}$ compared to the GR case, this is expected due to the inclusion of two extra parameters, $c_s^2$ and $\lambda_\mathrm{MG}$. The shift of $\Delta\chi^2_\mathrm{CMB} \approx 1$ is small relative to the additional degrees of freedom in the MG model. We conclude that the MG effects do not produce any significant improvement in the CMB goodness-of-fit.

\begin{figure*}
    \centering
    \includegraphics[width=0.32\linewidth]{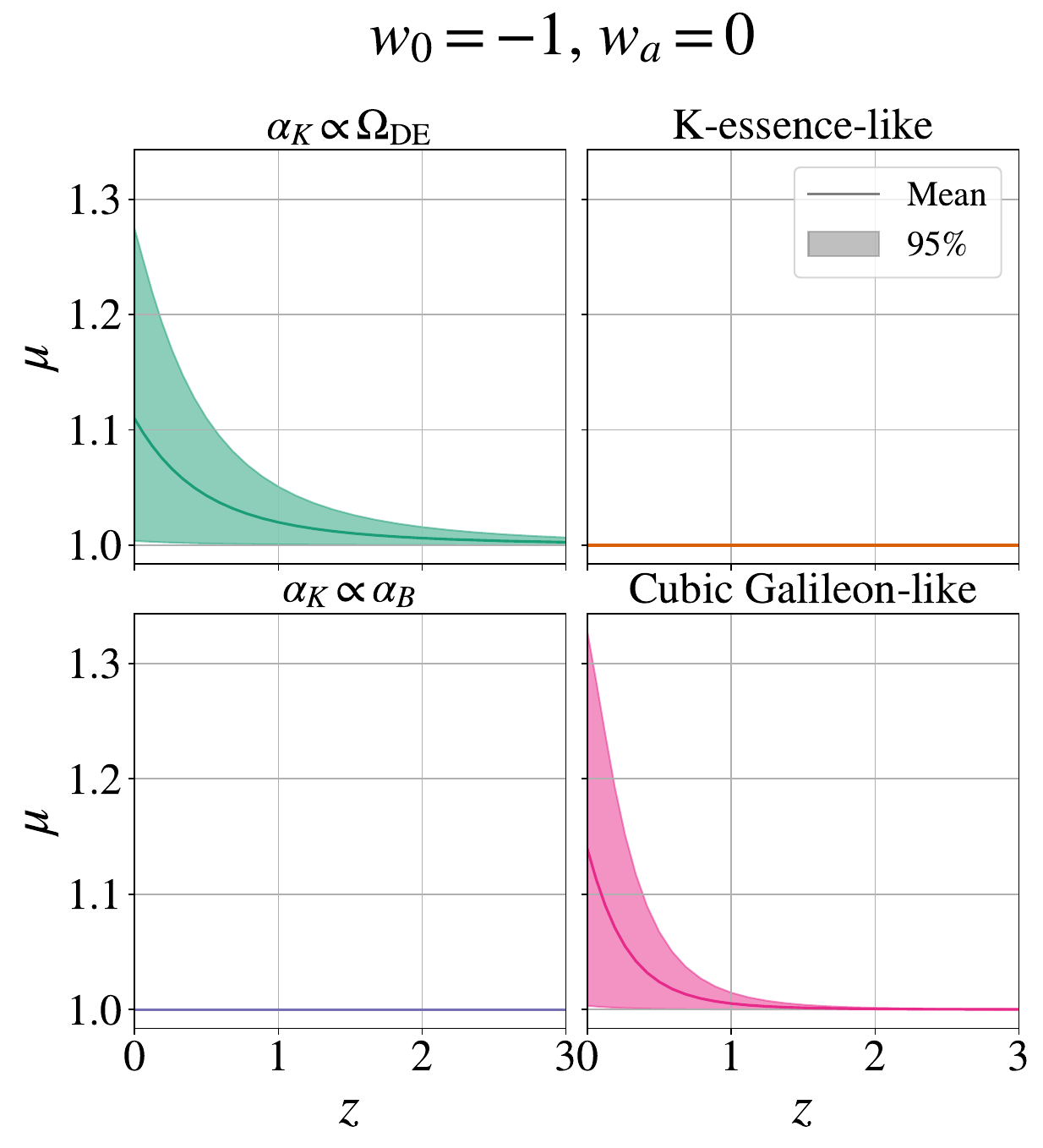}
    \includegraphics[width=0.32\linewidth]{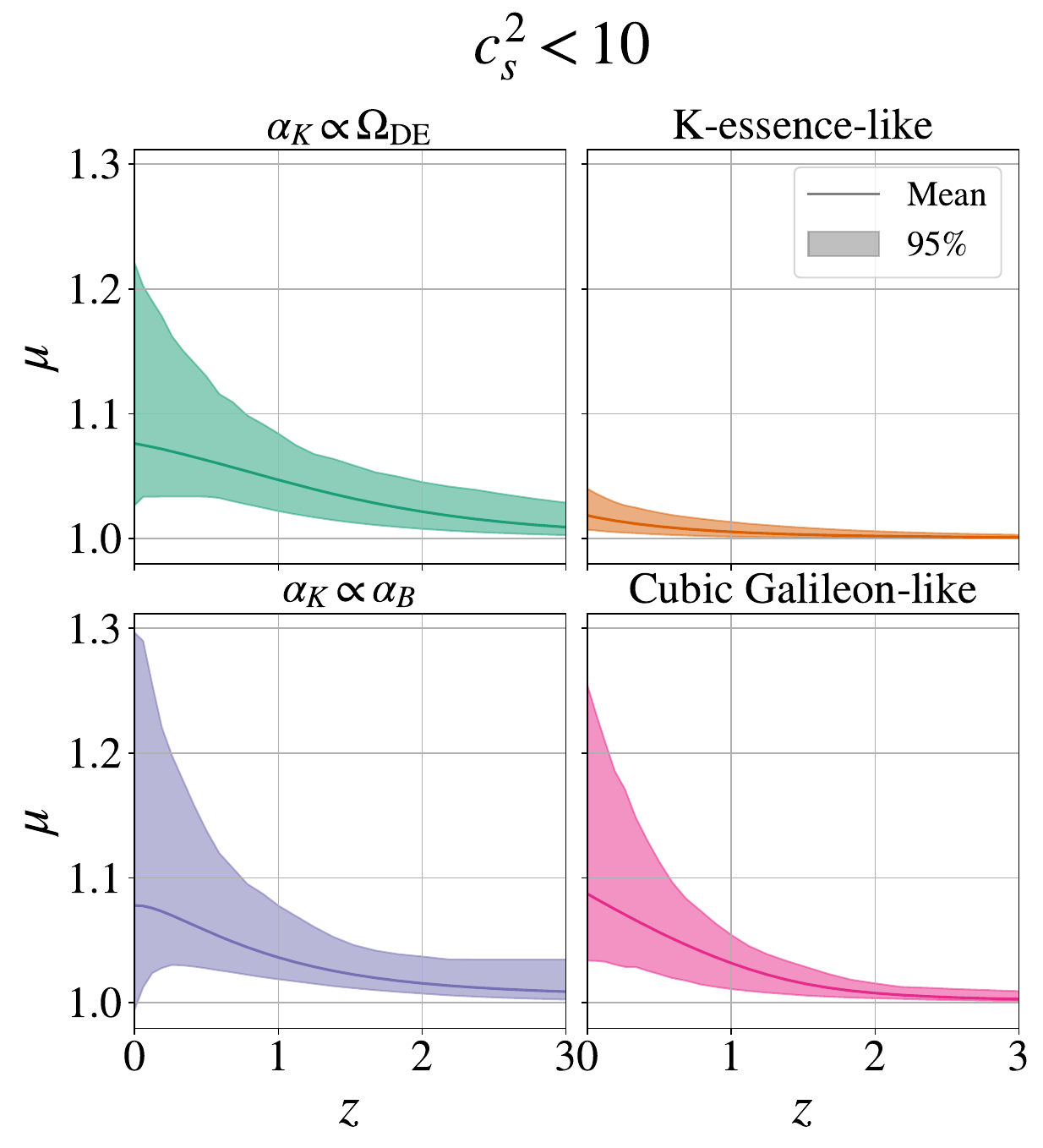}
    \includegraphics[width=0.32\linewidth]{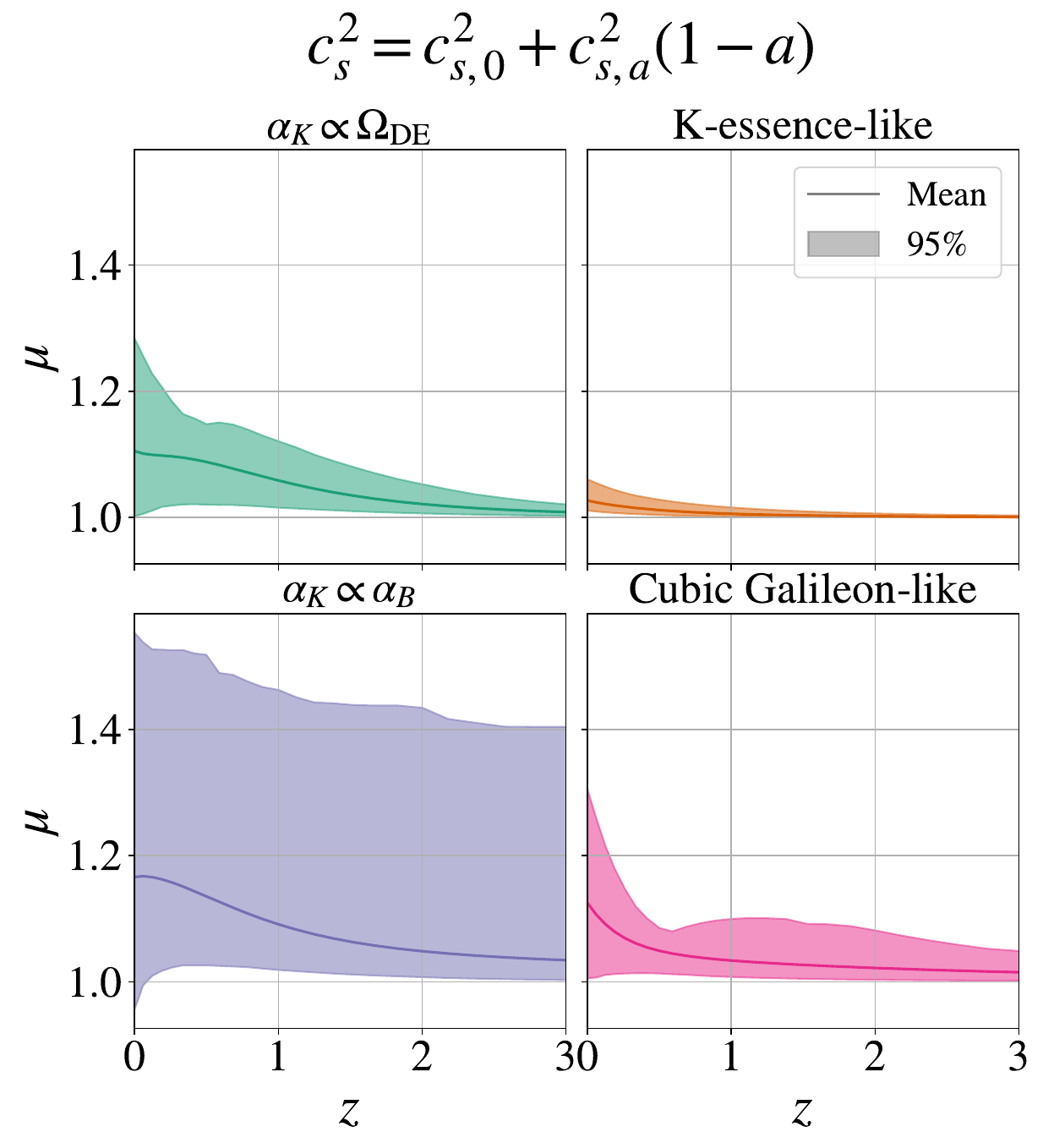}
    \caption{Same as Figure~\ref{fig:mu_constraint}, but with different priors on cosmological parameters: the left panel assumes a cosmological constant, the middle panel assumes a prior of $0 < c_s^2 < 10$, and the right panel assumes a dynamical sound speed of the form $c_s^2 = c_{s,0}^2 + c_{s,a}^2(1-a)$.}
    \label{fig:mu_constraints_cases}
\end{figure*}

To further test our hypothesis, we repeat our analysis while fixing the equation of state to $-1$, \textit{i.e.} setting $w_0 = -1$ and $w_a = 0$. The constraints on $\mu$ for this case are shown in the left panel of Figure~\ref{fig:mu_constraints_cases}. As stated above, for the k-essence-like and $\alpha_{\rm K} \propto \alpha_{\rm B}$ parameterizations, a cosmological constant implies $\alpha_{\rm B} = 0$ at all times, and therefore $\mu = 1$. For the Cubic Galileon-like and $\alpha_{\rm K} \propto \Omega_\mathrm{DE}$ parameterizations, the evolution of $\mu$ is still observed, but the constraints are now compatible with GR at the $2\sigma$ confidence level. This result highlights the interplay between the background expansion history and modified gravity constraints in theoretical models.

We have confirmed that our results remain consistent when adopting a superluminal prior of $c_s^2 < 10$. As discussed above, higher values of $c_s^2$ decrease the maximum value of $\mu$, bringing the model closer to the GR phenomenology. The middle panel of Figure~\ref{fig:mu_constraints_cases} shows constraints on $\mu(a)$ under this prior. While the mean value of $\mu(a=1)$ drops slightly below $1.1$, the conclusions drawn from the subluminal prior case still hold.

Furthermore, we have tested the effect of a dynamical sound speed on our results. We find that the coefficient $c_{s,a}^2$ is compatible with zero in all cases. An important consequence of this, is that while we expect that models with a dynamical scalar field have a time evolving sound speed, this feature is not detected by the data. Therefore, treating $c_{s}^{2}$ as a constant parameter is sufficient in the context of the datasets we consider in this work. The right panel of Figure~\ref{fig:mu_constraints_cases} shows the constraints on $\mu(a)$ assuming a dynamical sound speed. While these constraints remain mostly unchanged, the 95\% upper bound on $\mu(a)$ increases significantly in the $\alpha_{\rm K} \propto \alpha_{\rm B}$ case. We have confirmed that these deviations are localized at redshifts $z \le 20$.

\begin{figure}
    \centering
    \includegraphics[width=\linewidth]{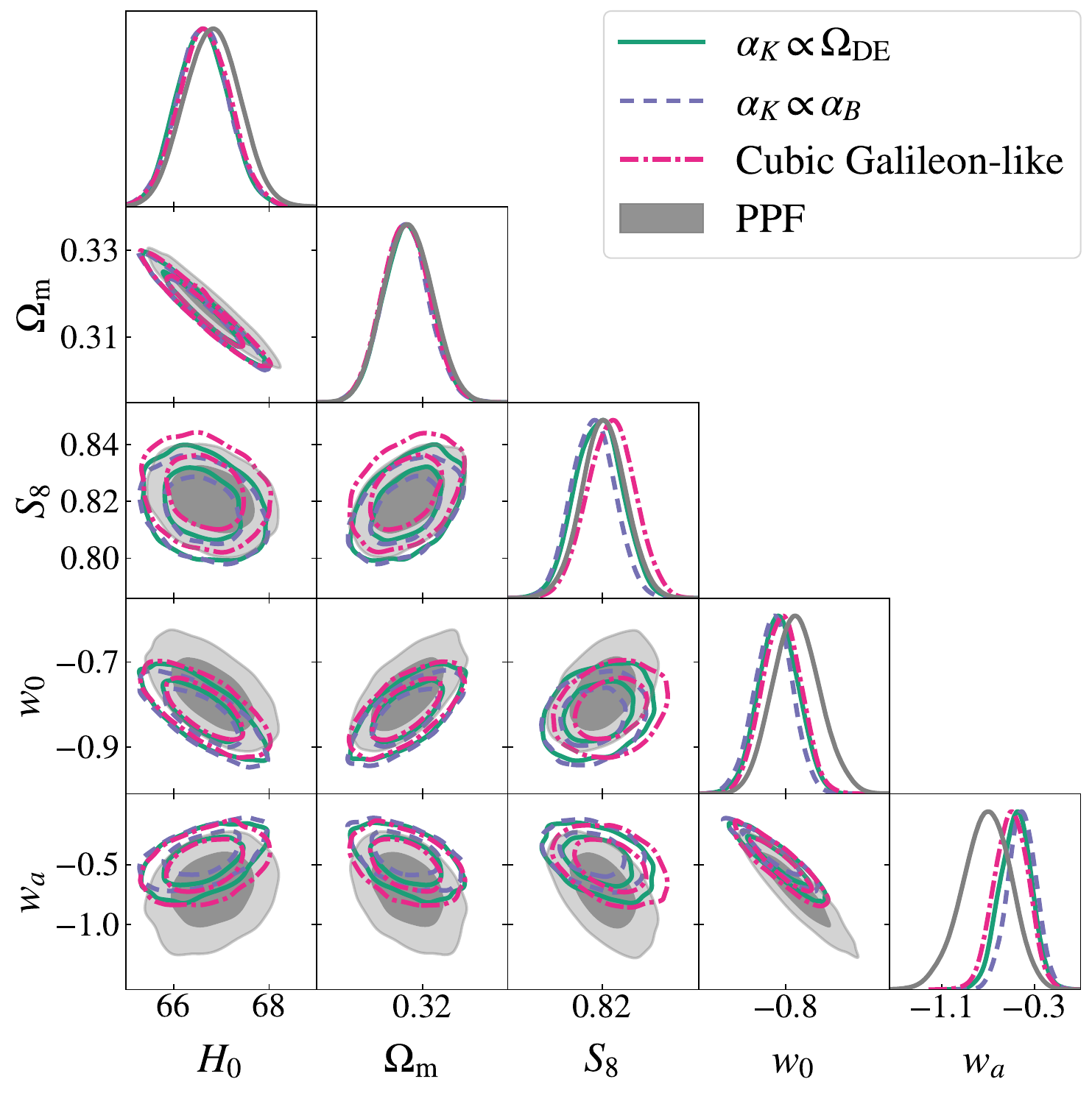}
    \caption{Same as Figure~\ref{fig:triangle_mg_subluminal}, but now for the dataset combination of CMBL+BAO+SN+CS Cosmic Shear.}
    \label{fig:triangle_mg_ds2}
\end{figure}

\begin{figure}
    \centering
    \includegraphics[width=\linewidth]{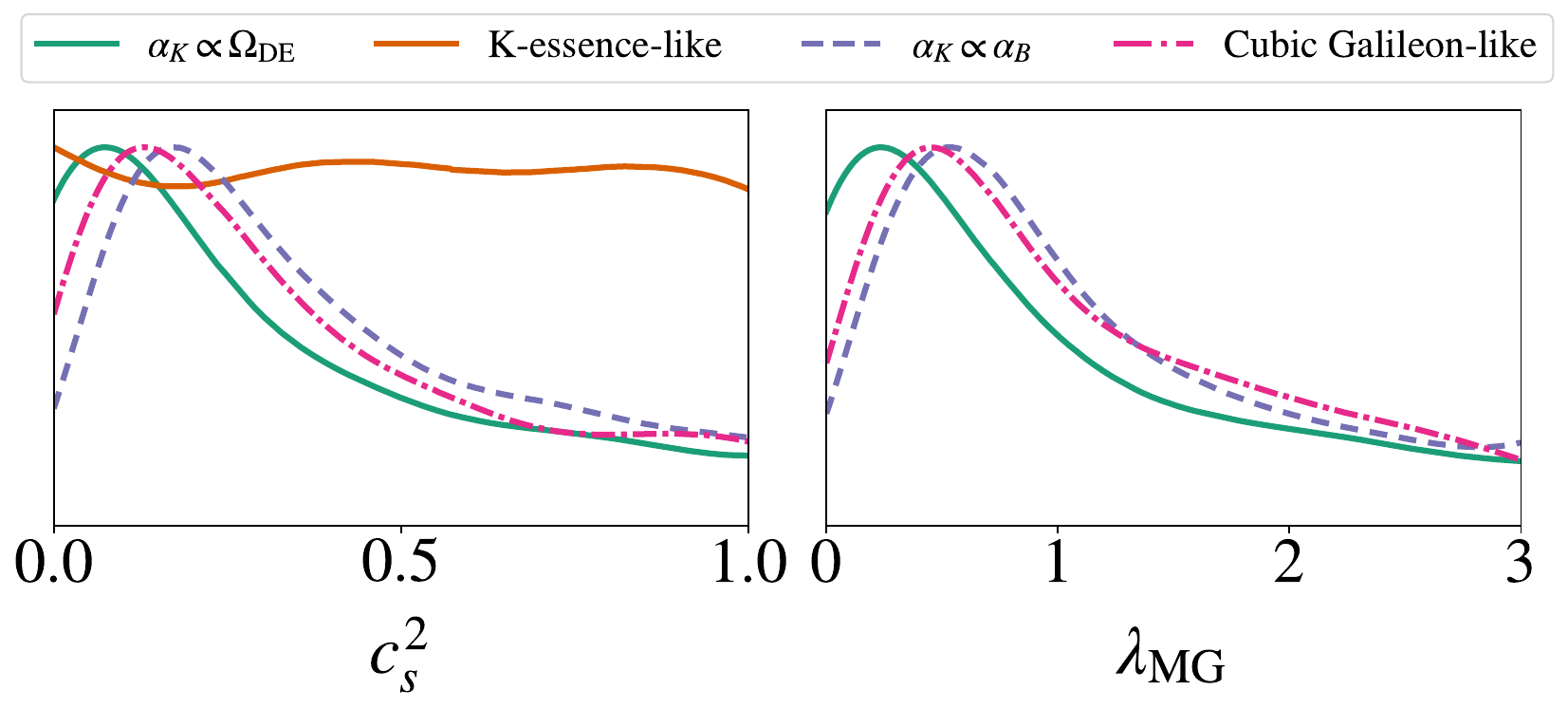}
    \caption{Same as Figure~\ref{fig:cs2_lambda_ds1}, but now for the dataset combination of CMBL+BAO+SN+CS Cosmic Shear.}
    \label{fig:cs2_lambda_ds2}
\end{figure}

\begin{figure}
    \centering
    \includegraphics[width=\linewidth]{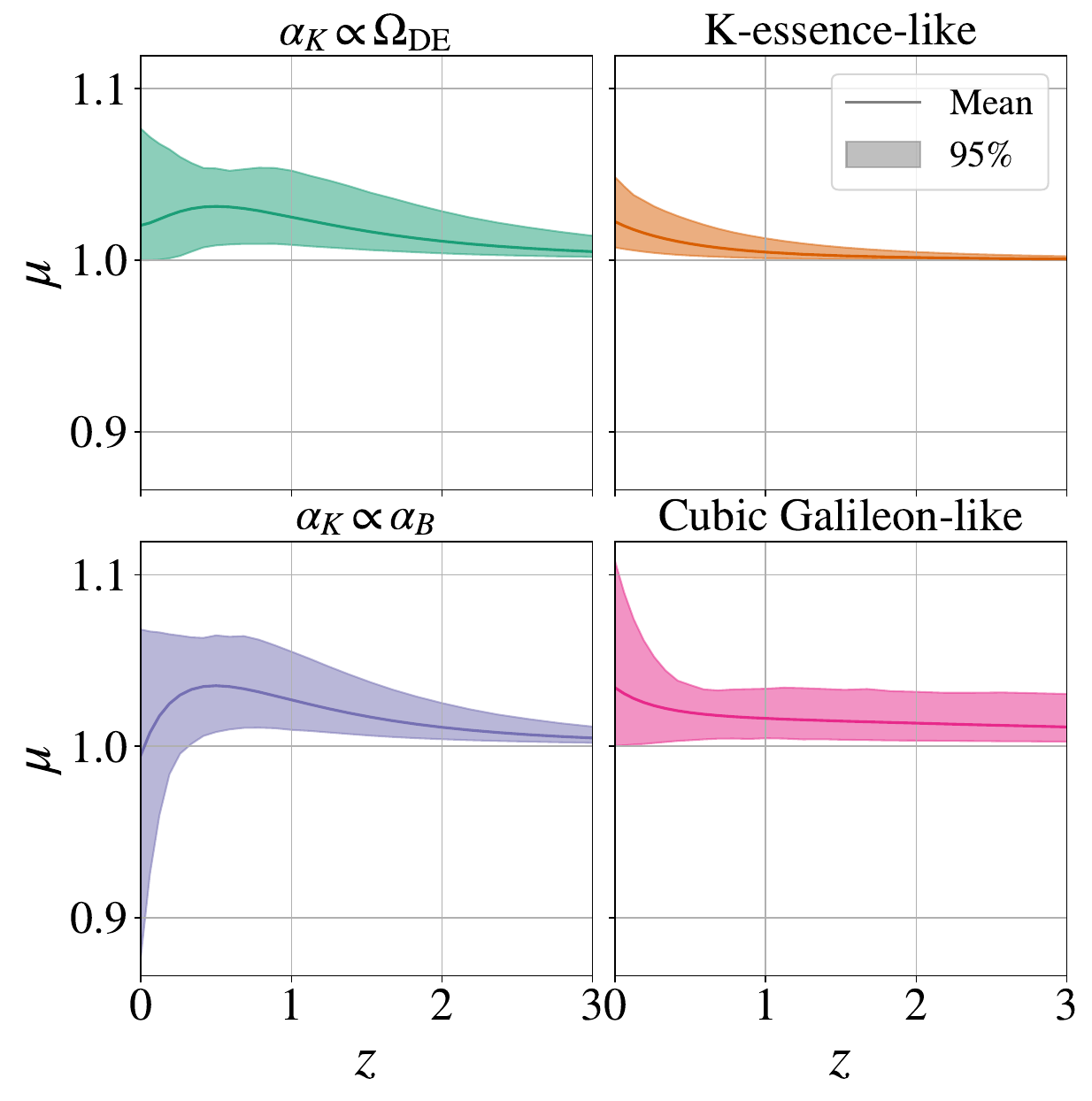}
    \caption{Same as Figure~\ref{fig:mu_constraint}, but now for the dataset combination of CMBL+BAO+SN+CS Cosmic Shear.}
    \label{fig:mu_constraint_ds2}
\end{figure}

We now turn our attention to the constraints for the CMBL+BAO+SN+CS dataset combination. Figure~\ref{fig:triangle_mg_ds2} shows confidence contours for the cosmological parameters, Figure~\ref{fig:cs2_lambda_ds2} shows the marginalized one-dimensional constraints on $c_s^2$ and $\lambda_\mathrm{MG}$, Figure~\ref{fig:mu_constraint_ds2} shows constraints on $\mu(a)$ and Table~\ref{tab:all_constraints} shows 1D marginalized constraints. Unlike the case without cosmic shear and CMB lensing, the constraints on $S_8$ obtained from the GR and MG cases now agree with one another. Furthermore, the posteriors of $w_0$ and $w_a$ are significantly shifted towards $\Lambda$CDM in the MG cases. The one-dimensional marginalized constraints, reported as means and 68\% confidence intervals, are:
\begin{itemize}
    \item $w_0 = -0.78 \pm 0.07, \; w_a = -0.72^{+0.22}_{-0.20}$ (GR),
    \item $w_0 = -0.82\pm 0.05, \; w_a = -0.46\pm 0.14$ (MG, $\alpha_{\rm K} \propto \Omega_\mathrm{DE}$),
    \item $w_0 = -0.83\pm 0.05, \; w_a = -0.41^{+0.13}_{-0.11}$ (MG, $\alpha_{\rm K} \propto \alpha_{\rm B}$),
    \item $w_0 = -0.81\pm 0.05, \; w_a = -0.50\pm 0.15$ (MG, Cubic Galileon-like).
\end{itemize}

These results suggest a weaker deviation from GR than in the case without LSS probes. Indeed, from Figure~\ref{fig:mu_constraint_ds2} and Table~\ref{tab:all_constraints}, we see that the mean value of $\mu_0$ decreases from $1.1$ to $1.02$ $\alpha_\mathrm{K} \propto \Omega_\mathrm{DE}$ and Cubic Galileon-like cases, and to $0.99$ in the $\alpha_\mathrm{K} \propto \alpha_\mathrm{B}$. These values represent a much weaker modification of gravity than in the case without DES-Y3 and CMB lensing. The k-essence-like case is practically unchanged. We conclude that large-scale structure probes are in tension with the previous results of $\mu > 1$, driving the resulting constraints towards GR. Due to the correlations between $\mu$ and the dark energy equation of state, the constraints on the latter are shifted towards a cosmological constant.

Even with additional information from large-scale structure, the constraints on the dark energy sound speed, shown in Figure~\ref{fig:cs2_lambda_ds2}, remain weak. For the cases where $w_{\rm DE}$ is allowed to be phantom, the posteriors peak at values around $c_s^2 \approx 0.2$, but are still compatible with the prior boundaries at 95\% confidence level. The constraints on $\lambda_\mathrm{MG}$ peak at values slightly below one and weakly disfavor higher values.

\begin{table*}[]
    \centering
    \begin{tabular}{|c|c|c|c|c|c|c|}
        \hline
         & $w_0w_a$CDM & $\alpha_{\rm K} \propto \Omega_\mathrm{DE}$ & $\alpha_{\rm K} \propto \alpha_{\rm B}$ & Cubic Galileon-like & K-essence-like & $w_0w_a$CDM $w \geq -1$  \\ 
        \hline
        & \multicolumn{6}{c|}{CMB+BAO+SN}\\
        \hline
        $\Omega_\mathrm{m}$ & $ 0.318\pm 0.006$ & $ 0.317\pm 0.006$ & $ 0.317\pm 0.005$ & $ 0.317\pm 0.006$ & $ 0.312\pm 0.005$ & $ 0.312\pm 0.006$ \\ \hline
        $H_0$ & $ 66.7\pm 0.6$ & $ 66.7\pm 0.6$ & $ 66.7\pm 0.5$ & $ 66.7\pm 0.6$ & $ 66.8\pm 0.5$ & $ 66.9\pm 0.6$ \\ \hline
        $S_8$ & $ 0.822\pm 0.011$ & $ 0.832\pm 0.013$ & $ 0.833\pm 0.013$ & $ 0.832\pm 0.013$ & $ 0.795\pm 0.010$ & $ 0.798\pm 0.010$ \\ \hline
        $w_0$ & $ -0.77\pm 0.06$ & $ -0.78\pm 0.06$ & $ -0.78\pm 0.05$ & $ -0.78\pm 0.06$ & $ -0.93\pm 0.03$ & $ -0.93\pm 0.03$ \\ \hline
        $w_a$ & $ -0.74^{+0.24}_{-0.22}$ & $ -0.66^{+0.22}_{-0.19}$ & $ -0.66^{+0.20}_{-0.17}$ & $ -0.67^{+0.22}_{-0.19}$ & $ -0.03^{+0.03}_{-0.06}$ & $ -0.04^{+0.04}_{-0.05}$ \\ \hline
        $\mu_0$ & $1$ & $1.10^{+0.09}_{-0.08}$ & $1.11^{+0.06}_{-0.08}$ & $1.12^{+0.10}_{-0.09}$ & $1.03 \pm 0.01$ & $1$ \\ \hline
        \hline
        & \multicolumn{6}{c|}{CMBL+BAO+SN+CS}\\
        \hline
        $\Omega_\mathrm{m}$ & $ 0.316\pm 0.006$ & $ 0.316\pm 0.006$ & $ 0.316\pm 0.005$ & $ 0.316\pm 0.006$ & $ 0.311 \pm 0.005$ & $ 0.312\pm 0.005$ \\ \hline
        $H_0$ & $ 66.8\pm 0.7$ & $ 66.6\pm 0.6$ & $ 66.6\pm 0.5$ & $ 66.6\pm 0.6$ & $ 67.0\pm 0.5$ & $ 67.0\pm 0.6$ \\ \hline
        $S_8$ & $ 0.821\pm 0.008$ & $ 0.819\pm 0.008$ & $ 0.817\pm 0.008$ & $ 0.823\pm 0.009$ & $ 0.803\pm 0.008$ & $ 0.804\pm 0.007$ \\ \hline
        $w_0$ & $ -0.78\pm 0.07$ & $ -0.82\pm 0.05$ & $ -0.83\pm 0.05$ & $ -0.81\pm 0.05$ & $ -0.94 \pm 0.03$ & $ -0.94\pm 0.03$ \\ \hline
        $w_a$ & $ -0.72^{+0.22}_{-0.20}$ & $ -0.46\pm 0.14$ & $ -0.41^{+0.13}_{-0.11}$ & $ -0.50\pm 0.15$ & $ -0.02^{+0.04}_{-0.05}$ & $ -0.04^{+0.04}_{-0.05}$ \\ \hline
        $\mu_0$ & $1$ & $1.02^{+0.02}_{-0.02}$ & $0.99^{+0.04}_{-0.04}$ & $1.03^{+0.03}_{-0.03}$ & $1.02 \pm 0.01$ &  $1$ \\ \hline
    \end{tabular}
    \caption{Constraints (mean and 68\% confidence intervals) for cosmological parameters $\Omega_m$, $H_0$, $S_8$, $w_0$, $w_a$ as well as $\mu_0$, the current value of the MG function $\mu(a)$ (see Equation~\ref{eq:mg_mu_sigma}).}
    \label{tab:all_constraints}
\end{table*}

\section{Conclusions}
\label{sec:conclusions}
Hints for dynamical dark energy from current surveys have motivated the investigation of new theoretical models that can consistently explain the observed signal in the $w_0w_a$ parameter space. These models, however, may differ substantially in their predictions for the growth of matter inhomogeneities. Future experiments, including galaxy surveys, CMB measurements, and 21-cm line intensity mapping, will provide a wealth of information about large-scale structure and will therefore be able to distinguish between candidate models for dark energy. In this context, the dark energy sound speed, usually assumed to be unity, provides a physically motivated parameter for testing the consistency of the $w_0w_a$ expansion history and growth of matter perturbations. While in GR the effects of a non-standard $c_s^2$ are small and concentrated at large scales, in modified gravity both the dark energy sound speed and equation of state are directly tied to changes in matter clustering at small scales.

In this work, we have investigated the direct relation between dark energy properties, including its equation of state $w_{\rm DE}$ and sound speed $c_s^2$, and the $\mu$ and $\Sigma$ functions that modify the growth of structure and light propagation. By using $c_s^2$ as an input parameter instead of the MG functions $\alpha_i$, we are able to solve for $\mu(a)$ and $\Sigma(a)$, provided additional parametrizations for $\alpha_{\rm K}$ are specified. We have tested four parametrizations, finding consistent results among three of them. This framework has the additional advantage of ensuring that the dark energy perturbations are stable, $c_s^2 > 0$ and $D_\mathrm{kin} > 0$. 

We find that the direct impact of the dark energy equation of state on matter perturbations increases the constraining power on $w_0$ and $w_a$ obtained from CMB and LSS probes. For the CMB+BAO+SN dataset combination, we find a preference for $\mu > 1$ at low redshifts at over the $2\sigma$ confidence level, consistent with previous results in the literature~\cite{Shah:2025vnt}. This preference is driven by the phantom dark energy equation of state preference in the data rather than by an improvement in the CMB fit. Furthermore, this preference persists if we assume a superluminal sound speed or a dynamical sound speed, but vanishes if we assume a $\Lambda$CDM background. This result highlights the correlations between background physics and the growth of structure, especially in modified gravity models or in dark energy models with an imperfect stress-energy tensor. When including information from cosmic shear and CMB lensing, we find that the strength of the deviations from GR significantly decreases, along with a shift in $w_0$ and $w_a$ towards a cosmological constant.

Overall, this theoretical framework can simultaneously test the consistency of the $\Lambda$CDM model at both background and perturbative levels. Our results can be further extended by including a model for the $\alpha_{\rm M}$ function, which controls the dark energy anisotropic stress. This framework can also be readily implemented in \textit{N}-body codes~\cite{mg_sims}, relaxing the need for the linear scale cuts used for cosmic shear in this work. In particular, the Comoving Lagrangian Acceleration (COLA) algorithm~\cite{Tassev:2013pn} has proven suitable for generating large simulation suites for machine learning models to learn the nonlinear corrections to the matter power spectrum~\cite{cola_wcdm, cola_w0wa}. In the future, with a suitable model for nonlinear gravitational effects, we aim to further extend our results by forecasting constraints for future galaxy surveys.

\acknowledgments

The authors thank Kazuya Koyama for useful comments. This study was financed in part by the Coordenação de Aperfeiçoamento de Pessoal de Nível Superior – Brasil (CAPES) – Finance Code 001. JR acknowledges financial support from CAPES. FTF acknowledges financial support from the National Scientific and Technological Research Council (CNPq, Brazil). This work made use of the CHE cluster, managed and funded by COSMO/CBPF/MCTI, with financial support from CNPq, FINEP and FAPERJ. 

\appendix

\section{Results for K-essence-like Parameterization}
\label{app:kessence_results}
\begin{figure*}
    \centering
    \includegraphics[width=0.48\linewidth]{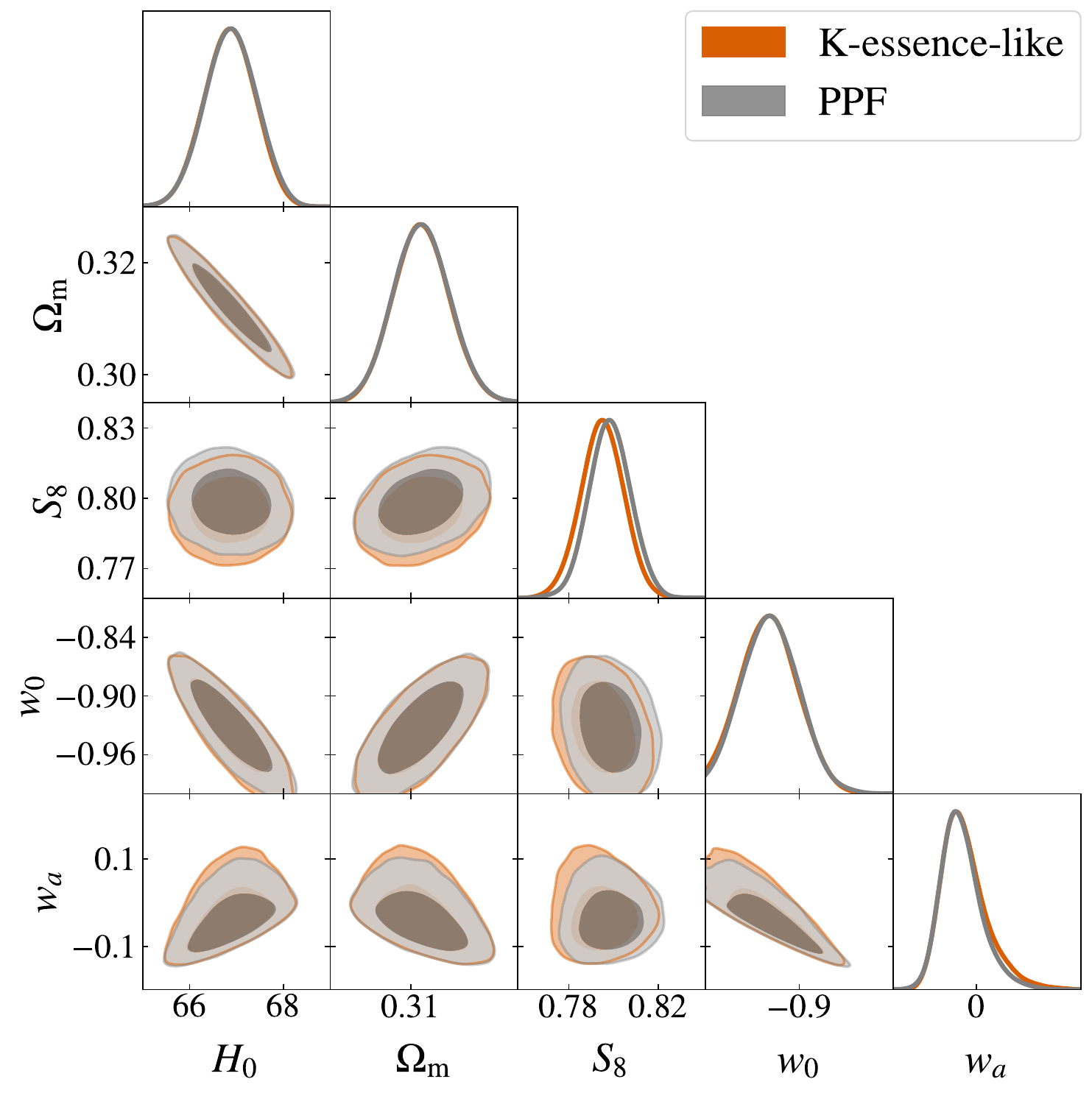}
    \includegraphics[width=0.48\linewidth]{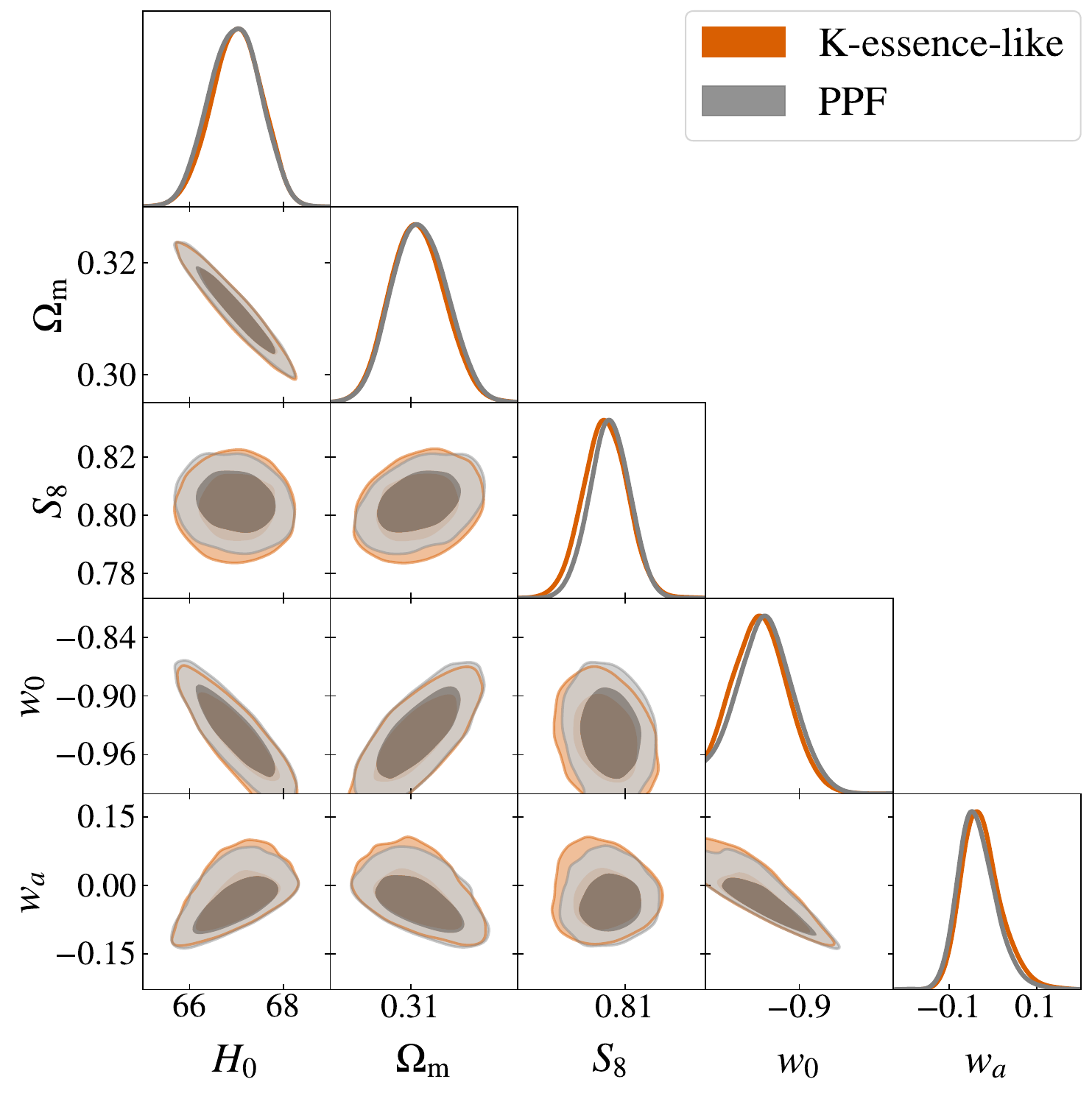}
    \caption{Cosmological parameter posterior confidence contours (68\% and 95\% credible regions). The left plot shows the posteriors for the dataset combination CMB+BAO+SN, while the right plot shows the posteriors for CMBL+BAO+SN+CS. The gray contours shows the GR case and the orange contours show the k-essence-like MG case.}
    \label{fig:triangle_kessence}
\end{figure*}

In this Appendix, we show contours plots for the posteriors of the k-essence-like parametrization, which were not shown in the main text for the sake of conciseness. Figure~\ref{fig:triangle_kessence} shows the posterior confidence contours for the dataset combinations of CMB+BAO+SN and CMBL+BAO+SN+CS.

\begin{comment}
\section{Linear CMB Lensing}
\label{app:comparison_linear}

In this Appendix, we compare constraints obtained using linear CMB lensing versus those obtained using nonlinear CMB lensing assuming \texttt{HMCode2020}. Figure~\ref{fig:linear_lensing_comparison} shows constraints on $\mu(a)$ on the four MG parameterizations, for the dataset combination CMBL+BAO+SN+CS, where the top row shows constraints using nonlinear CMB lensing, and the bottom panel shows constraints using linear CMB lensing (\textit{i.e.} calculated from the linear matter power spectrum). We observe no significant difference in the results, suggesting that our conclusions are robust with respect to nonlinear effects in CMB lensing.

\begin{figure}
    \centering
    \includegraphics[width=\linewidth]{figures/compare_linear_mu.pdf}
    \caption{Constraints (mean and 95\% confidence level) on the modified gravity function $\mu(a)$ for the four parameterizations considered in this work, using the dataset combination CMBL+BAO+SN+CS. The top panel shows constraints using nonlinear CMB lensing, and the bottom panel shows constraints using linear CMB lensing.}
    \label{fig:linear_lensing_comparison}
\end{figure}
\end{comment}
\bibliography{refs_short}% Produces the bibliography via BibTeX.

@ARTICLE{bellini_sawicky,
       author = {{Bellini}, Emilio and {Sawicki}, Ignacy},
        title = "{Maximal freedom at minimum cost: linear large-scale structure in general modifications of gravity}",
      journal = {\jcap},
         year = 2014,
        month = jul,
       volume = {2014},
       number = {7},
          eid = {050},
        pages = {050},
          doi = {10.1088/1475-7516/2014/07/050},
archivePrefix = {arXiv},
       eprint = {1404.3713},
 primaryClass = {astro-ph.CO},
       adsurl = {https://ui.adsabs.harvard.edu/abs/2014JCAP...07..050B}
}

@ARTICLE{kessence,
       author = {{Chiba}, Takeshi and {Okabe}, Takahiro and {Yamaguchi}, Masahide},
        title = "{Kinetically driven quintessence}",
      journal = {\prd},
         year = 2000,
        month = jul,
       volume = {62},
       number = {2},
          eid = {023511},
        pages = {023511},
          doi = {10.1103/PhysRevD.62.023511},
archivePrefix = {arXiv},
       eprint = {astro-ph/9912463},
 primaryClass = {astro-ph},
       adsurl = {https://ui.adsabs.harvard.edu/abs/2000PhRvD..62b3511C}
}

@ARTICLE{kazuya_mg_phantom,
       author = {{Cataneo}, Matteo and {Koyama}, Kazuya},
        title = "{Non-parametric exploration of minimally coupled gravity with phantom crossing}",
      journal = {arXiv e-prints},
         year = 2025,
        month = dec,
          eid = {arXiv:2512.13691},
        pages = {arXiv:2512.13691},
          doi = {10.48550/arXiv.2512.13691},
archivePrefix = {arXiv},
       eprint = {2512.13691},
 primaryClass = {astro-ph.CO},
       adsurl = {https://ui.adsabs.harvard.edu/abs/2025arXiv251213691C}
}

@ARTICLE{kazuya_theoretical_priors_mg,
       author = {{Shah}, Neel and {Koyama}, Kazuya and {Noller}, Johannes},
        title = "{Dark energy constraints in light of theoretical priors}",
      journal = {arXiv e-prints},
         year = 2025,
        month = jul,
          eid = {arXiv:2507.19450},
        pages = {arXiv:2507.19450},
          doi = {10.48550/arXiv.2507.19450},
archivePrefix = {arXiv},
       eprint = {2507.19450},
 primaryClass = {astro-ph.CO},
       adsurl = {https://ui.adsabs.harvard.edu/abs/2025arXiv250719450S}
}

@ARTICLE{mg_sims,
       author = {{Brando}, Guilherme and {Koyama}, Kazuya and {Wands}, David and {Zumalac{\'a}rregui}, Miguel and {Sawicki}, Ignacy and {Bellini}, Emilio},
        title = "{Fully relativistic predictions in Horndeski gravity from standard Newtonian N-body simulations}",
      journal = {\jcap},
         year = 2021,
        month = sep,
       volume = {2021},
       number = {9},
          eid = {024},
        pages = {024},
          doi = {10.1088/1475-7516/2021/09/024},
archivePrefix = {arXiv},
       eprint = {2105.04491},
 primaryClass = {astro-ph.CO},
       adsurl = {https://ui.adsabs.harvard.edu/abs/2021JCAP...09..024B}
}

@ARTICLE{cmb_pheno_hard,
       author = {{Trendafilova}, Cynthia and {Khalife}, Ali Rida and {Galli}, Silvia},
        title = "{The end of easy phenomenology for CMB experiments: A case study in the dark sector}",
      journal = {\jcap},
         year = 2025,
        month = may,
       volume = {2025},
       number = {5},
          eid = {094},
        pages = {094},
          doi = {10.1088/1475-7516/2025/05/094},
archivePrefix = {arXiv},
       eprint = {2502.19383},
 primaryClass = {astro-ph.CO},
       adsurl = {https://ui.adsabs.harvard.edu/abs/2025JCAP...05..094T}
}

@ARTICLE{growth_mg,
       author = {{Brando}, Guilherme and {Koyama}, Kazuya and {Wands}, David},
        title = "{Relativistic corrections to the growth of structure in modified gravity}",
      journal = {\jcap},
         year = 2021,
        month = jan,
       volume = {2021},
       number = {1},
          eid = {013},
        pages = {013},
          doi = {10.1088/1475-7516/2021/01/013},
archivePrefix = {arXiv},
       eprint = {2006.11019},
 primaryClass = {astro-ph.CO},
       adsurl = {https://ui.adsabs.harvard.edu/abs/2021JCAP...01..013B}
}

@ARTICLE{galileon,
       author = {{de Felice}, Antonio and {Tsujikawa}, Shinji},
        title = "{Cosmology of a Covariant Galileon Field}",
      journal = {\prl},
         year = 2010,
        month = sep,
       volume = {105},
       number = {11},
          eid = {111301},
        pages = {111301},
          doi = {10.1103/PhysRevLett.105.111301},
archivePrefix = {arXiv},
       eprint = {1007.2700},
 primaryClass = {astro-ph.CO},
       adsurl = {https://ui.adsabs.harvard.edu/abs/2010PhRvL.105k1301D}
}

@ARTICLE{galileon_planck,
       author = {{Barreira}, Alexandre and {Li}, Baojiu and {Baugh}, Carlton M. and {Pascoli}, Silvia},
        title = "{The observational status of Galileon gravity after Planck}",
      journal = {\jcap},
         year = 2014,
        month = aug,
       volume = {2014},
       number = {8},
        pages = {059-059},
          doi = {10.1088/1475-7516/2014/08/059},
archivePrefix = {arXiv},
       eprint = {1406.0485},
 primaryClass = {astro-ph.CO},
       adsurl = {https://ui.adsabs.harvard.edu/abs/2014JCAP...08..059B}
}

@ARTICLE{mochi_class,
       author = {{Cataneo}, Matteo and {Bellini}, Emilio},
        title = "{mochi\_class: Modelling Optimisation to Compute Horndeski In class}",
      journal = {The Open Journal of Astrophysics},
         year = 2024,
        month = sep,
       volume = {7},
          eid = {76},
        pages = {76},
          doi = {10.33232/001c.123470},
archivePrefix = {arXiv},
       eprint = {2407.11968},
 primaryClass = {astro-ph.CO},
       adsurl = {https://ui.adsabs.harvard.edu/abs/2024OJAp....7E..76C}
}

@ARTICLE{desi_y1_bao,
       author = {{Adame}, A.~G. and {Aguilar}, J. and {Ahlen}, S. and {Alam}, S. and {Alexander}, D.~M. and {Alvarez}, M. and {Alves}, O. and {Anand}, A. and {Andrade}, U. and {Armengaud}, E. and {Avila}, S. and {Aviles}, A. and {Awan}, H. and {Bahr-Kalus}, B. and {Bailey}, S. and {Baltay}, C. and {Bault}, A. and {Behera}, J. and {BenZvi}, S. and {Bera}, A. and {Beutler}, F. and {Bianchi}, D. and {Blake}, C. and {Blum}, R. and {Brieden}, S. and {Brodzeller}, A. and {Brooks}, D. and {Buckley-Geer}, E. and {Burtin}, E. and {Calderon}, R. and {Canning}, R. and {Carnero Rosell}, A. and {Cereskaite}, R. and {Cervantes-Cota}, J.~L. and {Chabanier}, S. and {Chaussidon}, E. and {Chaves-Montero}, J. and {Chen}, S. and {Chen}, X. and {Claybaugh}, T. and {Cole}, S. and {Cuceu}, A. and {Davis}, T.~M. and {Dawson}, K. and {de la Macorra}, A. and {de Mattia}, A. and {Deiosso}, N. and {Dey}, A. and {Dey}, B. and {Ding}, Z. and {Doel}, P. and {Edelstein}, J. and {Eftekharzadeh}, S. and {Eisenstein}, D.~J. and {Elliott}, A. and {Fagrelius}, P. and {Fanning}, K. and {Ferraro}, S. and {Ereza}, J. and {Findlay}, N. and {Flaugher}, B. and {Font-Ribera}, A. and {Forero-S{\'a}nchez}, D. and {Forero-Romero}, J.~E. and {Frenk}, C.~S. and {Garcia-Quintero}, C. and {Gazta{\~n}aga}, E. and {Gil-Mar{\'\i}n}, H. and {Gontcho a Gontcho}, S. and {Gonzalez-Morales}, A.~X. and {Gonzalez-Perez}, V. and {Gordon}, C. and {Green}, D. and {Gruen}, D. and {Gsponer}, R. and {Gutierrez}, G. and {Guy}, J. and {Hadzhiyska}, B. and {Hahn}, C. and {Hanif}, M.~M.~S. and {Herrera-Alcantar}, H.~K. and {Honscheid}, K. and {Howlett}, C. and {Huterer}, D. and {Ir{\v{s}}i{\v{c}}}, V. and {Ishak}, M. and {Juneau}, S. and {Kara{\c{c}}ayl{\i}}, N.~G. and {Kehoe}, R. and {Kent}, S. and {Kirkby}, D. and {Kremin}, A. and {Krolewski}, A. and {Lai}, Y. and {Lan}, T.-W. and {Landriau}, M. and {Lang}, D. and {Lasker}, J. and {Le Goff}, J.~M. and {Le Guillou}, L. and {Leauthaud}, A. and {Levi}, M.~E. and {Li}, T.~S. and {Linder}, E. and {Lodha}, K. and {Magneville}, C. and {Manera}, M. and {Margala}, D. and {Martini}, P. and {Maus}, M. and {McDonald}, P. and {Medina-Varela}, L. and {Meisner}, A. and {Mena-Fern{\'a}ndez}, J. and {Miquel}, R. and {Moon}, J. and {Moore}, S. and {Moustakas}, J. and {Mueller}, E. and {Mu{\~n}oz-Guti{\'e}rrez}, A. and {Myers}, A.~D. and {Nadathur}, S. and {Napolitano}, L. and {Neveux}, R. and {Newman}, J.~A. and {Nguyen}, N.~M. and {Nie}, J. and {Niz}, G. and {Noriega}, H.~E. and {Padmanabhan}, N. and {Paillas}, E. and {Palanque-Delabrouille}, N. and {Pan}, J. and {Penmetsa}, S. and {Percival}, W.~J. and {Pieri}, M.~M. and {Pinon}, M. and {Poppett}, C. and {Porredon}, A. and {Prada}, F. and {P{\'e}rez-Fern{\'a}ndez}, A. and {P{\'e}rez-R{\`a}fols}, I. and {Rabinowitz}, D. and {Raichoor}, A. and {Ram{\'\i}rez-P{\'e}rez}, C. and {Ramirez-Solano}, S. and {Rashkovetskyi}, M. and {Ravoux}, C. and {Rezaie}, M. and {Rich}, J. and {Rocher}, A. and {Rockosi}, C. and {Roe}, N.~A. and {Rosado-Marin}, A. and {Ross}, A.~J. and {Rossi}, G. and {Ruggeri}, R. and {Ruhlmann-Kleider}, V. and {Samushia}, L. and {Sanchez}, E. and {Saulder}, C. and {Schlafly}, E.~F. and {Schlegel}, D. and {Schubnell}, M. and {Seo}, H. and {Shafieloo}, A. and {Sharples}, R. and {Silber}, J. and {Slosar}, A. and {Smith}, A. and {Sprayberry}, D. and {Tan}, T. and {Tarl{\'e}}, G. and {Taylor}, P. and {Trusov}, S. and {Ure{\~n}a-L{\'o}pez}, L.~A. and {Vaisakh}, R. and {Valcin}, D. and {Valdes}, F. and {Vargas-Maga{\~n}a}, M. and {Verde}, L. and {Walther}, M. and {Wang}, B. and {Wang}, M.~S. and {Weaver}, B.~A. and {Weaverdyck}, N. and {Wechsler}, R.~H. and {Weinberg}, D.~H. and {White}, M. and {Yu}, J. and {Yu}, Y. and {Yuan}, S. and {Y{\`e}che}, C. and {Zaborowski}, E.~A. and {Zarrouk}, P. and {Zhang}, H. and {Zhao}, C. and {Zhao}, R. and {Zhou}, R. and {Zhuang}, T.},
        title = "{DESI 2024 VI: cosmological constraints from the measurements of baryon acoustic oscillations}",
      journal = {\jcap},
         year = 2025,
        month = feb,
       volume = {2025},
       number = {2},
          eid = {021},
        pages = {021},
          doi = {10.1088/1475-7516/2025/02/021},
archivePrefix = {arXiv},
       eprint = {2404.03002},
 primaryClass = {astro-ph.CO},
       adsurl = {https://ui.adsabs.harvard.edu/abs/2025JCAP...02..021A}
}

@ARTICLE{desi_dr2_bao,
       author = {{Abdul Karim}, M. and {Aguilar}, J. and {Ahlen}, S. and {Alam}, S. and {Allen}, L. and {Prieto}, C. Allende and {Alves}, O. and {Anand}, A. and {Andrade}, U. and {Armengaud}, E. and {Aviles}, A. and {Bailey}, S. and {Baltay}, C. and {Bansal}, P. and {Bault}, A. and {Behera}, J. and {BenZvi}, S. and {Bianchi}, D. and {Blake}, C. and {Brieden}, S. and {Brodzeller}, A. and {Brooks}, D. and {Buckley-Geer}, E. and {Burtin}, E. and {Calderon}, R. and {Canning}, R. and {Rosell}, A. Carnero and {Carrilho}, P. and {Casas}, L. and {Castander}, F.~J. and {Charles}, M. and {Chaussidon}, E. and {Chaves-Montero}, J. and {Chebat}, D. and {Chen}, X. and {Claybaugh}, T. and {Cole}, S. and {Cooper}, A.~P. and {Cuceu}, A. and {Dawson}, K.~S. and {de la Macorra}, A. and {de Mattia}, A. and {Deiosso}, N. and {Della Costa}, J. and {Demina}, R. and {Dey}, A. and {Dey}, B. and {Ding}, Z. and {Doel}, P. and {Edelstein}, J. and {Eisenstein}, D.~J. and {Elbers}, W. and {Fagrelius}, P. and {Fanning}, K. and {Fern{\'a}ndez-Garc{\'\i}a}, E. and {Ferraro}, S. and {Font-Ribera}, A. and {Forero-Romero}, J.~E. and {Frenk}, C.~S. and {Garcia-Quintero}, C. and {Garrison}, L.~H. and {Gazta{\~n}aga}, E. and {Gil-Mar{\'\i}n}, H. and {Gontcho A Gontcho}, S. and {Gonzalez}, D. and {Gonzalez-Morales}, A.~X. and {Gordon}, C. and {Green}, D. and {Gutierrez}, G. and {Guy}, J. and {Hadzhiyska}, B. and {Hahn}, C. and {He}, S. and {Herbold}, M. and {Herrera-Alcantar}, H.~K. and {Ho}, M.-F. and {Honscheid}, K. and {Howlett}, C. and {Huterer}, D. and {Ishak}, M. and {Juneau}, S. and {Kamble}, N.~V. and {Kara{\c{c}}ayl{\i}}, N.~G. and {Kehoe}, R. and {Kent}, S. and {Kim}, A.~G. and {Kirkby}, D. and {Kisner}, T. and {Koposov}, S.~E. and {Kremin}, A. and {Krolewski}, A. and {Lahav}, O. and {Lamman}, C. and {Landriau}, M. and {Lang}, D. and {Lasker}, J. and {Le Goff}, J.~M. and {Le Guillou}, L. and {Leauthaud}, A. and {Levi}, M.~E. and {Li}, Q. and {Li}, T.~S. and {Lodha}, K. and {Lokken}, M. and {Lozano-Rodr{\'\i}guez}, F. and {Magneville}, C. and {Manera}, M. and {Martini}, P. and {Matthewson}, W.~L. and {Meisner}, A. and {Mena-Fern{\'a}ndez}, J. and {Menegas}, A. and {Mergulh{\~a}o}, T. and {Miquel}, R. and {Moustakas}, J. and {Mu{\~n}oz-Guti{\'e}rrez}, A. and {Mu{\~n}oz-Santos}, D. and {Myers}, A.~D. and {Nadathur}, S. and {Naidoo}, K. and {Napolitano}, L. and {Newman}, J.~A. and {Niz}, G. and {Noriega}, H.~E. and {Paillas}, E. and {Palanque-Delabrouille}, N. and {Pan}, J. and {Peacock}, J.~A. and {Pellejero Ibanez}, M. and {Percival}, W.~J. and {P{\'e}rez-Fern{\'a}ndez}, A. and {P{\'e}rez-R{\`a}fols}, I. and {Pieri}, M.~M. and {Poppett}, C. and {Prada}, F. and {Rabinowitz}, D. and {Raichoor}, A. and {Ram{\'\i}rez-P{\'e}rez}, C. and {Rashkovetskyi}, M. and {Ravoux}, C. and {Rich}, J. and {Rocher}, A. and {Rockosi}, C. and {Rohlf}, J. and {Rom{\'a}n-Herrera}, J.~O. and {Ross}, A.~J. and {Rossi}, G. and {Ruggeri}, R. and {Ruhlmann-Kleider}, V. and {Samushia}, L. and {Sanchez}, E. and {Sanders}, N. and {Schlegel}, D. and {Schubnell}, M. and {Seo}, H. and {Shafieloo}, A. and {Sharples}, R. and {Silber}, J. and {Sinigaglia}, F. and {Sprayberry}, D. and {Tan}, T. and {Tarl{\'e}}, G. and {Taylor}, P. and {Turner}, W. and {Ure{\~n}a-L{\'o}pez}, L.~A. and {Vaisakh}, R. and {Valdes}, F. and {Valogiannis}, G. and {Vargas-Maga{\~n}a}, M. and {Verde}, L. and {Walther}, M. and {Weaver}, B.~A. and {Weinberg}, D.~H. and {White}, M. and {Wolfson}, M. and {Y{\`e}che}, C. and {Yu}, J. and {Zaborowski}, E.~A. and {Zarrouk}, P. and {Zhai}, Z. and {Zhang}, H. and {Zhao}, C. and {Zhao}, G.~B. and {Zhou}, R. and {Zou}, H. and {DESI Collaboration}},
        title = "{DESI DR2 results. II. Measurements of baryon acoustic oscillations and cosmological constraints}",
      journal = {\prd},
         year = 2025,
        month = oct,
       volume = {112},
       number = {8},
          eid = {083515},
        pages = {083515},
          doi = {10.1103/tr6y-kpc6},
archivePrefix = {arXiv},
       eprint = {2503.14738},
 primaryClass = {astro-ph.CO},
       adsurl = {https://ui.adsabs.harvard.edu/abs/2025PhRvD.112h3515A}
}

@ARTICLE{desi_full_shape,
       author = {{Adame}, A.~G. and {Aguilar}, J. and {Ahlen}, S. and {Alam}, S. and {Alexander}, D.~M. and {Alvarez}, M. and {Alves}, O. and {Anand}, A. and {Andrade}, U. and {Armengaud}, E. and {Avila}, S. and {Aviles}, A. and {Awan}, H. and {Bailey}, S. and {Baltay}, C. and {Bault}, A. and {Behera}, J. and {BenZvi}, S. and {Beutler}, F. and {Bianchi}, D. and {Blake}, C. and {Blum}, R. and {Brieden}, S. and {Brodzeller}, A. and {Brooks}, D. and {Buckley-Geer}, E. and {Burtin}, E. and {Calderon}, R. and {Canning}, R. and {Carnero Rosell}, A. and {Cereskaite}, R. and {Cervantes-Cota}, J.~L. and {Chabanier}, S. and {Chaussidon}, E. and {Chaves-Montero}, J. and {Chen}, S. and {Chen}, X. and {Claybaugh}, T. and {Cole}, S. and {Cuceu}, A. and {Davis}, T.~M. and {Dawson}, K. and {de la Macorra}, A. and {de Mattia}, A. and {Deiosso}, N. and {Dey}, A. and {Dey}, B. and {Ding}, Z. and {Doel}, P. and {Edelstein}, J. and {Eftekharzadeh}, S. and {Eisenstein}, D.~J. and {Elliott}, A. and {Fagrelius}, P. and {Fanning}, K. and {Ferraro}, S. and {Ereza}, J. and {Findlay}, N. and {Flaugher}, B. and {Font-Ribera}, A. and {Forero-S{\'a}nchez}, D. and {Forero-Romero}, J.~E. and {Garcia-Quintero}, C. and {Garrison}, L.~H. and {Gazta{\~n}aga}, E. and {Gil-Mar{\'\i}n}, H. and {Gontcho}, S. Gontcho A. and {Gonzalez-Morales}, A.~X. and {Gonzalez-Perez}, V. and {Gordon}, C. and {Green}, D. and {Gruen}, D. and {Gsponer}, R. and {Gutierrez}, G. and {Guy}, J. and {Hadzhiyska}, B. and {Hahn}, C. and {Hanif}, M.~M.~S. and {Herrera-Alcantar}, H.~K. and {Honscheid}, K. and {Howlett}, C. and {Huterer}, D. and {Ir{\v{s}}i{\v{c}}}, V. and {Ishak}, M. and {Juneau}, S. and {Kara{\c{c}}ayl{\i}}, N.~G. and {Kehoe}, R. and {Kent}, S. and {Kirkby}, D. and {Kong}, H. and {Koposov}, S.~E. and {Kremin}, A. and {Krolewski}, A. and {Lai}, Y. and {Lan}, T.-W. and {Landriau}, M. and {Lang}, D. and {Lasker}, J. and {Le Goff}, J.~M. and {Le Guillou}, L. and {Leauthaud}, A. and {Levi}, M.~E. and {Li}, T.~S. and {Lodha}, K. and {Magneville}, C. and {Manera}, M. and {Margala}, D. and {Martini}, P. and {Maus}, M. and {McDonald}, P. and {Medina-Varela}, L. and {Meisner}, A. and {Mena-Fern{\'a}ndez}, J. and {Miquel}, R. and {Moon}, J. and {Moore}, S. and {Moustakas}, J. and {Mueller}, E. and {Mu{\~n}oz-Guti{\'e}rrez}, A. and {Myers}, A.~D. and {Nadathur}, S. and {Napolitano}, L. and {Neveux}, R. and {Newman}, J.~A. and {Nguyen}, N.~M. and {Nie}, J. and {Niz}, G. and {Noriega}, H.~E. and {Padmanabhan}, N. and {Paillas}, E. and {Palanque-Delabrouille}, N. and {Pan}, J. and {Penmetsa}, S. and {Percival}, W.~J. and {Pieri}, M.~M. and {Pinon}, M. and {Poppett}, C. and {Porredon}, A. and {Prada}, F. and {P{\'e}rez-Fern{\'a}ndez}, A. and {P{\'e}rez-R{\`a}fols}, I. and {Rabinowitz}, D. and {Raichoor}, A. and {Ram{\'\i}rez-P{\'e}rez}, C. and {Ramirez-Solano}, S. and {Rashkovetskyi}, M. and {Ravoux}, C. and {Rezaie}, M. and {Rich}, J. and {Rocher}, A. and {Rockosi}, C. and {Rodr{\'\i}guez-Mart{\'\i}nez}, F. and {Roe}, N.~A. and {Rosado-Marin}, A. and {Ross}, A.~J. and {Rossi}, G. and {Ruggeri}, R. and {Ruhlmann-Kleider}, V. and {Samushia}, L. and {Sanchez}, E. and {Saulder}, C. and {Schlafly}, E.~F. and {Schlegel}, D. and {Schubnell}, M. and {Seo}, H. and {Sharples}, R. and {Silber}, J. and {Slosar}, A. and {Smith}, A. and {Sprayberry}, D. and {Tan}, T. and {Tarl{\'e}}, G. and {Trusov}, S. and {Vaisakh}, R. and {Valcin}, D. and {Valdes}, F. and {Vargas-Maga{\~n}a}, M. and {Verde}, L. and {Walther}, M. and {Wang}, B. and {Wang}, M.~S. and {Weaver}, B.~A. and {Weaverdyck}, N. and {Wechsler}, R.~H. and {Weinberg}, D.~H. and {White}, M. and {Wilson}, M.~J. and {Yu}, J. and {Yu}, Y. and {Yuan}, S. and {Y{\`e}che}, C. and {Zaborowski}, E.~A. and {Zarrouk}, P. and {Zhang}, H. and {Zhao}, C. and {Zhao}, R. and {Zhou}, R. and {Zou}, H. and {The DESI collaboration}},
        title = "{DESI 2024 V: Full-Shape galaxy clustering from galaxies and quasars}",
      journal = {\jcap},
         year = 2025,
        month = sep,
       volume = {2025},
       number = {9},
          eid = {008},
        pages = {008},
          doi = {10.1088/1475-7516/2025/09/008},
archivePrefix = {arXiv},
       eprint = {2411.12021},
 primaryClass = {astro-ph.CO},
       adsurl = {https://ui.adsabs.harvard.edu/abs/2025JCAP...09..008A}
}

@ARTICLE{planck_2018_cmb,
       author = {{Planck Collaboration} and {Aghanim}, N. and {Akrami}, Y. and {Ashdown}, M. and {Aumont}, J. and {Baccigalupi}, C. and {Ballardini}, M. and {Banday}, A.~J. and {Barreiro}, R.~B. and {Bartolo}, N. and {Basak}, S. and {Battye}, R. and {Benabed}, K. and {Bernard}, J. -P. and {Bersanelli}, M. and {Bielewicz}, P. and {Bock}, J.~J. and {Bond}, J.~R. and {Borrill}, J. and {Bouchet}, F.~R. and {Boulanger}, F. and {Bucher}, M. and {Burigana}, C. and {Butler}, R.~C. and {Calabrese}, E. and {Cardoso}, J. -F. and {Carron}, J. and {Challinor}, A. and {Chiang}, H.~C. and {Chluba}, J. and {Colombo}, L.~P.~L. and {Combet}, C. and {Contreras}, D. and {Crill}, B.~P. and {Cuttaia}, F. and {de Bernardis}, P. and {de Zotti}, G. and {Delabrouille}, J. and {Delouis}, J. -M. and {Di Valentino}, E. and {Diego}, J.~M. and {Dor{\'e}}, O. and {Douspis}, M. and {Ducout}, A. and {Dupac}, X. and {Dusini}, S. and {Efstathiou}, G. and {Elsner}, F. and {En{\ss}lin}, T.~A. and {Eriksen}, H.~K. and {Fantaye}, Y. and {Farhang}, M. and {Fergusson}, J. and {Fernandez-Cobos}, R. and {Finelli}, F. and {Forastieri}, F. and {Frailis}, M. and {Fraisse}, A.~A. and {Franceschi}, E. and {Frolov}, A. and {Galeotta}, S. and {Galli}, S. and {Ganga}, K. and {G{\'e}nova-Santos}, R.~T. and {Gerbino}, M. and {Ghosh}, T. and {Gonz{\'a}lez-Nuevo}, J. and {G{\'o}rski}, K.~M. and {Gratton}, S. and {Gruppuso}, A. and {Gudmundsson}, J.~E. and {Hamann}, J. and {Handley}, W. and {Hansen}, F.~K. and {Herranz}, D. and {Hildebrandt}, S.~R. and {Hivon}, E. and {Huang}, Z. and {Jaffe}, A.~H. and {Jones}, W.~C. and {Karakci}, A. and {Keih{\"a}nen}, E. and {Keskitalo}, R. and {Kiiveri}, K. and {Kim}, J. and {Kisner}, T.~S. and {Knox}, L. and {Krachmalnicoff}, N. and {Kunz}, M. and {Kurki-Suonio}, H. and {Lagache}, G. and {Lamarre}, J. -M. and {Lasenby}, A. and {Lattanzi}, M. and {Lawrence}, C.~R. and {Le Jeune}, M. and {Lemos}, P. and {Lesgourgues}, J. and {Levrier}, F. and {Lewis}, A. and {Liguori}, M. and {Lilje}, P.~B. and {Lilley}, M. and {Lindholm}, V. and {L{\'o}pez-Caniego}, M. and {Lubin}, P.~M. and {Ma}, Y. -Z. and {Mac{\'\i}as-P{\'e}rez}, J.~F. and {Maggio}, G. and {Maino}, D. and {Mandolesi}, N. and {Mangilli}, A. and {Marcos-Caballero}, A. and {Maris}, M. and {Martin}, P.~G. and {Martinelli}, M. and {Mart{\'\i}nez-Gonz{\'a}lez}, E. and {Matarrese}, S. and {Mauri}, N. and {McEwen}, J.~D. and {Meinhold}, P.~R. and {Melchiorri}, A. and {Mennella}, A. and {Migliaccio}, M. and {Millea}, M. and {Mitra}, S. and {Miville-Desch{\^e}nes}, M. -A. and {Molinari}, D. and {Montier}, L. and {Morgante}, G. and {Moss}, A. and {Natoli}, P. and {N{\o}rgaard-Nielsen}, H.~U. and {Pagano}, L. and {Paoletti}, D. and {Partridge}, B. and {Patanchon}, G. and {Peiris}, H.~V. and {Perrotta}, F. and {Pettorino}, V. and {Piacentini}, F. and {Polastri}, L. and {Polenta}, G. and {Puget}, J. -L. and {Rachen}, J.~P. and {Reinecke}, M. and {Remazeilles}, M. and {Renzi}, A. and {Rocha}, G. and {Rosset}, C. and {Roudier}, G. and {Rubi{\~n}o-Mart{\'\i}n}, J.~A. and {Ruiz-Granados}, B. and {Salvati}, L. and {Sandri}, M. and {Savelainen}, M. and {Scott}, D. and {Shellard}, E.~P.~S. and {Sirignano}, C. and {Sirri}, G. and {Spencer}, L.~D. and {Sunyaev}, R. and {Suur-Uski}, A. -S. and {Tauber}, J.~A. and {Tavagnacco}, D. and {Tenti}, M. and {Toffolatti}, L. and {Tomasi}, M. and {Trombetti}, T. and {Valenziano}, L. and {Valiviita}, J. and {Van Tent}, B. and {Vibert}, L. and {Vielva}, P. and {Villa}, F. and {Vittorio}, N. and {Wandelt}, B.~D. and {Wehus}, I.~K. and {White}, M. and {White}, S.~D.~M. and {Zacchei}, A. and {Zonca}, A.},
        title = "{Planck 2018 results. VI. Cosmological parameters}",
      journal = {Astron. Astrophys.},
         year = 2020,
        month = sep,
       volume = {641},
          eid = {A6},
        pages = {A6},
          doi = {10.1051/0004-6361/201833910},
archivePrefix = {arXiv},
       eprint = {1807.06209},
 primaryClass = {astro-ph.CO},
       adsurl = {https://ui.adsabs.harvard.edu/abs/2020A&A...641A...6P}
}

@ARTICLE{planck_pr4_results,
       author = {{Tristram}, M. and {Banday}, A.~J. and {Douspis}, M. and {Garrido}, X. and {G{\'o}rski}, K.~M. and {Henrot-Versill{\'e}}, S. and {Hergt}, L.~T. and {Ili{\'c}}, S. and {Keskitalo}, R. and {Lagache}, G. and {Lawrence}, C.~R. and {Partridge}, B. and {Scott}, D.},
        title = "{Cosmological parameters derived from the final Planck data release (PR4)}",
      journal = {\aap},
         year = 2024,
        month = feb,
       volume = {682},
          eid = {A37},
        pages = {A37},
          doi = {10.1051/0004-6361/202348015},
archivePrefix = {arXiv},
       eprint = {2309.10034},
 primaryClass = {astro-ph.CO},
       adsurl = {https://ui.adsabs.harvard.edu/abs/2024A&A...682A..37T}
}

@ARTICLE{planck_pr4_lensing,
       author = {{Carron}, Julien and {Mirmelstein}, Mark and {Lewis}, Antony},
        title = "{CMB lensing from Planck PR4 maps}",
      journal = {\jcap},
         year = 2022,
        month = sep,
       volume = {2022},
       number = {9},
          eid = {039},
        pages = {039},
          doi = {10.1088/1475-7516/2022/09/039},
archivePrefix = {arXiv},
       eprint = {2206.07773},
 primaryClass = {astro-ph.CO},
       adsurl = {https://ui.adsabs.harvard.edu/abs/2022JCAP...09..039C}
}

@ARTICLE{planck_iswl,
       author = {{Carron}, Julien and {Lewis}, Antony and {Fabbian}, Giulio},
        title = "{Planck integrated Sachs-Wolfe-lensing likelihood and the CMB temperature}",
      journal = {\prd},
         year = 2022,
        month = nov,
       volume = {106},
       number = {10},
          eid = {103507},
        pages = {103507},
          doi = {10.1103/PhysRevD.106.103507},
archivePrefix = {arXiv},
       eprint = {2209.07395},
 primaryClass = {astro-ph.CO},
       adsurl = {https://ui.adsabs.harvard.edu/abs/2022PhRvD.106j3507C}
}

@ARTICLE{act_cmb,
       author = {{Aiola}, Simone and {Calabrese}, Erminia and {Maurin}, Lo{\"\i}c and {Naess}, Sigurd and {Schmitt}, Benjamin L. and {Abitbol}, Maximilian H. and {Addison}, Graeme E. and {Ade}, Peter A.~R. and {Alonso}, David and {Amiri}, Mandana and {Amodeo}, Stefania and {Angile}, Elio and {Austermann}, Jason E. and {Baildon}, Taylor and {Battaglia}, Nick and {Beall}, James A. and {Bean}, Rachel and {Becker}, Daniel T. and {Bond}, J. Richard and {Bruno}, Sarah Marie and {Calafut}, Victoria and {Campusano}, Luis E. and {Carrero}, Felipe and {Chesmore}, Grace E. and {Cho}, Hsiao-mei and {Choi}, Steve K. and {Clark}, Susan E. and {Cothard}, Nicholas F. and {Crichton}, Devin and {Crowley}, Kevin T. and {Darwish}, Omar and {Datta}, Rahul and {Denison}, Edward V. and {Devlin}, Mark J. and {Duell}, Cody J. and {Duff}, Shannon M. and {Duivenvoorden}, Adriaan J. and {Dunkley}, Jo and {D{\"u}nner}, Rolando and {Essinger-Hileman}, Thomas and {Fankhanel}, Max and {Ferraro}, Simone and {Fox}, Anna E. and {Fuzia}, Brittany and {Gallardo}, Patricio A. and {Gluscevic}, Vera and {Golec}, Joseph E. and {Grace}, Emily and {Gralla}, Megan and {Guan}, Yilun and {Hall}, Kirsten and {Halpern}, Mark and {Han}, Dongwon and {Hargrave}, Peter and {Hasselfield}, Matthew and {Helton}, Jakob M. and {Henderson}, Shawn and {Hensley}, Brandon and {Hill}, J. Colin and {Hilton}, Gene C. and {Hilton}, Matt and {Hincks}, Adam D. and {Hlo{\v{z}}ek}, Ren{\'e}e and {Ho}, Shuay-Pwu Patty and {Hubmayr}, Johannes and {Huffenberger}, Kevin M. and {Hughes}, John P. and {Infante}, Leopoldo and {Irwin}, Kent and {Jackson}, Rebecca and {Klein}, Jeff and {Knowles}, Kenda and {Koopman}, Brian and {Kosowsky}, Arthur and {Lakey}, Vincent and {Li}, Dale and {Li}, Yaqiong and {Li}, Zack and {Lokken}, Martine and {Louis}, Thibaut and {Lungu}, Marius and {MacInnis}, Amanda and {Madhavacheril}, Mathew and {Maldonado}, Felipe and {Mallaby-Kay}, Maya and {Marsden}, Danica and {McMahon}, Jeff and {Menanteau}, Felipe and {Moodley}, Kavilan and {Morton}, Tim and {Namikawa}, Toshiya and {Nati}, Federico and {Newburgh}, Laura and {Nibarger}, John P. and {Nicola}, Andrina and {Niemack}, Michael D. and {Nolta}, Michael R. and {Orlowski-Sherer}, John and {Page}, Lyman A. and {Pappas}, Christine G. and {Partridge}, Bruce and {Phakathi}, Phumlani and {Pisano}, Giampaolo and {Prince}, Heather and {Puddu}, Roberto and {Qu}, Frank J. and {Rivera}, Jesus and {Robertson}, Naomi and {Rojas}, Felipe and {Salatino}, Maria and {Schaan}, Emmanuel and {Schillaci}, Alessandro and {Sehgal}, Neelima and {Sherwin}, Blake D. and {Sierra}, Carlos and {Sievers}, Jon and {Sifon}, Cristobal and {Sikhosana}, Precious and {Simon}, Sara and {Spergel}, David N. and {Staggs}, Suzanne T. and {Stevens}, Jason and {Storer}, Emilie and {Sunder}, Dhaneshwar D. and {Switzer}, Eric R. and {Thorne}, Ben and {Thornton}, Robert and {Trac}, Hy and {Treu}, Jesse and {Tucker}, Carole and {Vale}, Leila R. and {Van Engelen}, Alexander and {Van Lanen}, Jeff and {Vavagiakis}, Eve M. and {Wagoner}, Kasey and {Wang}, Yuhan and {Ward}, Jonathan T. and {Wollack}, Edward J. and {Xu}, Zhilei and {Zago}, Fernando and {Zhu}, Ningfeng},
        title = "{The Atacama Cosmology Telescope: DR4 maps and cosmological parameters}",
      journal = {Journal of Cosmology and Astroparticle Physics},
         year = 2020,
        month = dec,
       volume = {2020},
       number = {12},
          eid = {047},
        pages = {047},
          doi = {10.1088/1475-7516/2020/12/047},
archivePrefix = {arXiv},
       eprint = {2007.07288},
 primaryClass = {astro-ph.CO},
       adsurl = {https://ui.adsabs.harvard.edu/abs/2020JCAP...12..047A}
}

@ARTICLE{act_dr6_results,
       author = {{Louis}, Thibaut and {La Posta}, Adrien and {Atkins}, Zachary and {Jense}, Hidde T. and {Abril-Cabezas}, Irene and {Addison}, Graeme E. and {Ade}, Peter A.~R. and {Aiola}, Simone and {Alford}, Tommy and {Alonso}, David and {Amiri}, Mandana and {An}, Rui and {Austermann}, Jason E. and {Barbavara}, Eleonora and {Battaglia}, Nicholas and {Battistelli}, Elia Stefano and {Beall}, James A. and {Bean}, Rachel and {Beheshti}, Ali and {Beringue}, Benjamin and {Bhandarkar}, Tanay and {Biermann}, Emily and {Bolliet}, Boris and {Bond}, J. Richard and {Calabrese}, Erminia and {Capalbo}, Valentina and {Carrero}, Felipe and {Chen}, Shi-Fan and {Chesmore}, Grace and {Cho}, Hsiao-mei and {Choi}, Steve K. and {Clark}, Susan E. and {Cothard}, Nicholas F. and {Coughlin}, Kevin and {Coulton}, William and {Crichton}, Devin and {Crowley}, Kevin T. and {Darwish}, Omar and {Devlin}, Mark J. and {Dicker}, Simon and {Duell}, Cody J. and {Duff}, Shannon M. and {Duivenvoorden}, Adriaan J. and {Dunkley}, Jo and {Dunner}, Rolando and {Embil Villagra}, Carmen and {Fankhanel}, Max and {Farren}, Gerrit S. and {Ferraro}, Simone and {Foster}, Allen and {Freundt}, Rodrigo and {Fuzia}, Brittany and {Gallardo}, Patricio A. and {Garrido}, Xavier and {Gerbino}, Martina and {Giardiello}, Serena and {Gill}, Ajay and {Givans}, Jahmour and {Gluscevic}, Vera and {Goldstein}, Samuel and {Golec}, Joseph E. and {Gong}, Yulin and {Guan}, Yilun and {Halpern}, Mark and {Harrison}, Ian and {Hasselfield}, Matthew and {Healy}, Erin and {Henderson}, Shawn and {Hensley}, Brandon and {Herv{\'\i}as-Caimapo}, Carlos and {Hill}, J. Colin and {Hilton}, Gene C. and {Hilton}, Matt and {Hincks}, Adam D. and {Hlo{\v{z}}ek}, Ren{\'e}e and {Ho}, Shuay-Pwu Patty and {Hood}, John and {Hornecker}, Erika and {Huber}, Zachary B. and {Hubmayr}, Johannes and {Huffenberger}, Kevin M. and {Hughes}, John P. and {Ikape}, Margaret and {Irwin}, Kent and {Isopi}, Giovanni and {Joshi}, Neha and {Keller}, Ben and {Kim}, Joshua and {Knowles}, Kenda and {Koopman}, Brian J. and {Kosowsky}, Arthur and {Kramer}, Darby and {Kusiak}, Aleksandra and {Lagu{\"e}}, Alex and {Lakey}, Victoria and {Lee}, Eunseong and {Li}, Yaqiong and {Li}, Zack and {Limon}, Michele and {Lokken}, Martine and {Lungu}, Marius and {MacCrann}, Niall and {MacInnis}, Amanda and {Madhavacheril}, Mathew S. and {Maldonado}, Diego and {Maldonado}, Felipe and {Mallaby-Kay}, Maya and {Marques}, Gabriela A. and {van Marrewijk}, Joshiwa and {McCarthy}, Fiona and {McMahon}, Jeff and {Mehta}, Yogesh and {Menanteau}, Felipe and {Moodley}, Kavilan and {Morris}, Thomas W. and {Mroczkowski}, Tony and {Naess}, Sigurd and {Namikawa}, Toshiya and {Nati}, Federico and {Nerval}, Simran K. and {Newburgh}, Laura and {Nicola}, Andrina and {Niemack}, Michael D. and {Nolta}, Michael R. and {Orlowski-Scherer}, John and {Pagano}, Luca and {Page}, Lyman A. and {Pandey}, Shivam and {Partridge}, Bruce and {Perez Sarmiento}, Karen and {Prince}, Heather and {Puddu}, Roberto and {Qu}, Frank J. and {Ragavan}, Damien C. and {Ried Guachalla}, Bernardita and {Rogers}, Keir K. and {Rojas}, Felipe and {Sakuma}, Tai and {Schaan}, Emmanuel and {Schmitt}, Benjamin L. and {Sehgal}, Neelima and {Shaikh}, Shabbir and {Sherwin}, Blake D. and {Sierra}, Carlos and {Sievers}, Jon and {Sif{\'o}n}, Crist{\'o}bal and {Simon}, Sara and {Sonka}, Rita and {Spergel}, David N. and {Staggs}, Suzanne T. and {Storer}, Emilie and {Surrao}, Kristen and {Switzer}, Eric R. and {Tampier}, Niklas and {Thornton}, Robert and {Trac}, Hy and {Tucker}, Carole and {Ullom}, Joel and {Vale}, Leila R. and {Van Engelen}, Alexander and {Van Lanen}, Jeff and {Vargas}, Cristian and {Vavagiakis}, Eve M. and {Wagoner}, Kasey and {Wang}, Yuhan and {Wenzl}, Lukas and {Wollack}, Edward J. and {Zheng}, Kaiwen and {The Atacama Cosmology Telescope collaboration}},
        title = "{The Atacama Cosmology Telescope: DR6 power spectra, likelihoods and {\ensuremath{\Lambda}}CDM parameters}",
      journal = {\jcap},
         year = 2025,
        month = nov,
       volume = {2025},
       number = {11},
          eid = {062},
        pages = {062},
          doi = {10.1088/1475-7516/2025/11/062},
archivePrefix = {arXiv},
       eprint = {2503.14452},
 primaryClass = {astro-ph.CO},
       adsurl = {https://ui.adsabs.harvard.edu/abs/2025JCAP...11..062L}
}

@ARTICLE{spt_cmb,
       author = {{Balkenhol}, L. and {Dutcher}, D. and {Spurio Mancini}, A. and {Doussot}, A. and {Benabed}, K. and {Galli}, S. and {Ade}, P.~A.~R. and {Anderson}, A.~J. and {Ansarinejad}, B. and {Archipley}, M. and {Bender}, A.~N. and {Benson}, B.~A. and {Bianchini}, F. and {Bleem}, L.~E. and {Bouchet}, F.~R. and {Bryant}, L. and {Camphuis}, E. and {Carlstrom}, J.~E. and {Cecil}, T.~W. and {Chang}, C.~L. and {Chaubal}, P. and {Chichura}, P.~M. and {Chou}, T. -L. and {Coerver}, A. and {Crawford}, T.~M. and {Cukierman}, A. and {Daley}, C. and {de Haan}, T. and {Dibert}, K.~R. and {Dobbs}, M.~A. and {Everett}, W. and {Feng}, C. and {Ferguson}, K.~R. and {Foster}, A. and {Gambrel}, A.~E. and {Gardner}, R.~W. and {Goeckner-Wald}, N. and {Gualtieri}, R. and {Guidi}, F. and {Guns}, S. and {Halverson}, N.~W. and {Hivon}, E. and {Holder}, G.~P. and {Holzapfel}, W.~L. and {Hood}, J.~C. and {Huang}, N. and {Knox}, L. and {Korman}, M. and {Kuo}, C. -L. and {Lee}, A.~T. and {Lowitz}, A.~E. and {Lu}, C. and {Millea}, M. and {Montgomery}, J. and {Nakato}, Y. and {Natoli}, T. and {Noble}, G.~I. and {Novosad}, V. and {Omori}, Y. and {Padin}, S. and {Pan}, Z. and {Paschos}, P. and {Prabhu}, K. and {Quan}, W. and {Rahimi}, M. and {Rahlin}, A. and {Reichardt}, C.~L. and {Rouble}, M. and {Ruhl}, J.~E. and {Schiappucci}, E. and {Smecher}, G. and {Sobrin}, J.~A. and {Stark}, A.~A. and {Stephen}, J. and {Suzuki}, A. and {Tandoi}, C. and {Thompson}, K.~L. and {Thorne}, B. and {Tucker}, C. and {Umilta}, C. and {Vieira}, J.~D. and {Wang}, G. and {Whitehorn}, N. and {Wu}, W.~L.~K. and {Yefremenko}, V. and {Young}, M.~R. and {Zebrowski}, J.~A. and {SPT-3G Collaboration}},
        title = "{Measurement of the CMB temperature power spectrum and constraints on cosmology from the SPT-3G 2018 T T , T E , and E E dataset}",
      journal = {Phys. Rev. D},
         year = 2023,
        month = jul,
       volume = {108},
       number = {2},
          eid = {023510},
        pages = {023510},
          doi = {10.1103/PhysRevD.108.023510},
archivePrefix = {arXiv},
       eprint = {2212.05642},
 primaryClass = {astro-ph.CO},
       adsurl = {https://ui.adsabs.harvard.edu/abs/2023PhRvD.108b3510B}
}

@ARTICLE{spt_3g_2025,
       author = {{Camphuis}, E. and {Quan}, W. and {Balkenhol}, L. and {Khalife}, A.~R. and {Ge}, F. and {Guidi}, F. and {Huang}, N. and {Lynch}, G.~P. and {Omori}, Y. and {Trendafilova}, C. and {Anderson}, A.~J. and {Ansarinejad}, B. and {Archipley}, M. and {Barry}, P.~S. and {Benabed}, K. and {Bender}, A.~N. and {Benson}, B.~A. and {Bianchini}, F. and {Bleem}, L.~E. and {Bouchet}, F.~R. and {Bryant}, L. and {Campitiello}, M.~G. and {Carlstrom}, J.~E. and {Chang}, C.~L. and {Chaubal}, P. and {Chichura}, P.~M. and {Chokshi}, A. and {Chou}, T.-L. and {Coerver}, A. and {Crawford}, T.~M. and {Daley}, C. and {de Haan}, T. and {Dibert}, K.~R. and {Dobbs}, M.~A. and {Doohan}, M. and {Doussot}, A. and {Dutcher}, D. and {Everett}, W. and {Feng}, C. and {Ferguson}, K.~R. and {Fichman}, K. and {Foster}, A. and {Galli}, S. and {Gambrel}, A.~E. and {Gardner}, R.~W. and {Goeckner-Wald}, N. and {Gualtieri}, R. and {Guns}, S. and {Halverson}, N.~W. and {Hivon}, E. and {Holder}, G.~P. and {Holzapfel}, W.~L. and {Hood}, J.~C. and {Hryciuk}, A. and {K{\'e}ruzor{\'e}}, F. and {Knox}, L. and {Korman}, M. and {Kornoelje}, K. and {Kuo}, C.-L. and {Levy}, K. and {Lowitz}, A.~E. and {Lu}, C. and {Maniyar}, A. and {Martsen}, E.~S. and {Menanteau}, F. and {Millea}, M. and {Montgomery}, J. and {Nakato}, Y. and {Natoli}, T. and {Noble}, G.~I. and {Ouellette}, A. and {Pan}, Z. and {Paschos}, P. and {Phadke}, K.~A. and {Pollak}, A.~W. and {Prabhu}, K. and {Raghunathan}, S. and {Rahimi}, M. and {Rahlin}, A. and {Reichardt}, C.~L. and {Rouble}, M. and {Ruhl}, J.~E. and {Schiappucci}, E. and {Simpson}, A. and {Sobrin}, J.~A. and {Stark}, A.~A. and {Stephen}, J. and {Tandoi}, C. and {Thorne}, B. and {Umilta}, C. and {Vieira}, J.~D. and {Vitrier}, A. and {Wan}, Y. and {Whitehorn}, N. and {Wu}, W.~L.~K. and {Young}, M.~R. and {Zebrowski}, J.~A.},
        title = "{SPT-3G D1: CMB temperature and polarization power spectra and cosmology from 2019 and 2020 observations of the SPT-3G Main field}",
      journal = {arXiv e-prints},
         year = 2025,
        month = jun,
          eid = {arXiv:2506.20707},
        pages = {arXiv:2506.20707},
          doi = {10.48550/arXiv.2506.20707},
archivePrefix = {arXiv},
       eprint = {2506.20707},
 primaryClass = {astro-ph.CO},
       adsurl = {https://ui.adsabs.harvard.edu/abs/2025arXiv250620707C}
}

@ARTICLE{sh0es_hubble,
       author = {{Galbany}, L. and {de Jaeger}, T. and {Riess}, A.~G. and {M{\"u}ller-Bravo}, T.~E. and {Dhawan}, S. and {Phan}, K. and {Stritzinger}, M.~D. and {Karamehmetoglu}, E. and {Leibundgut}, B. and {Burns}, C. and {Peterson}, E. and {D'Arcy Kenworthy}, W. and {Johansson}, J. and {Maguire}, K. and {Jha}, S.~W.},
        title = "{An updated measurement of the Hubble constant from near-infrared observations of Type Ia supernovae}",
      journal = {Astron. Astrophys.},
         year = 2023,
        month = nov,
       volume = {679},
          eid = {A95},
        pages = {A95},
          doi = {10.1051/0004-6361/202244893},
archivePrefix = {arXiv},
       eprint = {2209.02546},
 primaryClass = {astro-ph.CO},
       adsurl = {https://ui.adsabs.harvard.edu/abs/2023A&A...679A..95G}
}

@ARTICLE{local_dist_network,
       author = {{H0DN Collaboration} and {Casertano}, Stefano and {Anand}, Gagandeep and {Anderson}, Richard I. and {Beaton}, Rachael and {Bhardwaj}, Anupam and {Blakeslee}, John P. and {Boubel}, Paula and {Breuval}, Louise and {Brout}, Dillon and {Cantiello}, Michele and {Cruz Reyes}, Mauricio and {Cs{\"o}rnyei}, Geza and {de Jaeger}, Thomas and {Dhawan}, Suhail and {Di Valentino}, Eleonora and {Galbany}, Llu{\'\i}s and {Gil-Mar{\'\i}n}, H{\'e}ctor and {Graczyk}, Dariusz and {Huang}, Caroline and {Jensen}, Joseph B. and {Kervella}, Pierre and {Leibundgut}, Bruno and {Lengen}, Bastian and {Li}, Siyang and {Macri}, Lucas and {{\"O}z{\"u}lker}, Emre and {Pesce}, Dominic W. and {Riess}, Adam and {Romaniello}, Martino and {Said}, Khaled and {Sch{\"o}neberg}, Nils and {Scolnic}, Dan and {Sicignano}, Teresa and {Skowron}, Dorota M. and {Uddin}, Syed A. and {Verde}, Licia and {Nota}, Antonella},
        title = "{The Local Distance Network: a community consensus report on the measurement of the Hubble constant at 1\% precision}",
      journal = {arXiv e-prints},
         year = 2025,
        month = oct,
          eid = {arXiv:2510.23823},
        pages = {arXiv:2510.23823},
          doi = {10.48550/arXiv.2510.23823},
archivePrefix = {arXiv},
       eprint = {2510.23823},
 primaryClass = {astro-ph.CO},
       adsurl = {https://ui.adsabs.harvard.edu/abs/2025arXiv251023823H}
}

@ARTICLE{pantheonplus,
       author = {{Scolnic}, Dan and {Brout}, Dillon and {Carr}, Anthony and {Riess}, Adam G. and {Davis}, Tamara M. and {Dwomoh}, Arianna and {Jones}, David O. and {Ali}, Noor and {Charvu}, Pranav and {Chen}, Rebecca and {Peterson}, Erik R. and {Popovic}, Brodie and {Rose}, Benjamin M. and {Wood}, Charlotte M. and {Brown}, Peter J. and {Chambers}, Ken and {Coulter}, David A. and {Dettman}, Kyle G. and {Dimitriadis}, Georgios and {Filippenko}, Alexei V. and {Foley}, Ryan J. and {Jha}, Saurabh W. and {Kilpatrick}, Charles D. and {Kirshner}, Robert P. and {Pan}, Yen-Chen and {Rest}, Armin and {Rojas-Bravo}, Cesar and {Siebert}, Matthew R. and {Stahl}, Benjamin E. and {Zheng}, WeiKang},
        title = "{The Pantheon+ Analysis: The Full Data Set and Light-curve Release}",
      journal = {Astrophys. J.},
         year = 2022,
        month = oct,
       volume = {938},
       number = {2},
          eid = {113},
        pages = {113},
          doi = {10.3847/1538-4357/ac8b7a},
archivePrefix = {arXiv},
       eprint = {2112.03863},
 primaryClass = {astro-ph.CO},
       adsurl = {https://ui.adsabs.harvard.edu/abs/2022ApJ...938..113S}
}

@ARTICLE{pantheon,
       author = {{Scolnic}, D.~M. and {Jones}, D.~O. and {Rest}, A. and {Pan}, Y.~C. and {Chornock}, R. and {Foley}, R.~J. and {Huber}, M.~E. and {Kessler}, R. and {Narayan}, G. and {Riess}, A.~G. and {Rodney}, S. and {Berger}, E. and {Brout}, D.~J. and {Challis}, P.~J. and {Drout}, M. and {Finkbeiner}, D. and {Lunnan}, R. and {Kirshner}, R.~P. and {Sanders}, N.~E. and {Schlafly}, E. and {Smartt}, S. and {Stubbs}, C.~W. and {Tonry}, J. and {Wood-Vasey}, W.~M. and {Foley}, M. and {Hand}, J. and {Johnson}, E. and {Burgett}, W.~S. and {Chambers}, K.~C. and {Draper}, P.~W. and {Hodapp}, K.~W. and {Kaiser}, N. and {Kudritzki}, R.~P. and {Magnier}, E.~A. and {Metcalfe}, N. and {Bresolin}, F. and {Gall}, E. and {Kotak}, R. and {McCrum}, M. and {Smith}, K.~W.},
        title = "{The Complete Light-curve Sample of Spectroscopically Confirmed SNe Ia from Pan-STARRS1 and Cosmological Constraints from the Combined Pantheon Sample}",
      journal = {Astrophys. J.},
         year = 2018,
        month = jun,
       volume = {859},
       number = {2},
          eid = {101},
        pages = {101},
          doi = {10.3847/1538-4357/aab9bb},
archivePrefix = {arXiv},
       eprint = {1710.00845},
 primaryClass = {astro-ph.CO},
       adsurl = {https://ui.adsabs.harvard.edu/abs/2018ApJ...859..101S}
}

@ARTICLE{desy5,
       author = {{M{\"o}ller}, A. and {Smith}, M. and {Sako}, M. and {Sullivan}, M. and {Vincenzi}, M. and {Wiseman}, P. and {Armstrong}, P. and {Asorey}, J. and {Brout}, D. and {Carollo}, D. and {Davis}, T.~M. and {Frohmaier}, C. and {Galbany}, L. and {Glazebrook}, K. and {Kelsey}, L. and {Kessler}, R. and {Lewis}, G.~F. and {Lidman}, C. and {Malik}, U. and {Nichol}, R.~C. and {Scolnic}, D. and {Tucker}, B.~E. and {Abbott}, T.~M.~C. and {Aguena}, M. and {Allam}, S. and {Annis}, J. and {Bertin}, E. and {Bocquet}, S. and {Brooks}, D. and {Burke}, D.~L. and {Carnero Rosell}, A. and {Carrasco Kind}, M. and {Carretero}, J. and {Castander}, F.~J. and {Conselice}, C. and {Costanzi}, M. and {Crocce}, M. and {da Costa}, L.~N. and {De Vicente}, J. and {Desai}, S. and {Diehl}, H.~T. and {Doel}, P. and {Everett}, S. and {Ferrero}, I. and {Finley}, D.~A. and {Flaugher}, B. and {Friedel}, D. and {Frieman}, J. and {Garc{\'\i}a-Bellido}, J. and {Gerdes}, D.~W. and {Gruen}, D. and {Gruendl}, R.~A. and {Gschwend}, J. and {Gutierrez}, G. and {Herner}, K. and {Hinton}, S.~R. and {Hollowood}, D.~L. and {Honscheid}, K. and {James}, D.~J. and {Kuehn}, K. and {Kuropatkin}, N. and {Lahav}, O. and {March}, M. and {Marshall}, J.~L. and {Menanteau}, F. and {Miquel}, R. and {Morgan}, R. and {Palmese}, A. and {Paz-Chinch{\'o}n}, F. and {Pieres}, A. and {Plazas Malag{\'o}n}, A.~A. and {Romer}, A.~K. and {Roodman}, A. and {Sanchez}, E. and {Scarpine}, V. and {Schubnell}, M. and {Serrano}, S. and {Sevilla-Noarbe}, I. and {Suchyta}, E. and {Tarle}, G. and {Thomas}, D. and {To}, C. and {Varga}, T.~N.},
        title = "{The dark energy survey 5-yr photometrically identified type Ia supernovae}",
      journal = {Mon. Not. Roy. Astron. Soc.},
         year = 2022,
        month = aug,
       volume = {514},
       number = {4},
        pages = {5159-5177},
          doi = {10.1093/mnras/stac1691},
archivePrefix = {arXiv},
       eprint = {2201.11142},
 primaryClass = {astro-ph.CO},
       adsurl = {https://ui.adsabs.harvard.edu/abs/2022MNRAS.514.5159M}
}

@ARTICLE{desy5_cosmo,
       author = {{DES Collaboration} and {Abbott}, T.~M.~C. and {Acevedo}, M. and {Aguena}, M. and {Alarcon}, A. and {Allam}, S. and {Alves}, O. and {Amon}, A. and {Andrade-Oliveira}, F. and {Annis}, J. and {Armstrong}, P. and {Asorey}, J. and {Avila}, S. and {Bacon}, D. and {Bassett}, B.~A. and {Bechtol}, K. and {Bernardinelli}, P.~H. and {Bernstein}, G.~M. and {Bertin}, E. and {Blazek}, J. and {Bocquet}, S. and {Brooks}, D. and {Brout}, D. and {Buckley-Geer}, E. and {Burke}, D.~L. and {Camacho}, H. and {Camilleri}, R. and {Campos}, A. and {Carnero Rosell}, A. and {Carollo}, D. and {Carr}, A. and {Carretero}, J. and {Castander}, F.~J. and {Cawthon}, R. and {Chang}, C. and {Chen}, R. and {Choi}, A. and {Conselice}, C. and {Costanzi}, M. and {da Costa}, L.~N. and {Crocce}, M. and {Davis}, T.~M. and {DePoy}, D.~L. and {Desai}, S. and {Diehl}, H.~T. and {Dixon}, M. and {Dodelson}, S. and {Doel}, P. and {Doux}, C. and {Drlica-Wagner}, A. and {Elvin-Poole}, J. and {Everett}, S. and {Ferrero}, I. and {Fert{\'e}}, A. and {Flaugher}, B. and {Foley}, R.~J. and {Fosalba}, P. and {Friedel}, D. and {Frieman}, J. and {Frohmaier}, C. and {Galbany}, L. and {Garc{\'\i}a-Bellido}, J. and {Gatti}, M. and {Gaztanaga}, E. and {Giannini}, G. and {Glazebrook}, K. and {Graur}, O. and {Gruen}, D. and {Gruendl}, R.~A. and {Gutierrez}, G. and {Hartley}, W.~G. and {Herner}, K. and {Hinton}, S.~R. and {Hollowood}, D.~L. and {Honscheid}, K. and {Huterer}, D. and {Jain}, B. and {James}, D.~J. and {Jeffrey}, N. and {Kasai}, E. and {Kelsey}, L. and {Kent}, S. and {Kessler}, R. and {Kim}, A.~G. and {Kirshner}, R.~P. and {Kovacs}, E. and {Kuehn}, K. and {Lahav}, O. and {Lee}, J. and {Lee}, S. and {Lewis}, G.~F. and {Li}, T.~S. and {Lidman}, C. and {Lin}, H. and {Malik}, U. and {Marshall}, J.~L. and {Martini}, P. and {Mena-Fern{\'a}ndez}, J. and {Menanteau}, F. and {Miquel}, R. and {Mohr}, J.~J. and {Mould}, J. and {Muir}, J. and {M{\"o}ller}, A. and {Neilsen}, E. and {Nichol}, R.~C. and {Nugent}, P. and {Ogando}, R.~L.~C. and {Palmese}, A. and {Pan}, Y. -C. and {Paterno}, M. and {Percival}, W.~J. and {Pereira}, M.~E.~S. and {Pieres}, A. and {Plazas Malag{\'o}n}, A.~A. and {Popovic}, B. and {Porredon}, A. and {Prat}, J. and {Qu}, H. and {Raveri}, M. and {Rodr{\'\i}guez-Monroy}, M. and {Romer}, A.~K. and {Roodman}, A. and {Rose}, B. and {Sako}, M. and {Sanchez}, E. and {Sanchez Cid}, D. and {Schubnell}, M. and {Scolnic}, D. and {Sevilla-Noarbe}, I. and {Shah}, P. and {Allyn. Smith}, J. and {Smith}, M. and {Soares-Santos}, M. and {Suchyta}, E. and {Sullivan}, M. and {Suntzeff}, N. and {Swanson}, M.~E.~C. and {S{\'a}nchez}, B.~O. and {Tarle}, G. and {Taylor}, G. and {Thomas}, D. and {To}, C. and {Toy}, M. and {Troxel}, M.~A. and {Tucker}, B.~E. and {Tucker}, D.~L. and {Uddin}, S.~A. and {Vincenzi}, M. and {Walker}, A.~R. and {Weaverdyck}, N. and {Wechsler}, R.~H. and {Weller}, J. and {Wester}, W. and {Wiseman}, P. and {Yamamoto}, M. and {Yuan}, F. and {Zhang}, B. and {Zhang}, Y.},
        title = "{The Dark Energy Survey: Cosmology Results With \raisebox{-0.5ex}\textasciitilde1500 New High-redshift Type Ia Supernovae Using The Full 5-year Dataset}",
      journal = {arXiv e-prints},
         year = 2024,
        month = jan,
          eid = {arXiv:2401.02929},
        pages = {arXiv:2401.02929},
          doi = {10.48550/arXiv.2401.02929},
archivePrefix = {arXiv},
       eprint = {2401.02929},
 primaryClass = {astro-ph.CO},
       adsurl = {https://ui.adsabs.harvard.edu/abs/2024arXiv240102929D}
}

@ARTICLE{des_dovekie,
       author = {{Popovic}, B. and {Shah}, P. and {Kenworthy}, W.~D. and {Kessler}, R. and {Davis}, T.~M. and {Goobar}, A. and {Scolnic}, D. and {Vincenzi}, M. and {Wiseman}, P. and {Chen}, R. and {Charleton}, E. and {Acevedo}, M. and {Armstrong}, P. and {Boyd}, B.~M. and {Brout}, D. and {Camilleri}, R. and {Frieman}, J. and {Galbany}, L. and {Grayling}, M. and {Kelsey}, L. and {Rose}, B. and {S{\'a}nchez}, B. and {Lee}, J. and {M{\"o}ller}, A. and {Smith}, M. and {Sullivan}, M. and {Shiamtanis}, N. and {Alarcon}, A. and {Allam}, S.~S. and {Andrade-Oliveira}, F. and {Avila}, S. and {Bacon}, D. and {Blazek}, J. and {Bocquet}, S. and {Brooks}, D. and {Burke}, D.~L. and {Carnero Rosell}, A. and {Carretero}, J. and {Cawthon}, R. and {da Costa}, L.~N. and {da Silva Pereira}, M.~E. and {Diehl}, H.~T. and {Dodelson}, S. and {Doel}, P. and {Everett}, S. and {Frohmaier}, C. and {Garc{\'\i}a-Bellido}, J. and {Gruen}, D. and {Gutierrez}, G. and {Herner}, K. and {Hinton}, S.~R. and {Hollowood}, D.~L. and {Honscheid}, K. and {Huterer}, D. and {James}, D.~J. and {Jeffrey}, N. and {Kuehn}, K. and {Lahav}, O. and {Lee}, S. and {Lidman}, C. and {Marshall}, J.~L. and {Mena-Fern{\'a}ndez}, J. and {Menanteau}, F. and {Miquel}, R. and {Muir}, J. and {Myles}, J. and {Ogando}, R.~L.~C. and {Paterno}, M. and {Plazas Malag{\'o}n}, A.~A. and {Porredon}, A. and {Prat}, J. and {Nichol}, R.~C. and {Romer}, A.~K. and {Roodman}, A. and {Sanchez}, E. and {Sanchez Cid}, D. and {Sevilla-Noarbe}, I. and {Suchyta}, E. and {Swanson}, M.~E.~C. and {To}, C. and {Tucker}, D.~L. and {Walker}, A.~R. and {Weaverdyck}, N.},
        title = "{The Dark Energy Survey Supernova Program: A Reanalysis Of Cosmology Results And Evidence For Evolving Dark Energy With An Updated Type Ia Supernova Calibration}",
      journal = {arXiv e-prints},
         year = 2025,
        month = nov,
          eid = {arXiv:2511.07517},
        pages = {arXiv:2511.07517},
          doi = {10.48550/arXiv.2511.07517},
archivePrefix = {arXiv},
       eprint = {2511.07517},
 primaryClass = {astro-ph.CO},
       adsurl = {https://ui.adsabs.harvard.edu/abs/2025arXiv251107517P}
}

@ARTICLE{union3,
       author = {{Rubin}, David and {Aldering}, Greg and {Betoule}, Marc and {Fruchter}, Andy and {Huang}, Xiaosheng and {Kim}, Alex G. and {Lidman}, Chris and {Linder}, Eric and {Perlmutter}, Saul and {Ruiz-Lapuente}, Pilar and {Suzuki}, Nao},
        title = "{Union Through UNITY: Cosmology with 2,000 SNe Using a Unified Bayesian Framework}",
      journal = {arXiv e-prints},
         year = 2023,
        month = nov,
          eid = {arXiv:2311.12098},
        pages = {arXiv:2311.12098},
          doi = {10.48550/arXiv.2311.12098},
archivePrefix = {arXiv},
       eprint = {2311.12098},
 primaryClass = {astro-ph.CO},
       adsurl = {https://ui.adsabs.harvard.edu/abs/2023arXiv231112098R}
}

@ARTICLE{union3-1,
       author = {{Hoyt}, Taylor J. and {Rubin}, David and {Aldering}, Greg and {Perlmutter}, Saul and {Cuceu}, Andrei and {Gupta}, Ravi},
        title = "{Union3.1: Self-consistent Measurements of Host Galaxy Properties for 2000 Type Ia Supernovae}",
      journal = {arXiv e-prints},
         year = 2026,
        month = jan,
          eid = {arXiv:2601.19424},
        pages = {arXiv:2601.19424},
          doi = {10.48550/arXiv.2601.19424},
archivePrefix = {arXiv},
       eprint = {2601.19424},
 primaryClass = {astro-ph.CO},
       adsurl = {https://ui.adsabs.harvard.edu/abs/2026arXiv260119424H}
}

@ARTICLE{kids-1000-shear,
       author = {{Asgari}, Marika and {Lin}, Chieh-An and {Joachimi}, Benjamin and {Giblin}, Benjamin and {Heymans}, Catherine and {Hildebrandt}, Hendrik and {Kannawadi}, Arun and {St{\"o}lzner}, Benjamin and {Tr{\"o}ster}, Tilman and {van den Busch}, Jan Luca and {Wright}, Angus H. and {Bilicki}, Maciej and {Blake}, Chris and {de Jong}, Jelte and {Dvornik}, Andrej and {Erben}, Thomas and {Getman}, Fedor and {Hoekstra}, Henk and {K{\"o}hlinger}, Fabian and {Kuijken}, Konrad and {Miller}, Lance and {Radovich}, Mario and {Schneider}, Peter and {Shan}, HuanYuan and {Valentijn}, Edwin},
        title = "{KiDS-1000 cosmology: Cosmic shear constraints and comparison between two point statistics}",
      journal = {Astron. Astrophys.},
         year = 2021,
        month = jan,
       volume = {645},
          eid = {A104},
        pages = {A104},
          doi = {10.1051/0004-6361/202039070},
archivePrefix = {arXiv},
       eprint = {2007.15633},
 primaryClass = {astro-ph.CO},
       adsurl = {https://ui.adsabs.harvard.edu/abs/2021A&A...645A.104A}
}

@ARTICLE{hsc-shear,
       author = {{Hamana}, Takashi and {Shirasaki}, Masato and {Miyazaki}, Satoshi and {Hikage}, Chiaki and {Oguri}, Masamune and {More}, Surhud and {Armstrong}, Robert and {Leauthaud}, Alexie and {Mandelbaum}, Rachel and {Miyatake}, Hironao and {Nishizawa}, Atsushi J. and {Simet}, Melanie and {Takada}, Masahiro and {Aihara}, Hiroaki and {Bosch}, James and {Komiyama}, Yutaka and {Lupton}, Robert and {Murayama}, Hitoshi and {Strauss}, Michael A. and {Tanaka}, Masayuki},
        title = "{Cosmological constraints from cosmic shear two-point correlation functions with HSC survey first-year data}",
      journal = {Publications of the Astronomical Society of Japan},
         year = 2020,
        month = feb,
       volume = {72},
       number = {1},
          eid = {16},
        pages = {16},
          doi = {10.1093/pasj/psz138},
archivePrefix = {arXiv},
       eprint = {1906.06041},
 primaryClass = {astro-ph.CO},
       adsurl = {https://ui.adsabs.harvard.edu/abs/2020PASJ...72...16H}
}

@ARTICLE{des_y1_3x2,
       author = {{Abbott}, T.~M.~C. and {Abdalla}, F.~B. and {Alarcon}, A. and {Aleksi{\'c}}, J. and {Allam}, S. and {Allen}, S. and {Amara}, A. and {Annis}, J. and {Asorey}, J. and {Avila}, S. and {Bacon}, D. and {Balbinot}, E. and {Banerji}, M. and {Banik}, N. and {Barkhouse}, W. and {Baumer}, M. and {Baxter}, E. and {Bechtol}, K. and {Becker}, M.~R. and {Benoit-L{\'e}vy}, A. and {Benson}, B.~A. and {Bernstein}, G.~M. and {Bertin}, E. and {Blazek}, J. and {Bridle}, S.~L. and {Brooks}, D. and {Brout}, D. and {Buckley-Geer}, E. and {Burke}, D.~L. and {Busha}, M.~T. and {Campos}, A. and {Capozzi}, D. and {Carnero Rosell}, A. and {Carrasco Kind}, M. and {Carretero}, J. and {Castander}, F.~J. and {Cawthon}, R. and {Chang}, C. and {Chen}, N. and {Childress}, M. and {Choi}, A. and {Conselice}, C. and {Crittenden}, R. and {Crocce}, M. and {Cunha}, C.~E. and {D'Andrea}, C.~B. and {da Costa}, L.~N. and {Das}, R. and {Davis}, T.~M. and {Davis}, C. and {De Vicente}, J. and {DePoy}, D.~L. and {DeRose}, J. and {Desai}, S. and {Diehl}, H.~T. and {Dietrich}, J.~P. and {Dodelson}, S. and {Doel}, P. and {Drlica-Wagner}, A. and {Eifler}, T.~F. and {Elliott}, A.~E. and {Elsner}, F. and {Elvin-Poole}, J. and {Estrada}, J. and {Evrard}, A.~E. and {Fang}, Y. and {Fernandez}, E. and {Fert{\'e}}, A. and {Finley}, D.~A. and {Flaugher}, B. and {Fosalba}, P. and {Friedrich}, O. and {Frieman}, J. and {Garc{\'\i}a-Bellido}, J. and {Garcia-Fernandez}, M. and {Gatti}, M. and {Gaztanaga}, E. and {Gerdes}, D.~W. and {Giannantonio}, T. and {Gill}, M.~S.~S. and {Glazebrook}, K. and {Goldstein}, D.~A. and {Gruen}, D. and {Gruendl}, R.~A. and {Gschwend}, J. and {Gutierrez}, G. and {Hamilton}, S. and {Hartley}, W.~G. and {Hinton}, S.~R. and {Honscheid}, K. and {Hoyle}, B. and {Huterer}, D. and {Jain}, B. and {James}, D.~J. and {Jarvis}, M. and {Jeltema}, T. and {Johnson}, M.~D. and {Johnson}, M.~W.~G. and {Kacprzak}, T. and {Kent}, S. and {Kim}, A.~G. and {King}, A. and {Kirk}, D. and {Kokron}, N. and {Kovacs}, A. and {Krause}, E. and {Krawiec}, C. and {Kremin}, A. and {Kuehn}, K. and {Kuhlmann}, S. and {Kuropatkin}, N. and {Lacasa}, F. and {Lahav}, O. and {Li}, T.~S. and {Liddle}, A.~R. and {Lidman}, C. and {Lima}, M. and {Lin}, H. and {MacCrann}, N. and {Maia}, M.~A.~G. and {Makler}, M. and {Manera}, M. and {March}, M. and {Marshall}, J.~L. and {Martini}, P. and {McMahon}, R.~G. and {Melchior}, P. and {Menanteau}, F. and {Miquel}, R. and {Miranda}, V. and {Mudd}, D. and {Muir}, J. and {M{\"o}ller}, A. and {Neilsen}, E. and {Nichol}, R.~C. and {Nord}, B. and {Nugent}, P. and {Ogando}, R.~L.~C. and {Palmese}, A. and {Peacock}, J. and {Peiris}, H.~V. and {Peoples}, J. and {Percival}, W.~J. and {Petravick}, D. and {Plazas}, A.~A. and {Porredon}, A. and {Prat}, J. and {Pujol}, A. and {Rau}, M.~M. and {Refregier}, A. and {Ricker}, P.~M. and {Roe}, N. and {Rollins}, R.~P. and {Romer}, A.~K. and {Roodman}, A. and {Rosenfeld}, R. and {Ross}, A.~J. and {Rozo}, E. and {Rykoff}, E.~S. and {Sako}, M. and {Salvador}, A.~I. and {Samuroff}, S. and {S{\'a}nchez}, C. and {Sanchez}, E. and {Santiago}, B. and {Scarpine}, V. and {Schindler}, R. and {Scolnic}, D. and {Secco}, L.~F. and {Serrano}, S. and {Sevilla-Noarbe}, I. and {Sheldon}, E. and {Smith}, R.~C. and {Smith}, M. and {Smith}, J. and {Soares-Santos}, M. and {Sobreira}, F. and {Suchyta}, E. and {Tarle}, G. and {Thomas}, D. and {Troxel}, M.~A. and {Tucker}, D.~L. and {Tucker}, B.~E. and {Uddin}, S.~A. and {Varga}, T.~N. and {Vielzeuf}, P. and {Vikram}, V. and {Vivas}, A.~K. and {Walker}, A.~R. and {Wang}, M. and {Wechsler}, R.~H. and {Weller}, J. and {Wester}, W. and {Wolf}, R.~C. and {Yanny}, B. and {Yuan}, F. and {Zenteno}, A. and {Zhang}, B. and {Zhang}, Y. and {Zuntz}, J. and {Dark Energy Survey Collaboration}},
        title = "{Dark Energy Survey year 1 results: Cosmological constraints from galaxy clustering and weak lensing}",
      journal = {Phys. Rev. D},
         year = 2018,
        month = aug,
       volume = {98},
       number = {4},
          eid = {043526},
        pages = {043526},
          doi = {10.1103/PhysRevD.98.043526},
archivePrefix = {arXiv},
       eprint = {1708.01530},
 primaryClass = {astro-ph.CO},
       adsurl = {https://ui.adsabs.harvard.edu/abs/2018PhRvD..98d3526A}
}

@ARTICLE{des_y3_extensions,
       author = {{Abbott}, T.~M.~C. and {Aguena}, M. and {Alarcon}, A. and {Alves}, O. and {Amon}, A. and {Andrade-Oliveira}, F. and {Annis}, J. and {Avila}, S. and {Bacon}, D. and {Baxter}, E. and {Bechtol}, K. and {Becker}, M.~R. and {Bernstein}, G.~M. and {Birrer}, S. and {Blazek}, J. and {Bocquet}, S. and {Brandao-Souza}, A. and {Bridle}, S.~L. and {Brooks}, D. and {Burke}, D.~L. and {Camacho}, H. and {Campos}, A. and {Carnero Rosell}, A. and {Carrasco Kind}, M. and {Carretero}, J. and {Castander}, F.~J. and {Cawthon}, R. and {Chang}, C. and {Chen}, A. and {Chen}, R. and {Choi}, A. and {Conselice}, C. and {Cordero}, J. and {Costanzi}, M. and {Crocce}, M. and {da Costa}, L.~N. and {Pereira}, M.~E.~S. and {Davis}, C. and {Davis}, T.~M. and {DeRose}, J. and {Desai}, S. and {Di Valentino}, E. and {Diehl}, H.~T. and {Dodelson}, S. and {Doel}, P. and {Doux}, C. and {Drlica-Wagner}, A. and {Eckert}, K. and {Eifler}, T.~F. and {Elsner}, F. and {Elvin-Poole}, J. and {Everett}, S. and {Fang}, X. and {Farahi}, A. and {Ferrero}, I. and {Fert{\'e}}, A. and {Flaugher}, B. and {Fosalba}, P. and {Friedel}, D. and {Friedrich}, O. and {Frieman}, J. and {Garc{\'\i}a-Bellido}, J. and {Gatti}, M. and {Giani}, L. and {Giannantonio}, T. and {Giannini}, G. and {Gruen}, D. and {Gruendl}, R.~A. and {Gschwend}, J. and {Gutierrez}, G. and {Hamaus}, N. and {Harrison}, I. and {Hartley}, W.~G. and {Herner}, K. and {Hinton}, S.~R. and {Hollowood}, D.~L. and {Honscheid}, K. and {Huang}, H. and {Huff}, E.~M. and {Huterer}, D. and {Jain}, B. and {James}, D.~J. and {Jarvis}, M. and {Jeffrey}, N. and {Jeltema}, T. and {Kovacs}, A. and {Krause}, E. and {Kuehn}, K. and {Kuropatkin}, N. and {Lahav}, O. and {Lee}, S. and {Leget}, P.-F. and {Lemos}, P. and {Leonard}, C.~D. and {Liddle}, A.~R. and {Lima}, M. and {Lin}, H. and {MacCrann}, N. and {Marshall}, J.~L. and {McCullough}, J. and {Mena-Fern{\'a}ndez}, J. and {Menanteau}, F. and {Miquel}, R. and {Miranda}, V. and {Mohr}, J.~J. and {Muir}, J. and {Myles}, J. and {Nadathur}, S. and {Navarro-Alsina}, A. and {Nichol}, R.~C. and {Ogando}, R.~L.~C. and {Omori}, Y. and {Palmese}, A. and {Pandey}, S. and {Park}, Y. and {Paterno}, M. and {Paz-Chinch{\'o}n}, F. and {Percival}, W.~J. and {Pieres}, A. and {Plazas Malag{\'o}n}, A.~A. and {Porredon}, A. and {Prat}, J. and {Raveri}, M. and {Rodriguez-Monroy}, M. and {Rogozenski}, P. and {Rollins}, R.~P. and {Romer}, A.~K. and {Roodman}, A. and {Rosenfeld}, R. and {Ross}, A.~J. and {Rykoff}, E.~S. and {Samuroff}, S. and {S{\'a}nchez}, C. and {Sanchez}, E. and {Sanchez}, J. and {Sanchez Cid}, D. and {Scarpine}, V. and {Scolnic}, D. and {Secco}, L.~F. and {Sevilla-Noarbe}, I. and {Sheldon}, E. and {Shin}, T. and {Smith}, M. and {Soares-Santos}, M. and {Suchyta}, E. and {Tabbutt}, M. and {Tarle}, G. and {Thomas}, D. and {To}, C. and {Troja}, A. and {Troxel}, M.~A. and {Tutusaus}, I. and {Varga}, T.~N. and {Vincenzi}, M. and {Walker}, A.~R. and {Weaverdyck}, N. and {Wechsler}, R.~H. and {Weller}, J. and {Yanny}, B. and {Yin}, B. and {Zhang}, Y. and {Zuntz}, J. and {DES Collaboration}},
        title = "{Dark Energy Survey Year 3 results: Constraints on extensions to {\ensuremath{\Lambda}} CDM with weak lensing and galaxy clustering}",
      journal = {\prd},
         year = 2023,
        month = apr,
       volume = {107},
       number = {8},
          eid = {083504},
        pages = {083504},
          doi = {10.1103/PhysRevD.107.083504},
archivePrefix = {arXiv},
       eprint = {2207.05766},
 primaryClass = {astro-ph.CO},
       adsurl = {https://ui.adsabs.harvard.edu/abs/2023PhRvD.107h3504A}
}

@ARTICLE{des_y3_3x2,
       author = {{Abbott}, T.~M.~C. and {Aguena}, M. and {Alarcon}, A. and {Allam}, S. and {Alves}, O. and {Amon}, A. and {Andrade-Oliveira}, F. and {Annis}, J. and {Avila}, S. and {Bacon}, D. and {Baxter}, E. and {Bechtol}, K. and {Becker}, M.~R. and {Bernstein}, G.~M. and {Bhargava}, S. and {Birrer}, S. and {Blazek}, J. and {Brandao-Souza}, A. and {Bridle}, S.~L. and {Brooks}, D. and {Buckley-Geer}, E. and {Burke}, D.~L. and {Camacho}, H. and {Campos}, A. and {Carnero Rosell}, A. and {Carrasco Kind}, M. and {Carretero}, J. and {Castander}, F.~J. and {Cawthon}, R. and {Chang}, C. and {Chen}, A. and {Chen}, R. and {Choi}, A. and {Conselice}, C. and {Cordero}, J. and {Costanzi}, M. and {Crocce}, M. and {da Costa}, L.~N. and {da Silva Pereira}, M.~E. and {Davis}, C. and {Davis}, T.~M. and {De Vicente}, J. and {DeRose}, J. and {Desai}, S. and {Di Valentino}, E. and {Diehl}, H.~T. and {Dietrich}, J.~P. and {Dodelson}, S. and {Doel}, P. and {Doux}, C. and {Drlica-Wagner}, A. and {Eckert}, K. and {Eifler}, T.~F. and {Elsner}, F. and {Elvin-Poole}, J. and {Everett}, S. and {Evrard}, A.~E. and {Fang}, X. and {Farahi}, A. and {Fernandez}, E. and {Ferrero}, I. and {Fert{\'e}}, A. and {Fosalba}, P. and {Friedrich}, O. and {Frieman}, J. and {Garc{\'\i}a-Bellido}, J. and {Gatti}, M. and {Gaztanaga}, E. and {Gerdes}, D.~W. and {Giannantonio}, T. and {Giannini}, G. and {Gruen}, D. and {Gruendl}, R.~A. and {Gschwend}, J. and {Gutierrez}, G. and {Harrison}, I. and {Hartley}, W.~G. and {Herner}, K. and {Hinton}, S.~R. and {Hollowood}, D.~L. and {Honscheid}, K. and {Hoyle}, B. and {Huff}, E.~M. and {Huterer}, D. and {Jain}, B. and {James}, D.~J. and {Jarvis}, M. and {Jeffrey}, N. and {Jeltema}, T. and {Kovacs}, A. and {Krause}, E. and {Kron}, R. and {Kuehn}, K. and {Kuropatkin}, N. and {Lahav}, O. and {Leget}, P. -F. and {Lemos}, P. and {Liddle}, A.~R. and {Lidman}, C. and {Lima}, M. and {Lin}, H. and {MacCrann}, N. and {Maia}, M.~A.~G. and {Marshall}, J.~L. and {Martini}, P. and {McCullough}, J. and {Melchior}, P. and {Mena-Fern{\'a}ndez}, J. and {Menanteau}, F. and {Miquel}, R. and {Mohr}, J.~J. and {Morgan}, R. and {Muir}, J. and {Myles}, J. and {Nadathur}, S. and {Navarro-Alsina}, A. and {Nichol}, R.~C. and {Ogando}, R.~L.~C. and {Omori}, Y. and {Palmese}, A. and {Pandey}, S. and {Park}, Y. and {Paz-Chinch{\'o}n}, F. and {Petravick}, D. and {Pieres}, A. and {Plazas Malag{\'o}n}, A.~A. and {Porredon}, A. and {Prat}, J. and {Raveri}, M. and {Rodriguez-Monroy}, M. and {Rollins}, R.~P. and {Romer}, A.~K. and {Roodman}, A. and {Rosenfeld}, R. and {Ross}, A.~J. and {Rykoff}, E.~S. and {Samuroff}, S. and {S{\'a}nchez}, C. and {Sanchez}, E. and {Sanchez}, J. and {Sanchez Cid}, D. and {Scarpine}, V. and {Schubnell}, M. and {Scolnic}, D. and {Secco}, L.~F. and {Serrano}, S. and {Sevilla-Noarbe}, I. and {Sheldon}, E. and {Shin}, T. and {Smith}, M. and {Soares-Santos}, M. and {Suchyta}, E. and {Swanson}, M.~E.~C. and {Tabbutt}, M. and {Tarle}, G. and {Thomas}, D. and {To}, C. and {Troja}, A. and {Troxel}, M.~A. and {Tucker}, D.~L. and {Tutusaus}, I. and {Varga}, T.~N. and {Walker}, A.~R. and {Weaverdyck}, N. and {Wechsler}, R. and {Weller}, J. and {Yanny}, B. and {Yin}, B. and {Zhang}, Y. and {Zuntz}, J. and {DES Collaboration}},
        title = "{Dark Energy Survey Year 3 results: Cosmological constraints from galaxy clustering and weak lensing}",
      journal = {Phys. Rev. D},
         year = 2022,
        month = jan,
       volume = {105},
       number = {2},
          eid = {023520},
        pages = {023520},
          doi = {10.1103/PhysRevD.105.023520},
archivePrefix = {arXiv},
       eprint = {2105.13549},
 primaryClass = {astro-ph.CO},
       adsurl = {https://ui.adsabs.harvard.edu/abs/2022PhRvD.105b3520A}
}

@ARTICLE{kids_legacy,
       author = {{St{\"o}lzner}, Benjamin and {Wright}, Angus H. and {Asgari}, Marika and {Heymans}, Catherine and {Hildebrandt}, Hendrik and {Hoekstra}, Henk and {Joachimi}, Benjamin and {Kuijken}, Konrad and {Li}, Shun-Sheng and {Mahony}, Constance and {Reischke}, Robert and {Yoon}, Mijin and {Bilicki}, Maciej and {Burger}, Pierre and {Chisari}, Nora Elisa and {Dvornik}, Andrej and {Georgiou}, Christos and {Giblin}, Benjamin and {Harnois-D{\'e}raps}, Joachim and {Jalan}, Priyanka and {William}, Anjitha John and {Joudaki}, Shahab and {Lesci}, Giorgio Francesco and {Linke}, Laila and {Loureiro}, Arthur and {Maturi}, Matteo and {Moscardini}, Lauro and {Napolitano}, Nicola R. and {Porth}, Lucas and {Radovich}, Mario and {Tr{\"o}ster}, Tilman and {Valentijn}, Edwin and {von Wietersheim-Kramsta}, Maximilian and {Wittje}, Anna and {Yan}, Ziang and {Zhang}, Yun-Hao},
        title = "{KiDS-Legacy: Consistency of cosmic shear measurements and joint cosmological constraints with external probes}",
      journal = {\aap},
         year = 2025,
        month = oct,
       volume = {702},
          eid = {A169},
        pages = {A169},
          doi = {10.1051/0004-6361/202554893},
archivePrefix = {arXiv},
       eprint = {2503.19442},
 primaryClass = {astro-ph.CO},
       adsurl = {https://ui.adsabs.harvard.edu/abs/2025A&A...702A.169S}
}

@ARTICLE{growth_modification_s8,
       author = {{Lin}, Meng-Xiang and {Jain}, Bhuvnesh and {Raveri}, Marco and {Baxter}, Eric J. and {Chang}, Chihway and {Gatti}, Marco and {Lee}, Sujeong and {Muir}, Jessica},
        title = "{Late time modification of structure growth and the S$_{8}$ tension}",
      journal = {Phys. Rev. D},
         year = 2024,
        month = mar,
       volume = {109},
       number = {6},
          eid = {063523},
        pages = {063523},
          doi = {10.1103/PhysRevD.109.063523},
archivePrefix = {arXiv},
       eprint = {2308.16183},
 primaryClass = {astro-ph.CO},
       adsurl = {https://ui.adsabs.harvard.edu/abs/2024PhRvD.109f3523L}
}

@article{Kodama:1984ziu,
    author = "Kodama, Hideo and Sasaki, Misao",
    title = "{Cosmological Perturbation Theory}",
    doi = "10.1143/PTPS.78.1",
    journal = "Prog. Theor. Phys. Suppl.",
    volume = "78",
    pages = "1--166",
    year = "1984"
}

\end{document}